\documentclass[aps,prb,twocolumn,superscriptaddress,10pt]{revtex4-2}
\pdfoutput=1 
\usepackage[utf8]{inputenc}
\usepackage[english]{babel}
\usepackage[T1]{fontenc}
\usepackage[pdftex]{graphicx}
\usepackage{amsmath}
\usepackage{cancel}
\usepackage{amsthm}
\usepackage{amssymb}
\usepackage{braket}
\usepackage{xcolor}
\usepackage{dcolumn}
\usepackage{bbm}
\usepackage{bbold}
\usepackage[normalem]{ulem}
\usepackage{footmisc}
\usepackage{xcolor}
\usepackage{booktabs}
\usepackage{comment}
\usepackage{multirow}
\usepackage[colorlinks=true,linkcolor=blue,citecolor=blue,urlcolor=blue,unicode]{hyperref}

\newcommand{\XAGP}{X^{\text{AGP}}}
\newcommand{\XAGPgen}{X^{\text{AGPgen}}}
\newcommand{\Xex}{X_{\text{ex}}}
\newcommand{\Xbo}{X_{\text{bo}}}
\newcommand{\Xbi}{X_{\text{bi}}}
\newcommand{\Xloc}{X_{\text{loc}}}
\newcommand{\Xstep}{X_{\text{step}}}

\newcommand{\sv}[1]{\vec \sigma_{#1}}

\begin{document}
\title{Finite-size generators for weak integrability
breaking perturbations in the Heisenberg chain}
\author{Sara Vanovac}
\affiliation{Department of Physics and Institute for Quantum Information and Matter,
California Institute of Technology, Pasadena, California 91125, USA}
\author{Federica Maria Surace}
\affiliation{Department of Physics and Institute for Quantum Information and Matter,
California Institute of Technology, Pasadena, California 91125, USA}
\author{Olexei Motrunich}
\affiliation{Department of Physics and Institute for Quantum Information and Matter,
California Institute of Technology, Pasadena, California 91125, USA}

\begin{abstract}
An integrable model perturbed by special ``weak integrability-breaking'' perturbations thermalizes at timescales much longer than predicted by Fermi’s golden rule. Recently, a systematic construction of such perturbations based on the so-called long-range deformations of integrable chains was formulated. These perturbations, obtained as truncations of the long-range deformations in some small parameter expansions, can be viewed as produced by unitary rotations of the short-range integrable models. For infinite systems, several ``generators'' (extensive local, boosted, and bilocal operators)  of weak perturbations are known.
The main aim of this work is to understand the appropriate generators in finite systems with periodic boundaries since simple counterparts to boosted and bilocal operators are not known in such cases. We approach this by studying the structure of the adiabatic gauge potential (AGP), a proxy for such generators in finite chains, which was originally introduced as a very sensitive measure of quantum chaos. We prove an exact relation between the AGPs for the boosted and bilocal classes of generators and note that the counterpart to boost does not seem to have a closed analytic form in finite systems but shows quasi-locality nonetheless. We also introduce and study strictly local variants of weak integrability-breaking perturbations.
\end{abstract}

\maketitle

\section{Introduction}
\label{sec:intro}
Understanding thermalization of quantum many-body systems is a fundamental question in quantum statistical mechanics. 
The idea that a closed, isolated, generic quantum many-body system evolving via unitary dynamics from a specific initial state will approach thermal equilibrium for local observables can be understood from the eigenstate thermalization hypothesis (ETH), proposed in Refs.~\cite{Deutsch1991ETH, Srednicki1994ETH}. Systems that obey ETH (or that violate it only in a small fraction of the eigenstates, known as quantum many-body scars \cite{Papic2021review, Moudgalya2021review, Chandran2022review}) are generally referred to as chaotic, while many-body localized and integrable systems violate ETH \cite{Polkovnikov2011,eisert2015quantum,DAlessio2016,Deutsch_2018,santos2018nonequilibrium}.
Integrable systems, which do not thermalize in the usual sense, typically possess a large number of linearly independent and mutually commuting conserved charges $([H_0, Q_\alpha] = [Q_\alpha, Q_\beta] = 0 )$, which are assumed to be extensive local operators.
In this case, thermalization is described by the generalized Gibbs ensemble (GGE) \cite{Rigol2007,Rigol2009,Cassidy2011,Kollar2011,Gring2012,Ilievski2015,Langen2015, Essler_2016, Vidmar_2016}, which preserves the information about the initial expectation values of the conserved charges. 

One of the most important questions in the field of quantum chaos is to understand the behavior of quantum many-body systems under perturbations. In classical systems, the
Kolmogorov-Arnold-Moser (KAM) theorem states that under small perturbations integrable systems can remain stable for sufficiently long times \cite{arnol2013mathematical}. It is a natural question, still largely unanswered, if an analog of the KAM theorem exists for quantum systems \cite{Brandino2015,KAM2022}. The evolution of quantum integrable systems under small integrability breaking perturbations has been actively explored in recent years
\cite{Rabson2004,Jung2006, Mazets2008,Marcuzzi2013,Essler2014,Brandino2015, Bertini2015,Mierzejewski2015, Bertini2016,Tang2018, Friedman2020, LeBlond2021, Bastianello2021, Bulchandani,Durnin2021,kuo2023energy}. Such systems with broken integrability are expected to thermalize as predicted by ETH. However, approximately integrable dynamics may persist for long times before eventual thermalization.

In particular, it has been observed that quantum integrable systems whose integrability is broken by so-called {\it weak integrability-breaking perturbations} have much longer thermalization times than predicted by the Fermi's golden rule \cite{Szasz2021, Kurlov2022, Surace2023}. This unusually long thermalization time has been attributed to the existence of {\it approximate} integrals of motion or {\it quasi-conserved} quantities which commute with the perturbed Hamiltonian up to corrections of order $\lambda^2$, where $\lambda$ is the perturbation strength. 
While these weak integrability-breaking perturbations are a special, non-generic class, they can be of practical importance as they might represent simple, physically motivated perturbations: A prominent example is the spin-1/2 Heisenberg chain perturbed by next-neighbor SU$(2)$-symmetric interaction \cite{Jung2006, Jung2006_2, Kurlov2022, Surace2023, Orlov_2023}.

A systematic approach to constructing weak integrability breaking perturbations was developed and applied in Refs.~\cite{Pozsgay2020, Surace2023}.
This approach is based on long-range deformations of integrable spin chains \cite{Bargheer_2008, Bargheer_2009,Pozsgay2020} which were introduced in the context of AdS/CFT. These deformations can be thought of as unitary rotations of the Hamiltonian and the conserved quantities that preserve the integrability of the model. The idea explored in Ref.~\cite{Surace2023} is that truncating these deformations at finite order produces approximate conserved quantities or quasi-integrals of motion of the perturbed model; hence, the integrability of the original model is effectively broken only at higher order in the perturbation strength. More specifically, the construction is based on defining an operator that generates the long-range deformation. Three types of operators ({\it extensive local}, {\it boosted}, and {\it bilocal}) that lead to quasi-local deformations are currently known. It is still an open question whether all examples of weak integrability breaking are obtained using this approach. Additionally, while this method allows us to construct weak perturbations, it does not serve as a test if a given perturbation breaks integrability in a weak or a strong sense. Unless one can directly generate the perturbations following the procedure from one of the known generators, one cannot tell if a perturbation is weak or strong. 

Recently, a useful advance in this direction has been made by establishing a connection between the adiabatic gauge potential (AGP) \cite{KOLODRUBETZ20171,Sierant2019,Pandey2020,LeBlond2021} and the detection of integrability breaking. In fact, the scaling of the AGP with the system size has been successfully used to distinguish between weak and strong integrability breaking perturbations~\cite{Orlov_2023}.
The intuition is that the AGP plays the role of the generator in the construction of weak integrability breaking perturbations. While the analytic generators are typically defined in the limit of infinite systems, the AGP is obtained numerically from the exact diagonalization for finite system sizes. 
This aspect suggests promising insights into the definition of the generators: In fact, only one of the aforementioned three classes of generators (extensive local) is well-defined for finite systems with periodic boundaries, while the other two classes (boosted and bilocal operators) are only defined for infinite systems, leaving our understanding of the mechanism underlying weak integrability breaking still incomplete. Formulating long-range integrable deformations in finite chains with periodic boundary conditions (PBC) has been an open problem \cite{Gombor2022, Leeuw2023}, and studies of the corresponding AGPs viewed as proxies to the generators could contribute to this problem as well. 

Finally, understanding the properties of the generators can have important implications for the relaxation times and transport properties of such weakly-perturbed integrable systems.
Reference~\cite{Surace2023} presented suggestive arguments that for weak integrability breaking perturbations the relaxation rates become $O(\lambda^4)$ in the perturbation strength $\lambda$.
However, some of these arguments assumed existence of generators with nice locality properties. We therefore aim to study such properties of the AGPs in finite chains.
We expect that locality properties of the generators are also needed to understand transport properties under weak perturbations that have not been considered so far.

In the present work, we investigate the capability of the AGP as a tool for detecting various types of weak and strong integrability breaking perturbations.
We examine the scaling of the norm of AGP with system size for one of the most studied integrable spin chains -- the spin $1/2$-Heisenberg chain -- for different types of perturbations.
We consider both extensive translationally invariant perturbations (including many new examples extending the results of Ref.~\cite{Surace2023}) and strictly local (impurity-like) perturbations.
We note that the strictly local weak-integrability breaking perturbations that we consider represent novel classes that have not been studied before. In an effort to construct well defined equivalents of the boosted and bilocal operators for finite chains with periodic boundaries, and to provide useful insights for understanding thermalization times, we also study the operatorial content of the AGP. We focus in particular on locality properties of AGP, considering both the $k$-locality (support) and geometric locality (range) of the terms in the AGP. We find exact relations between the AGPs of different perturbations that explain some of the observed scalings and locality properties.

The paper is organized as follows.
In Section~\ref{sec:agp} we formally introduce AGP, different classes of weak integrability breaking, and our integrable model of choice: the spin-1/2 XXX Heisenberg chain. In Section~\ref{sec:AGPprops} we discuss some of the formal properties of AGP and introduce the quantities of interest for the subsequent numerical analysis.
In Sections~\ref{sec:transl} and \ref{sec:local} we present the results of characterizing the AGP respectively for extensive translationally invariant perturbations and for strictly local perturbations. In  Section~\ref{sec:conclusions} we summarize our findings and discuss future directions. In Appendices~\ref{app:sim_diag} and \ref{app:brute-force-Xdiag} we detail methodological developments generalizing the notion of AGP to resolve ambiguities and find nicest possible generators. In Appendix~\ref{app:oipr} we provide additional characterization of the AGPs using operator participation ratio, while in Appendix~\ref{app:add_data_for_agp_global} we present additional supporting data.

\section{Preliminaries: AGP and weak integrability breaking generators}
\label{sec:agp}
We start by considering an integrable Hamiltonian $H_0$ that commutes with a set of integrals of motion (IoMs or charges) $\{Q_\alpha^{(0)}\}$, which also mutually commute, i.e., $[H_0, Q_\alpha^{(0)}] = 0$ for every $\alpha = 1, 2, \dots$ and $[Q_\alpha^{(0)}, Q_\beta^{(0)}] = 0$ for every pair of positive integers $\alpha,\beta$. We then perturb $H_0$ with a perturbation $\lambda V$, where $\lambda$ is a small parameter and obtain the perturbed Hamiltonian $H$:
\begin{align}
H = H_0 + \lambda V ~.
\end{align}
Suppose there exists an operator $X$ such that
\begin{equation}
\label{eq:V}
V = i[X, H_0] ~.
\end{equation}
Then we can define a ``correction'' of an IoM $Q_\alpha^{(0)}$ as 
\begin{equation}
\label{eq:Q1}
Q_\alpha^{(1)} = i[X, Q_\alpha^{(0)}] ~.
\end{equation}
With this definition, we get
\begin{align}
& [H_0 + \lambda V, Q_\alpha^{(0)} + \lambda Q_\alpha^{(1)}] \nonumber\\
& \qquad = \lambda([H_0, Q_\alpha^{(1)}] - [Q_\alpha^{(0)}, V]) + O(\lambda^2) \nonumber\\
& \qquad = i\lambda([H_0, [X, Q_\alpha^{(0)}]] - [Q_\alpha^{(0)}, [X, H_0]]) + O(\lambda^2) \nonumber\\
& \qquad =O(\lambda^2) ~. \label{eq:quasiIom}
\end{align}

If $Q_\alpha^{(1)}$ is a local (extensive) operator, then Eq.~(\ref{eq:quasiIom}) implies that we can interpret the operator $Q_\alpha^{(0)} + \lambda Q_\alpha^{(1)}$ as a quasi-IoM that is conserved under $H$ up to corrections of order $\lambda^2$.
In this case, we say that $V$ is a weak integrability-breaking perturbation (to first order in $\lambda$), and $X$ is the generator.

\subsection{Adiabatic Gauge Potential}
To understand whether an operator $X$ that satisfies Eq.~(\ref{eq:V}) exists, it is convenient to consider an eigenbasis $\{\epsilon_n, \ket{n}\}$ of $H_0$:
We define $\XAGP \equiv \sum_{n,m} \XAGP_{nm} \ket{n}\bra{m}$, with matrix elements
\begin{equation}
\label{eq:AGPod}
\XAGP_{nm} = 
\begin{cases}
\frac{V_{nm}}{i (\epsilon_m - \epsilon_n)}, & \text{if $\epsilon_n \neq \epsilon_m$} \\
0, & \text{if $\epsilon_n = \epsilon_m$}
\end{cases} ~.
\end{equation}
We refer to this as AGP~\footnote{Compared to the standard AGP definition $\mathcal{A}$ in the literature~\cite{Pandey2020}, our operator has the opposite sign, $\XAGP = -\mathcal{A}$.}.
With this definition, we find 
\begin{equation}
\label{eq:offdiag}
i[\XAGP, H_0] = \sum_{n, m: \epsilon_n \neq \epsilon_m} V_{nm} \ket{n}\!\bra{m}  = V - V_{\text{diag}} ~.
\end{equation}
We thus recover Eq.~(\ref{eq:V}) up to an operator $V_{\text{diag}} = \sum_{n, m: \epsilon_n = \epsilon_m} V_{nm} \ket{n}\!\bra{m}$ that has block-diagonal structure set by the eigenspaces (possibly multi-dimensional due to degeneracies) of $H_0$.

The norm of the operator $\XAGP$ is
generally expected to grow exponentially with the system size. This is a consequence of the vanishing denominators in Eq.~(\ref{eq:AGPod}): The gaps $\epsilon_{n+1} - \epsilon_n$ between consecutive levels typically scale as $D^{-1}$ (where $D$ is the Hilbert space dimension), while the matrix elements $V_{nm}$ of a generic perturbation typically scale as $D^{-1/2}$. Therefore, the corresponding matrix elements of $\XAGP$ are expected to scale as $D^{1/2}$. With this exponential scaling in system size, it may seem hopeless to find an extensive local operator $Q_\alpha^{(1)}$ from Eq.~(\ref{eq:Q1}).

There are, however, some notable exceptions to this exponential scaling. For example, we can consider $X$ that is extensive local by construction and define $V$ and $Q_\alpha^{(1)}$ from Eqs.~({\ref{eq:V}) and ({\ref{eq:Q1}) using $X$: Then both $V$ and $Q_\alpha^{(1)}$ are also extensive local by construction and $V$ is a weak integrability-breaking perturbation. In this case $\XAGP_{nm} = X_{nm}$ for $\epsilon_n \neq \epsilon_m$; the issue of very small denominators is resolved by the matrix elements $V_{nm}$ being special (non-generic) and correspondingly small in this case.}

Extensive local $X$ are not the only class of operators that generate weak integrability-breaking perturbations in the sense of Eq~(\ref{eq:quasiIom}). References~\cite{Bargheer_2008, Bargheer_2009, Pozsgay2020} showed that two other non-trivial types of operators generate good quasi-IoMs in infinite systems.
We will review their construction in the following section.

\subsection{Generators of weak integrability-breaking perturbations in an infinite chain from long-range deformations}
\label{sec:generators}
In this section, we focus on the infinite chain and introduce three currently known classes of operators that can be used to generate weak perturbations as in Eqs.~(\ref{eq:V}) and (\ref{eq:Q1}).
These classes have been introduced in the context of long-range integrable deformations of integrable models \cite{Bargheer_2008,Bargheer_2009,Pozsgay2020, Szasz2021,Surace2023} and have been used to define weak integrability-breaking perturbations for finite (but arbitrary) orders in $\lambda$. 
Here we will only consider the truncation to the linear order and we refer to previous works for the general case.

In order to define the different classes of generators, it is useful to write the IoMs as sums of local operators
\begin{equation}
Q_\alpha^{(0)} = \sum_j q_{\alpha,j}^{(0)} ~,
\end{equation}
where $j$ labels the lattice sites and the {\it charge density} operator $q_{\alpha,j}^{(0)}$ has finite support around $j$.
The commutation relations $[Q_\alpha^{(0)}, Q_\beta^{(0)}] = 0$ imply the existence of continuity equations of the form
\begin{equation}
\label{eq:current}
i[Q_\alpha^{(0)}, q_{\beta,j}^{(0)}] = J_{\beta\alpha, j} - J_{\beta\alpha, j+1} ~,
\end{equation}
where the {\it generalized current} $J_{\beta\alpha, j}$ is a local operator with finite support around $j$.

In the Heisenberg model context below, $Q_2^{(0)}$ usually refers to the Hamiltonian itself, while $Q_{\alpha \geq 3}^{(0)}$ are the non-trivial IoMs; we will use the convention $H_0 = 2Q_2^{(0)}$.

\subsubsection{Extensive perturbations}
We now consider various classes of operators $X$ that generate an extensive local translationally-invariant perturbation $V = i[X, H_0]$, i.e., a perturbation that is a sum over lattice sites of local operators.

\textbf{Extensive local generators}.
The operator $X$ can be an arbitrary translationally-invariant operator of the form
\begin{equation}
\label{eq:ex}
X_\text{ex} = \sum_j O_j ~.
\end{equation}
Then $V_{\text{ex}} = i[X_{\text{ex}}, H_0]$ (and similarly for any IoM correction $Q_\alpha^{(1)}$) is also a translationally-invariant sum of local operators.

\textbf{Boosted generators.}
Given an IoM $Q_\beta^{(0)}$, we can define the boosted operator (using the same convention as in Ref.~\cite{Surace2023})
\begin{equation}
\label{eq:Xbo}
X_\text{bo} = -\mathcal{B}[Q_\beta^{(0)}] ~, \qquad \mathcal{B}[Q_\beta^{(0)}]\equiv \sum_j j q_{\beta,j}^{(0)} ~.
\end{equation}
With this definition, from Eq.~(\ref{eq:current}) we find
\begin{equation}
i[X_\text{bo}, Q_\alpha^{(0)}] = \sum_j J_{\beta\alpha, j} \equiv J_{\beta\alpha; \text{tot}} ~.
\label{eq:Jtot}
\end{equation}
This proves that the corrections to the IoMs, and, in particular, the Hamiltonian perturbation
\begin{equation}
V_\text{bo} = i[X_\text{bo}, H_0] = 2i[X_\text{bo}, Q_2^{(0)}] = 2J_{\beta,2;\text{tot}} ~,
\label{eq:Vbo_beta}
\end{equation}
are extensive local operators, as desired.
At times, we will refer to this perturbation also as $V_{\text{bo}}^\beta$.

\textbf{Bilocal generators.}
From two IoMs $Q_\beta^{(0)}$ and $Q_\gamma^{(0)}$ we define a bilocal operator as follows~\cite{Pozsgay2020, Surace2023}:
\begin{equation}
\label{eq:Xbi}
X_{\text{bi}} = [Q_\beta^{(0)}|Q_\gamma^{(0)}] \equiv \sum_{j<k} \{q_{\beta,j}^{(0)}, q_{\gamma,k}^{(0)} \} + \frac{1}{2} \sum_j \{ q_{\beta,j}^{(0)}, q_{\gamma,j}^{(0)} \} ~.
\end{equation}
Using Eq.~(\ref{eq:current}), we get
\begin{multline}
\label{eq:bilocal}
i\big[X_\text{bi} , Q_\alpha^{(0)}\big]
=\frac{1}{2}\sum_j\left(\{q_{\gamma,j}^{(0)}, J_{\beta\alpha,j}+J_{\beta\alpha,j+1}\}\right.
\\
\left. -\{q_{\beta,j}^{(0)}, J_{\gamma\alpha,j}+J_{\gamma\alpha,j+1}\}\right).
\end{multline}
We then find that, also in this case, the corrections to the IoMs, including the perturbation
\begin{equation}
V_\text{bi} = i[X_\text{bi}, H_0] = 2i[X_\text{bi}, Q_2^{(0)}] = 2i[[Q_\beta^{(0)}|Q_\gamma^{(0)}], Q_2^{(0)}] ~,
\label{eq:Vbi_betagamma}
\end{equation}
are extensive local operators.
At times, we refer to this perturbation also as $V_{\text{bi}}^{\beta\gamma}$.

\subsubsection{Local perturbations}
Here we consider some classes of operators that generate strictly local perturbations or ``local impurities,'' i.e., perturbations that have support on a finite region around a site $j_0$.

\textbf{Strictly local generators.} 
We first consider the case of a strictly local generator
\begin{equation}
X_{\text{loc}} = O_{j_0} ~,
\label{eq:Xloc}
\end{equation}
where $O_{j_0}$ has a finite support around the site $j_0$.
Clearly, all the corrections to the IoMs, as well as the Hamiltonian perturbation $V_{\text{loc}} = i[X_{\text{loc}}, H_0]$, are strictly local perturbations.

\textbf{Discontinuous step generators.} 
As a different class of a generator, given an IoM $Q_\beta^{(0)}$ we now define the step generator:
\begin{equation}
X_{\text{step}} = -\sum_{j \geq j_0} q_{\beta,j}^{(0)} ~.
\label{eq:Xstep}
\end{equation}
From Eq.~(\ref{eq:current}) we find
\begin{equation}
i[X_{\text{step}}, Q_\alpha^{(0)}] = J_{\beta\alpha,j_0} ~,
\end{equation}
which is a stricly local operator with support around $j_0$.
Therefore,
\begin{equation}
V_\text{step} = i[X_{\text{step}}, H_0] = 2i[X_{\text{step}}, Q_2^{(0)}] = 2J_{\beta,2;j_0}
\label{eq:Vstepbeta}
\end{equation}
is also a strictly local operator, which we will at times refer to as $V_\text{step}^\beta$.

\subsection{Weak integrability breaking generators in finite chains and $X^{\text{AGP}}$ as a proxie for $X_{\text{bo}/\text{bi}}$}
\label{sec:proxie}
The generators defined in Sec.~\ref{sec:generators} for infinite chains allow us to construct various types of weak integrability-breaking perturbations, with quasi-IoMs conserved up to $O(\lambda^2)$.
For these weak perturbations, the existence of quasi-IoMs satisfying $[H_0 + \lambda V, Q_\alpha^{(0)} + \lambda Q_\alpha^{(1)}] = O(\lambda^2)$ can be easily proved by a direct evaluation of the commutators: 
Since all the operators involved are extensive local, the results hold similarly for infinite and finite chains with periodic boundary conditions (provided that the system size is larger than the range of the terms in $Q_\alpha^{(0)}$ and $Q_\alpha^{(1)}$).

Despite this, with the exception of the extensive local and strictly local ones, the generators defined in Sec.~\ref{sec:generators} for an infinite chain do not have a simple counterpart for finite chains with periodic boundary conditions.
In these cases, there is no recipe for finding an explicit expression for an operator $X$ that satisfies Eq.~(\ref{eq:V}) in a finite chain.
We can nevertheless construct numerically the AGP defined in Sec.~\ref{sec:agp}: 
If such an operator $X$ exists, it has to coincide with $\XAGP$ up to an operator that has the block-diagonal structure set by the eigenspaces of $H_0$.

In this work, we would like to understand detailed properties of the extracted $\XAGP$ for such perturbations, which is important for a number of reasons we listed in the introduction. The numerical methods used to extract $\XAGP$ can be readily tested for perturbations $V_{\text{ex/loc}}$ generated by extensive or strictly local generators: 
In these cases, the generators $X_{\text{ex/loc}}$ are well-defined also for finite chains, providing an exact benchmark for our methods.
This allows us to understand subtleties of the numerical computation of $\XAGP$ and to determine conditions and additional methods for recovery of exact $X$.

\subsection{Heisenberg spin-1/2 chain: notation and review}
\label{subsec:Heischain}
In this work, we focus on the spin-1/2 Heisenberg chain (also known as the XXX model), which is a well known integrable model with the following Hamiltonian:
\begin{equation}
\label{eq:h0}
H_0 = \sum_{j=1}^L \vec{\sigma}_j \cdot \vec{\sigma}_{j+1} ~,
\end{equation}
where $\vec{\sigma}_j = (\sigma_j^x,\sigma_j^y,\sigma_j^z)$ is the vector of Pauli operators on site $j$.
$H_0$ is a translational, time reversal, inversion, and $SU(2)$ invariant sum of local operators acting on two sites.

The Hamiltonian $H_0$ commutes with an extensive set of IoMs $Q_\alpha^{(0)}$, $\alpha = 2, 3, \dots$.
We show here the local densities for the first few IoMs: 
\begin{align}
& q_{2,j}^{(0)} = \frac{1}{2} \vec{\sigma}_j \cdot \vec{\sigma}_{j+1} ~, \\
& q_{3,j}^{(0)} = -\frac{1}{2} (\vec{\sigma}_j\times \vec{\sigma}_{j+1}) \cdot \vec{\sigma}_{j+2} ~, \\
& q_{4,j}^{(0)} = [(\vec{\sigma}_{j-1} \times \vec{\sigma}_j) \times \vec{\sigma}_{j+1}] \cdot \vec{\sigma}_{j+2} - 2\vec{\sigma}_j \cdot \vec{\sigma}_{j+1} + \nonumber \\
& \qquad + \frac{1}{2} \vec{\sigma}_{j-1} \cdot \vec{\sigma}_{j+1} + \frac{1}{2} \vec{\sigma}_j \cdot \vec{\sigma}_{j+2} ~.
\end{align}
Note that $H_0 = 2Q_2^{(0)}$ and our conventions for $\{ Q_\alpha^{(0)}$\} are the same as in Refs.~\cite{Pozsgay2020, Surace2023} for easy reference.
Our choice for the density $q_{4,j}^{(0)}$ differs from Ref.~\cite{ Surace2023} to have an expression that is invariant under inversion at a point, here chosen to be bond center between $j$ and $j+1$.
This will be convenient when discussing lattice symmetry properties of $V$s and AGPs.

\section{Calculation and characterization of AGPs}
\label{sec:AGPprops}
In this section, we collect some observations about symmetry properties of AGPs and discuss practical considerations when numerically calculating and characterizing AGPs.

\subsection{Formal properties of $\XAGP$}
\label{subsec:propsAGP}
The AGP operator, Eq.~(\ref{eq:AGPod}), can be written as
\begin{align}
\XAGP[V] &= \sum_{n, m: \epsilon_n \neq \epsilon_m} \ket{n} \frac{\bra{n} V \ket{m}}{i(\epsilon_m - \epsilon_n)} \bra{m} \\
&= \sum_{\nu, \mu: \epsilon_\nu \neq \epsilon_\mu} \frac{P_\nu V P_\mu}{i(\epsilon_\mu - \epsilon_\nu)} ~,
\end{align}
where, e.g., Greek letter $\nu$ refers to a full eigenspace associated with an eigenvalue $\epsilon_\nu$ of $H_0$ and $P_\nu = \sum_{n, \epsilon_n = \epsilon_\nu} \ket{n}\bra{n}$ is the corresponding projector.
The block-diagonal part of $V$ with respect to the blocks defined by the eigenspaces of $H_0$ is $V_{\text{diag}} = \sum_\nu P_\nu V P_\nu$, while the AGP captures the off-diagonal part $V - V_{\text{diag}}$, see Eq.~(\ref{eq:offdiag}).
We note also that $(\XAGP)_{\text{diag}} = 0$ by construction.

The above operator form of $\XAGP$ immediately shows the basis-independence of the numerically calculated $\XAGP$ when dealing with degenerate energies and also allows deducing its symmetries from the properties of $H_0$ and $V$.
Specifically, suppose $H_0$ has a unitary symmetry $U$, $U H_0 U^\dagger = H_0$; it follows that $U P_\nu U^\dagger = P_\nu, \forall \nu$.
Further, suppose that $U V U^\dagger = u V$, e.g., $u = \pm 1$ for $V$ even/odd under $U$; we obtain $U \XAGP U^\dagger = u \XAGP$, i.e., $\XAGP$ has the same transformation properties under $U$ as $V$.
For example, in many cases both $H_0$ and $V$ have the lattice translation symmetry and the spin $SU(2)$ symmetry, hence the same is true about the calculated $\XAGP$.
For any other unitary symmetry of $H_0$, say lattice inversion, the symmetry of $\XAGP$ follows that of $V$.
On the other hand, for an anti-unitary symmetry $\Theta$ like the physical time reversal, $\Theta H_0 \Theta^{-1} = H_0$, $\Theta i \Theta^{-1} = -i$ (transforming physical spins as $\Theta \vec{\sigma}_j \Theta^{-1} = -\vec{\sigma}_j$), if $\Theta V \Theta^{-1} = \theta V$, then $\Theta \XAGP \Theta^{-1} = -\theta \XAGP$; that is, $\XAGP$ transforms oppositely under the time reversal compared to $V$.

Using similar arguments, for any integral of motion $Q_\alpha^{(0)}$ we have $Q_\alpha^{(0)} P_\nu = P_\nu Q_\alpha^{(0)}$, and hence
\begin{equation}
\XAGP\left[V + \sum_\alpha c_\alpha Q_\alpha^{(0)} \right] = \XAGP[V] ~;
\label{eq:Xagp_V}
\end{equation}
that is, adding any linear combination of IoMs to $V$ does not change the calculated AGP.

Finally, we note that
\begin{equation}
\label{eq:XAGPdiag}
\left[ \XAGP + \sum_\nu P_\nu A_\nu P_\nu, H_0 \right] = [\XAGP, H_0] ~,
\end{equation}
where $A_\nu = \sum_{n, n': \epsilon_n =  \epsilon_{n'} = \epsilon_\nu} a_{nn'} \ket{n}\bra{n'} = P_\nu A_\nu P_\nu$ is an arbitrary operator acting in the eigenspace of $H_0$ corresponding to the eigenvalue $\epsilon_\nu$.
That is, as far as solving for $X$ satisfying $i[X, H_0] = V$ is concerned, we can always add to $X$ such an arbitrary block-diagonal operator in the eigenbasis of $H_0$ (and the converse is also true, since $\sum_\nu P_\nu A_\nu P_\nu$ captures all operators that commute with $H_0$).
Recalling $(\XAGP)_{\text{diag}} = 0$,
in particular, $\XAGP$ might be not the simplest form (e.g., as far as representation in terms of local operators is concerned), although in some cases natural symmetry conditions on $X$ can enforce $A_\nu = 0$, removing this ambiguity in extracting $X$.

\subsection{Regularized AGP formulation and AGP norm}
Calculation of the AGP matrix elements involves energy denominators that can be exponentially small in system size.
Since such energy spacings have some inherent randomness to them (e.g., going from one pair of levels to another or from one system size to the next), the corresponding contributions to AGP properties can be very noisy.
To reduce such noise, Ref.~\cite{Pandey2020} introduced a regularized version of the AGP defined by the following replacement, for each $\XAGP_{nm}$ with energy denominator $\omega_{mn} = \epsilon_m - \epsilon_n$,
\begin{equation}
\label{eq:freg}
\frac{1}{\omega_{mn}} \quad \longrightarrow \quad f_{\text{reg.}}(\omega_{mn}) = \frac{\omega_{mn}}{\omega_{mn}^2 + \mu^2} ~,
\end{equation}
where $\mu$ is a small energy scale on the order of the energy level spacing (precise choices will be described later).
The regularized version of the AGP also reasonably treats possible degeneracies in the spectrum of $H_0$:
In numerical calculations, a degenerate pair of levels often appears as non-degenerate due to roundoff errors, but the regularized AGP version will properly suppress the corresponding matrix elements as long as $\mu$ is much larger than the roundoff error.
For some calculations when comparing with analytical results, to avoid any numerical dependence on the scale $\mu$, we also use a step regulator
\begin{equation}
f_{\text{reg.; step}}(\omega_{mn}) = 
\begin{cases}
\frac{1}{\omega_{mn}}, & \text{if $|\epsilon_m - \epsilon_n| > \mu$} \\
0, & \text{otherwise}
\end{cases} ~,
\label{eq:fregstep}
\end{equation}
with $\mu$ chosen to be much smaller than the typical level spacing but much larger than the typical roundoff error for degenerate eigenvalues (such separation appears to still work for our system sizes). Note that all our statements about the formal symmetry properties of $\XAGP$ continue to hold also for the regularized versions, since the regulators depend only on the energies and not on the eigenbasis choices for degenerate eigenvalues.

For our initial characterization of an AGP, we compute its Frobenius norms using convention
\begin{equation}
\label{eq:agp_norm}
\| \XAGP \|^2 \equiv \frac{1}{D} \sum_{n,m} |\XAGP_{nm}|^2 ~,
\end{equation}
where $D$ is the total Hilbert space dimension, e.g., $D = 2^L$ for a spin-1/2 chain of size $L$.
For an interacting system, the spacing between energy levels scales as $\epsilon_{n+1} - \epsilon_n \sim 1/D$, assuming non-degenerate levels for simplicity.
On the other hand, the corresponding off-diagonal matrix elements of a generic local operator scale as $V_{nm} \sim 1/D^{1/2}$, ignoring any polynomial factors in $L$.
Hence, $\XAGP_{n,n+1} \sim D^{1/2}$ in this case, and the AGP norm is at least $\| \XAGP \|^2 > \frac{1}{D} \sum_n |\XAGP_{n,n+1}|^2 \sim D \sim e^{\kappa L}$ with $\kappa = \ln 2$ (the last equation is for the spin-1/2 chain).
Such a scaling of the AGP norm is a signature that the perturbation is indeed a generic one, which we call ``strong perturbation.''

In contrast, for the special weak perturbations we expect only polynomial scaling of the AGP norm with the system size.
This is manifest when $\XAGP$ is explicitly known, like in the case of a perturbation generated using an extensive local generator $\Xex$:
By thinking in the Pauli string basis we readily obtain $\| \Xex \|^2 \sim L$.

We also expect polynomial scaling for perturbations generated using $\Xbo$ and $\Xbi$, where $\XAGP$ in finite chains is not known and hence the precise scaling is also not known; these cases will be the object of our studies. More generally, scaling of the AGP norm can be used to detect if a perturbation is weak even when no prior knowledge about its generator is available. 

Previous studies of the AGP norms for perturbations of the Heisenberg chain~\cite{Pandey2020, Orlov_2023} considered $\XAGP$ restricted to the $S^z_{\text{tot}} = 0$ sector and calculated the corresponding norm defined by Eq.~(\ref{eq:agp_norm}) with $n, m$ restricted to this sector and $D$ replaced by the dimension of this sector:
\begin{equation}
\label{eq:agp_norm_Sztot0}
\| \XAGP \|^2_{S^z_{\text{tot}} = 0} \equiv \frac{1}{D_{S^z_{\text{tot}} = 0}} \sum_{n, m: S^z_{\text{tot}} = 0} |\XAGP_{nm}|^2 ~.
\end{equation}
In order to access larger system sizes and stay consistent with the previous studies, the AGP norms reported in Figs.~\ref{fig:agp-both-global} and \ref{fig:agp-both-local} are computed in the zero magnetization sector as given by Eq.~(\ref{eq:agp_norm_Sztot0}).
Such restrictions do not change the exponential scaling of the AGP norm in the strong perturbation cases.
As far as the power of the polynomial scaling in the weak perturbation cases, we expect that it will not change either. 

A non-rigorous argument is as follows.
First, $V$ does not connect different $S^z_{\text{tot}}$ sectors, so thinking about the sectors separately and then combining them is appropriate.
Second, we expect that the increase in the number of contributions to the AGP norm when calculated in the entire Hilbert space (i.e., including all sectors) is roughly compensated by the corresponding increase in the Hilbert space dimension used in the denominator.
We numerically verify that $\| \XAGP \|^2_{S^z_{\text{tot}} = 0}$ gives correct scaling for the perturbations generated by $\Xex$ where analytical scaling in the entire Hilbert space is known.
For all of the weak perturbations studied, using somewhat smaller sizes, we also compare the AGP norm in the full Hilbert space to the one restricted to the $S^z_{\text{tot}} = 0$ sector and confirm that the scaling is not affected by this restriction.

\subsection{Characterization of the AGP in the Pauli string basis}
\label{subsec:agp_PauliS}
We would like to study the structure of AGPs in more detail, beyond just their norm scalings.
To this end, in each case we also calculate the full AGP (i.e., including all $S^z_{\text{tot}}$ sectors) and expand it over the Pauli string basis
\begin{equation}
\XAGP = \sum_{S=[\rho_1,\rho_2,\dots,\rho_L]} c_{\rho_1\rho_2\dots\rho_L} \sigma_1^{\rho_1} \sigma_2^{\rho_2} \dots \sigma_L^{\rho_L} ~,
\label{eq:PauliSbasis}
\end{equation}
$\rho_j \in \{0, x, y, z \}$ with $\sigma^0$ denoting the identity $2 \times 2$ matrix.
Symmetries impose constraints on the Pauli string amplitudes.
For example, if $\XAGP$ is even under the time reversal, then it can contain only strings with even numbers of non-trivial Pauli's, while if it is odd, then only odd number of Pauli's.
Other symmetry quantum numbers like under the lattice inversion impose specific relations on the Pauli string amplitudes, which we watch for in our analysis.

We can also express the AGP norm as $\| \XAGP \|^2 = \sum_S |c_S|^2$, and hence we can characterize the relative importance of each Pauli string $S$ by the corresponding contribution to the AGP norm, defining the corresponding weight (fraction of the norm) as
\begin{equation}
w_S = \frac{|c_S|^2}{\| \XAGP \|^2} ~, \qquad \sum_S w_S = 1 ~.
\label{eq:wS}
\end{equation}

For a given Pauli string $S=[\rho_1,\rho_2,\dots,\rho_L]$, we define its non-trivial support as
\begin{equation}
\text{supp}(S) = \{j, \rho_j \neq 0\} ~.
\end{equation}
We will often refer to the size of $\text{supp}(S)$ as simply support of $S$.
In the simplest characterization, we will measure the weight of all Pauli strings with a given support size $k$,
\begin{equation}
\label{eq:wsuppk}
w[\text{support}~k] = \sum_{S,~ |\text{supp}(S)| = k} w_S ~.
\end{equation}

When studying the spatial structure of the contributing Pauli strings, we will define the spatial range of $S$ in the PBC chain of size $L$ as 
\begin{equation}
\text{range}(S) = L - \max\{h: \exists j, \rho_j \!=\! \rho_{j+1} \!=\! \dots \!=\! \rho_{j+h-1} \!=\! 0 \},
\end{equation}
i.e., $L$ minus the size of the largest ``hole'' (segment of consecutive identity matrices) in the string.
For a short spatially localized string $S = [0, \dots, 0, \rho_i, \rho_{i+1}, \dots, \rho_{i+r-1}, 0, \dots, 0]$ with non-trivial $\rho_i$ and $\rho_{i+r-1}$ (while $\rho_\ell$} can be arbitrary for $i < \ell < i+r-1$), $r < L/2$, the above definition coincides with the natural notion of the range, $\text{range}(S) = r$, while it also works reasonably for arbitrary strings in PBC.
In the corresponding spatial characterization, we will measure the weight of all Pauli strings with a given range $r$,~
\begin{equation}
\label{eq:wranger}
w[\text{range}~r] = \sum_{S,~ \text{range}(S) = r} w_S ~.
\end{equation}

\section{AGP studies of extensive translationally invariant perturbations}
\label{sec:transl}
In this section, we present all of the extensive perturbations studied, a number of which have not appeared in the literature before, including a few strong integrability-breaking ones. While our main focus is to understand the generators of boost-type perturbations, studying weak perturbations with extensive local and bilocal generators provides additional insight into our original question. We first consider those perturbations that preserve the translational invariance of the original Hamiltonian as opposed to perturbations that are strictly local, which we will discuss in a later section. Since the strong perturbations, which we will discuss first, are very sensitive to small energy denominators we used the regularized AGP and introduced the regulator cutoff following Ref.~\cite{Pandey2020}. On the other hand, we have confirmed numerically that the extracted AGPs for weak perturbations are not sensitive to the choice of the cutoff, showing only small quantitative variation for a range of $\mu$. In fact, we obtain similar results also using the step regulator, Eq.~(\ref{eq:fregstep}), that for judicious choices of $\mu$ is independent of $\mu$ corresponding to using precisely the original AGP definition, which is preferred in some contexts.

\subsubsection{Strong integrability-breaking perturbations}
We start with generic extensive perturbations. Since most generic perturbations will break integrability strongly, this allows us to establish a baseline to which we compare the behavior of weak perturbations later on. Perturbations that break integrability strongly will all behave similarly. Perturbations of the type $\sum_j \vec{\sigma}_j \cdot \vec{\sigma}_{j+m}$ with $m={2,3,4,5}$ were previously studied in Ref.~\cite{Orlov_2023}, where the authors found that all except the case $m=2$ are strong perturbations that break the integrability. We will discuss why the second-neighbor perturbation is special in the next section.

The two extensive strong integrability-breaking perturbations that we study here are:
\begin{align}
& V_{\text{s},1} = \sum_j \vec{\sigma}_j \cdot \vec{\sigma}_{j+3} \label{eq:Vs1}~; \\
& V_{\text{s},2} = \sum_j (-1)^j \vec{\sigma}_j \cdot \vec{\sigma}_{j+1}\label{eq:Vs2} ~.
\end{align}

$V_{\text{s},1}$ has the same symmetry properties as the Hamiltonian and the corresponding $\XAGP$ is expected to be odd under the time reversal, so only Pauli strings with odd support have non-zero contributions to $\XAGP$.

On the other hand, $V_{\text{s},2}$ breaks the translation symmetry and is well defined only for an even number of sites; it is even under the time reversal, so the corresponding $\XAGP$ is also odd under the time reversal.
Additionally, $V_{\text{s},2}$ was recently considered in Ref.~\cite{DeNardis_2021} while studying the robustness of the superdiffusion in the Heisenberg spin chain to integrability-breaking perturbations. 
From the finite-size results in Ref.~\cite{DeNardis_2021}, the super-diffusion appeared to persist at least for small perturbation strength, which was suggestive of some robustness to the integrability-breaking.
However, the measure we will present here shows that $V_{\text{s},2}$ is actually a strong perturbation that breaks integrability.

\subsubsection{Weak perturbations generated by $\Xex$}
\label{sec:weak-Xex-global}
Perturbations generated by $\Xex$ have exactly known generators and thus establish a baseline for the methods we employ. As we will show in Sec.~\ref{subsec:agp_PauliS2}, our ability to recover an exact $X$ depends on the symmetries of $V$, in particular on whether or not $X_{\text{diag}}$ is zero.
We present here three perturbations generated from extensive local $\Xex$ with different $\Theta$ (time reversal) and $I$ (lattice inversion) symmetries:
\begin{align}
V_{\text{ex},1} &= 2\sum_j (\vec{\sigma}_j \times \vec{\sigma}_{j+3}) \cdot (\vec{\sigma}_{j+1} - \vec{\sigma}_{j+2}) ~, \label{eq:Vex1}\\
V_{\text{ex},2} &= 2\sum_j \big[ 4 \sv{j} \cdot (\sv{j+3} - \sv{j+2}) \notag\\
&\qquad\qquad + ((\sv{j}\times \sv{j+1})\times\sv{j+2})\cdot\sv{j+4} \notag\\
&\qquad\qquad + ((\sv{j}\times \sv{j+2})\times\sv{j+3})\cdot \sv{j+4} \notag\\
&\qquad\qquad - 2 ((\sv{j}\times \sv{j+1})\times\sv{j+3})\cdot\sv{j+4} \notag\\
&\qquad\qquad + 2 ((\sv{j}\times \sv{j+2})\times\sv{j+1})\cdot \sv{j+3} \big] ~, \label{eq:Vex2}\\
V_{\text{ex},3} &= 2\sum_j \big[ ((\sv{j}\times \sv{j+1})\times \sv{j+2})\cdot \sv{j+4} \notag\\
&\qquad\qquad - ((\sv{j}\times \sv{j+2})\times \sv{j+3})\cdot \sv{j+4} \big] ~. \label{eq:Vex3}
\end{align}
It is easy to see that $V_{\text{ex},1}$ is odd under $\Theta$ and even under $I$, $V_{\text{ex},2}$ is even under both $\Theta$ and $I$, and $V_{\text{ex},3}$ is even under $\Theta$ but odd under $I$.
The corresponding generators that work on any finite PBC chain are
\begin{align}
X_{\text{ex},1} &= \sum_j \vec{\sigma}_j \cdot \vec{\sigma}_{j+2} ~, \\
X_{\text{ex},2} &= \sum_j (\vec{\sigma}_j \times \vec{\sigma}_{j+3}) \cdot (\vec{\sigma}_{j+1} - \vec{\sigma}_{j+2}) ~, \label{eq:Xex2}\\
X_{\text{ex},3} &= \sum_j (\vec{\sigma}_j \times \vec{\sigma}_{j+3}) \cdot (\vec{\sigma}_{j+1} + \vec{\sigma}_{j+2}) ~.
\end{align}
These are examples with an exact $X$ that we sometimes refer to. Since the generator is known, we can directly compare it to the result from our procedure that tries to recover it by computing AGP numerically and expanding in the Pauli string basis.
This is not true for the boost and bilocal cases as we will explain next.

\subsubsection{Weak perturbations generated by $\Xbo$}
Our main interest are perturbations generated by $\Xbo$ defined in Eq.~(\ref{eq:Xbo}).
Following our discussion in the previous section, we know that adding a linear combination of IOMs to $V$ does not change the $\XAGP$, see Eq.~(\ref{eq:Xagp_V}).
We take advantage of this fact and use it to present particularly simple perturbations that have boosted operators as their generator:
\begin{align}
V_{\text{bo},1} &= \sum_j \vec{\sigma}_j \cdot \vec{\sigma}_{j+2} = 2J_{3,2;\text{tot}} + 8 Q_2^{(0)} + Q_4^{(0)} ~, \label{eq:Vbo1} \\
V_{\text{bo},2} &= \sum_j (\vec{\sigma}_j \times \vec{\sigma}_{j+3}) \cdot (\vec{\sigma}_{j+1} + \vec{\sigma}_{j+2}) \notag \\
&= 2J_{4,2;\text{tot}} + 12 Q_3^{(0)} + \frac{2}{3} Q_5^{(0)} ~. \label{eq:Vbo2}
\end{align}
Here the currents $J_{\beta,2;\text{tot}}$ are defined as in Eq.~(\ref{eq:Jtot}), while the IoM $Q_5^{(0)}$ is from Ref.~\cite{Surace2023}.
Hence, up to additions of IoMs, these perturbations in an infinite chain are indeed obtained using boosted generators:
With our convention $H_0 = 2Q_2^{(0)}$, the infinite-chain $X_{\text{bo},1} = -\mathcal{B}[Q_3^{(0)}]$ and $X_{\text{bo},2} = -\mathcal{B}[Q_4^{(0)}]$, as in Eq.~(\ref{eq:Xbo}).
As explained in Sec.~\ref{subsec:propsAGP}, in a finite PBC chain, e.g., $V_{\text{bo},1}$ and $2J_{3,2;\text{tot}}$ have identical AGP (the IoMs only contribute to the block-diagonal part $V_\text{diag}$), which we have verified as further checks on our numerical routines, and we will not be making distinction between these from now on. 

Perturbation $V_{\text{bo},1}$, which has the same symmetries as $H_0$ and is the simplest such perturbation of the Heisenberg chain, has a long history in the literature. Its weak integrability-breaking character was noticed already in Refs.~\cite{Jung2006, Jung2006_2} with the finding of the first quasi-IoM, further bolstered in Ref.~\cite{Kurlov2022} finding a few more quasi-IoMs. Its generation by $\Xbo$ was pointed out in Refs.~\cite{Surace2023}~and~\cite{Orlov_2023}.Reference~\cite{Orlov_2023} also studied the corresponding AGP norm finding $\| \XAGP[V_{\text{bo},1}] \|^2 \sim L \log L$ scaling with $L$. In this paper we further examine this scaling and its implications and will study the spatial structure of the AGP.

On the other hand, $V_{\text{bo},2}$ is odd under both $\Theta$ and $I$ symmetries, which are the same symmetries as for $Q_3^{(0)}$, and is the simplest such perturbation. 
To our knowledge, it has not been identified as a weak integrability-breaking perturbation before. By studying both $V_{\text{bo},1}$ and $V_{\text{bo},2}$, we hope to learn about general properties of any $V_{\text{bo}}$-type perturbation.

\subsubsection{Weak perturbations generated by $\Xbi$}
We consider two perturbations generated by the bilocal operator:
\begin{align}
\label{eq:V_bi1}
& V_{\text{bi},1} = i[[Q_2^{(0)}|Q_3^{(0)}], H_0] = \sum_j \Big[ 2\vec{\sigma}_j \cdot \vec{\sigma}_{j+2} - 2\vec{\sigma}_j \cdot \vec{\sigma}_{j+3} \notag \\
+ & (\vec{\sigma}_j \cdot \vec{\sigma}_{j+1})(\vec{\sigma}_{j+2} \cdot \vec{\sigma}_{j+3}) +  (\vec{\sigma}_j \cdot \vec{\sigma}_{j+3})(\vec{\sigma}_{j+1} \cdot \vec{\sigma}_{j+2}) \Big],\\
& V_{\text{bi},2} = i[[Q_2^{(0)}|Q_4^{(0)}], H_0] \notag\\
&\quad= \sum_j \Big[ 2(\vec{\sigma}_j \times \vec{\sigma}_{j+3}) \cdot (\vec{\sigma}_{j+1} + \vec{\sigma}_{j+2}) \notag \\
&\quad -2(\vec{\sigma}_j \times \vec{\sigma}_{j+4}) \cdot (\vec{\sigma}_{j+1} + \vec{\sigma}_{j+3})  \notag \\
&\quad +\{[(\sv{j+3}\times \sv{j+4})\times \sv{j+1}]\times \sv{j}\}\cdot \sv{j+2} \notag\\
&\quad + \{[(\sv{j+1}\times \sv{j})\times \sv{j+3}]\times \sv{j+2}\}\cdot \sv{j+4} \notag\\
&\quad +2\{[(\sv{j}\times \sv{j+1})\times \sv{j+2}]\times \sv{j+3}\}\cdot \sv{j+4} \notag\\
&\quad +2\{[(\sv{j+1}\times \sv{j+4})\times \sv{j+2}]\times \sv{j+3}\}\cdot \sv{j} \Big] ~.
\label{eq:Vbi2}
\end{align}
Perturbation $V_{\text{bi},1}$ was obtained in Ref.~\cite{Surace2023} and is symmetric under both $\Theta$ and $I$ symmetries.
On the other hand, $V_{\text{bi},2}$ has not been considered before; it is odd under both $\Theta$ and $I$.
By studying $V_{\text{bi},1}$ and $V_{\text{bi},2}$, we learn about general properties of any $V_{\text{bi}}$-type perturbation.

\begin{figure*}[ht]
    \centering
    \begin{minipage}[c]{0.52\linewidth}
        \includegraphics[width=\linewidth]{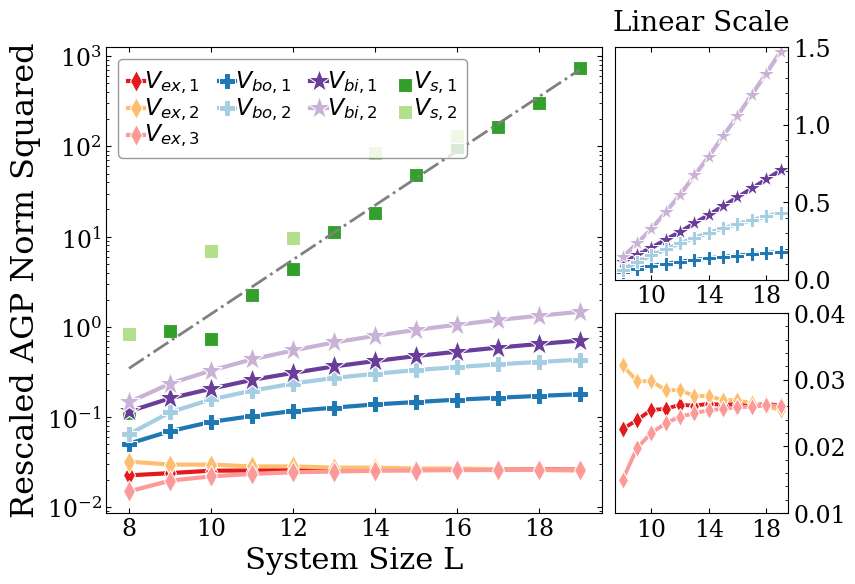}
    \end{minipage}
    \hfill
    \begin{minipage}[c]{0.47\linewidth}
        {\renewcommand{\arraystretch}{1.05}
        \begin{tabular}{c c  c c  c c }
        \centering
            $V$ & Eq.~reference & $\Theta$ & $I$ & ~$\|\XAGP\|^2/L$ & fit param. \\
            \midrule
            \midrule
            $V_{\text{bo},1}$ & Eq.(\ref{eq:Vbo1}) & even & even & $a+bL^p$ &  $p=0.03$ \\
            $V_{\text{bo},2}$ & Eq.(\ref{eq:Vbo2}) & odd & odd& $a+bL^p$ & $p=0.03$  \\
            \hline
            \hline
            $V_{\text{bi},1}$  &Eq.(\ref{eq:V_bi1})  &  even & even & $a+bL^p$ &  $p=1.32$  \\
            $V_{\text{bi},2} $ & Eq.(\ref{eq:Vbi2})&  odd & odd   &$a+bL^p$  &  $p=1.546$ \\
            \midrule
            \midrule
            $ V_{\text{ex},1}$& Eq.(\ref{eq:Vex1}) & odd  & even & const. &    \\
            $V_{\text{ex},2}$ & Eq.(\ref{eq:Vex2})  & even & even & const.  &    \\
            $V_{\text{ex},3}$ & Eq.(\ref{eq:Vex3}) & even & odd  & const. &   \\
            \midrule
            \midrule
            $V_{\text{s},1}$& Eq.(\ref{eq:Vs1}) & even & even &$a+be^{\kappa L}$ & $\kappa=0.77$    \\
            $V_{\text{s},2}$& Eq.(\ref{eq:Vs2}) &  even & & $a+be^{\kappa L}$ &$\kappa=0.66$  \\
            \hline
        \end{tabular}}
    \end{minipage}
    \caption{Left, main panel: The rescaled AGP norm squared, $\| \XAGP \|^2_{S^z_\text{tot}=0}/L$, as a function of the system size $L$ for the Heisenberg Hamiltonian perturbed by the extensive perturbations given in the table on the right.
    The grey line shows the exponential fit $\| \XAGP \|^2_{S^z_\text{tot}=0}/L \varpropto e^{\kappa L}$ with $\kappa \approx \ln{2}$. 
    Left, small panels: The weak integrability breaking perturbations only and on a linear scale; top: perturbations with known boosted and bilocal generators in the infinite system; bottom: perturbations with known extensive local generators on any finite PBC chain. For easier visualization, some of the perturbations have been rescaled by a constant ($L$-independent) prefactor.
    Right, table: Listing of the perturbations studied with reference to equations in the main text; quantum numbers under time reversal $\Theta$ and inversion $I$ symmetries; and their observed AGP norm scaling in the $S^z_{\text{tot}} = 0$ sector.
    (For $V_{\text{s},2}$, since it breaks translation symmetry, one needs to separate inversion in site or bond center, and we simply do not mark the corresponding $I$.)
    }
    \label{fig:agp-both-global}
\end{figure*}

\subsection{Scaling of the AGP norms}
\label{sec:global-agp-scaling}

We compute the regularized $\XAGP$ for each of the perturbations listed above.
We consider the Frobenius norm of $\XAGP$ restricted to the $S^z_{\text{tot}} = 0$ sector defined in Eq.~(\ref{eq:agp_norm_Sztot0}), regularized as in Eq.~(\ref{eq:freg}). 
Figure~\ref{fig:agp-both-global} shows the corresponding rescaled regularized AGP norm, $\| \XAGP \|^2 / L$, computed in the $S^z_{\text{tot}} = 0$ sector with regulator $\mu = L/D_{S^z_{\text{tot}}=0}$; the table in the figure reports the scaling fits in each case.

For the strong perturbations $V_{\text{s},1}$ and $V_{\text{s},2}$, we observe exponential scaling of the AGP norm $\sim e^{\kappa L}$, as predicted by ETH.
Figure~\ref{fig:agp-both-global} shows the exponential line given by $e^{\kappa L}$ with $\kappa = \ln 2 = 0.6931$ tracking the overall growth of these AGP norms.
Independent direct fits for $V_{\text{s},1}$ and $V_{\text{s},2}$ give $\kappa = 0.77$ and $0.66$ respectively, which is consistent with previous findings and expectations for generic perturbations~\cite{Pandey2020}.
The numerical results for $V_{\text{s},1}$ are in agreement with those in Ref.~\cite{Orlov_2023}, while the results for $V_{\text{s},2}$ show that this perturbation, considered specifically in Ref.~\cite{DeNardis_2021}, is also a strong integrability-breaking perturbation.

Turning to perturbations $V_{\text{ex},1}$, $V_{\text{ex},2}$, and $V_{\text{ex},3}$, we find that the AGP norm squared scales proportionally to $L$ in all these cases, as expected for local extensive generators.
The numerically calculated $\| \XAGP \|^2/L$ saturates to a fixed number in each case, as we show in more detail in a small panel in Fig.~\ref{fig:agp-both-global}.
Since we are computing the norm in the fixed $S^z_{\text{tot}} = 0$ sector, it does not exactly correspond to the AGP norm in the full Hilbert space.
However, this does not qualitatively change the scaling of the norm with the system size.
Furthermore, as we will explain in Sec.~\ref{subsubsec:AGPVex}, only for $V_{\text{ex},2}$ we can recover exactly $\XAGP = X_{\text{ex},2}$ in the full Hilbert space, while for $V_{\text{ex},1}$ and $V_{\text{ex},3}$ the recovery of the corresponding $X_{\text{ex},1}$ and $X_{\text{ex},3}$ suffers from the missing diagonal part in the calculated $\XAGP$. 
Nevertheless, we see that such obstacles to recovering exact generators do not change the qualitative AGP norm scaling properties.

We now consider perturbations $V_{\text{bo},1}$ and $V_{\text{bo},2}$. 
Naively, in these cases the norm of an $\XAGP$ approximating an $X_\text{bo}$ is expected to scale as $L^3$:
In an infinite system, $X_{\text{bo}} \sim \mathcal{B}[Q_\beta^{(0)}]$, and a naive restriction to a chain of length $L$, e.g., $\sum_{j=1}^L (j-j_0) q_{\beta,j}^{(0)}$, including any possible offset $j_0 \in [1,\dots,L]$ and ignoring that such expressions are ill-defined in PBC, would give $\| \Xbo \|^2 \sim L^3$, and the rescaled behavior corresponding to analysis in Fig.~\ref{fig:agp-both-global} would be $\| \Xbo \|^2/L \sim L^2$.
We instead find that $\| \XAGP \|^2/L$ scales as $L^p$ with $p<1$ for both $V_{\text{bo},1}$ and $V_{\text{bo},2}$ cases.
Our fits of $\| \XAGP \|^2/L = a + b L^p$ find that $p$ is fairly small in both cases, $p = 0.03$ for both.
The curves are also consistent with $\| \XAGP \|^2/L = c + d \log(L)$.
Logarithmic scaling of the AGP norm for $V_{\text{bo},1}$ was proposed before in~\cite{Orlov_2023}.
Here we confirm that such weaker-than-linear scaling of $\| \XAGP \|^2/L$ holds for both boost-generated perturbations.
This scaling will be important for understanding bilocal-generated and step-generated perturbations later.

Finally, we discuss the  bilocal generators $V_{\text{bi},1}$ and $V_{\text{bi},2}$.
Here, the norm of the $\XAGP$ obtained from $X_{\text{bi}}$ is expected to scale as $L^2$ if we naively restrict summations in Eq.~(\ref{eq:Xbi}) to run over $j, k \in [1, \dots, L]$.
We find that the rescaled norm squared of $\XAGP$ scales as $aL^p$ with $p\approx 1.32$ for $V_{\text{bi},1}$ and $p=1.546$ for $V_{\text{bi},2}$.
Note that these fits were obtained from system sizes $L=8$ to $19$.
Our analysis in Sec.~\ref{subsubsec:AGPVbi} suggests $p=1$ in this case, similar to the naive expectation, and the discrepancy is likely due to the rather limited range of available sizes.

\begin{figure}[ht]
\centering
\includegraphics[width=0.49\linewidth]{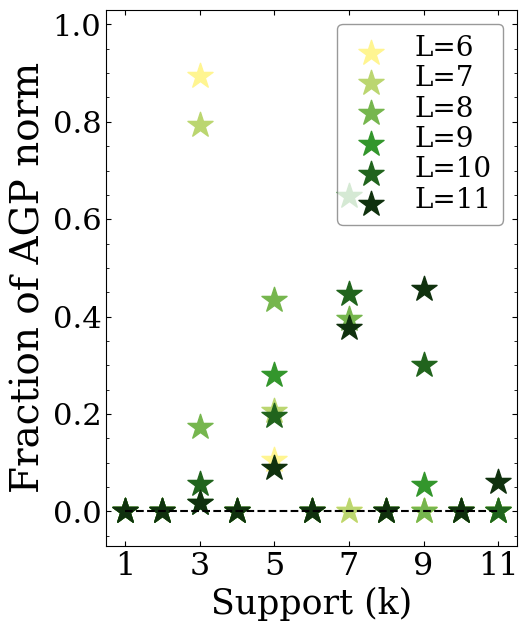}
\includegraphics[width=0.49\linewidth]{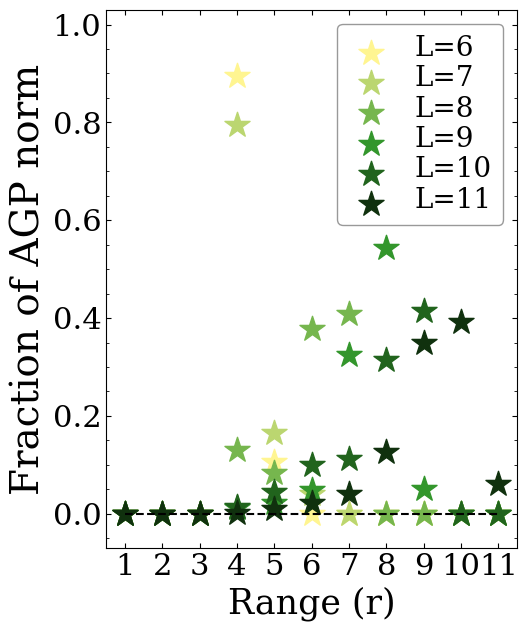}
\caption{Visualization of the structure of $\XAGP$ for $V_{\text{s},1}$ in terms of the decomposition in the Pauli string basis. 
Left: Fraction of the total AGP norm of strings with support $k$ as given in Eq.~(\ref{eq:wsuppk}).
Right: Fraction of the total AGP norm of strings with range $r$ as given in Eq.~(\ref{eq:wranger}).
The AGP of chaotic perturbations appears to be a highly non-local operator, which refers to its spread over the Pauli strings of all supports and ranges.
This will serve as a reference for the rest of the plots we will present in this study.}
\label{fig:agp_vs1}
\end{figure}

\subsection{Characterization of the AGPs in the Pauli string basis}
\label{subsec:agp_PauliS2}

\subsubsection{AGP for strong perturbations}
\label{subsubsec:AGPVs}
Strong integrability-breaking perturbations exhibit universal behavior.
As shown in Sec.~\ref{sec:global-agp-scaling}, norms of $\XAGP$ for generic non-integrable perturbations scale exponentially with the system size. We expect that the
$\XAGP$ operator is highly non-local in the Pauli string basis for such perturbations.

To confirm the highly non-local structure of the $\XAGP$ for the strong perturbations, we examine their decomposition in Pauli operators.
For $V_{\text{s},1}$, the weights for a given support and range, as defined in Eqs.~(\ref{eq:wsuppk}) and (\ref{eq:wranger}), are plotted in Fig.~\ref{fig:agp_vs1}.
Since $H_0$ and $V_{\text{s},1}$ are even under $\Theta$ and even under $I$, $\XAGP$ must be odd under $\Theta$ and even $I$.
Correspondingly, in Fig.~\ref{fig:agp_vs1} we observe that only Pauli strings with an odd support $k = 3, 5, 7, \dots$ and with correct inversion symmetry contribute. 

Crucially, we note that the weight is significantly spread over all supports and ranges for a fixed system size, and moves to larger supports and ranges as we increase the system size. We conclude that $\XAGP$ for a generic strong integrability-breaking perturbation is extended in space and highly non-local.

In addition, in Appendix~\ref{app:oipr}, Fig.~\ref{fig:OIPR}, we show comparison of the operator participation ratios (OPRs) for all of the perturbations in this section. The OPR of strong integrability-breaking perturbations captures this non-locality well.

\subsubsection{AGP for weak perturbations generated by $\Xex$}
\label{subsubsec:AGPVex}
\begin{figure}[ht]
\centering
\includegraphics[width=0.49\linewidth]{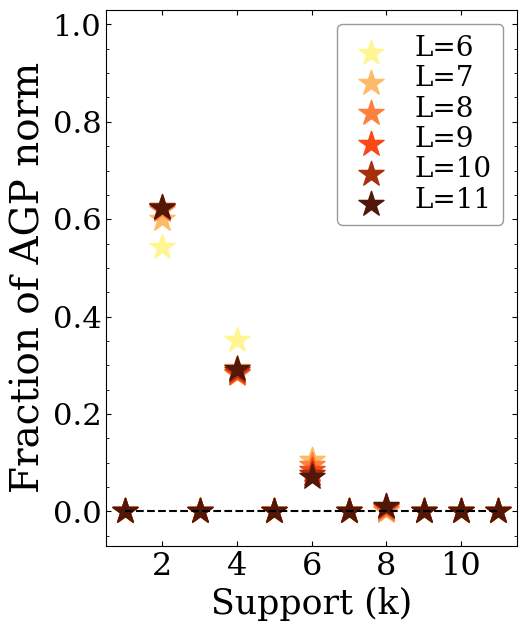}
\includegraphics[width=0.49\linewidth]{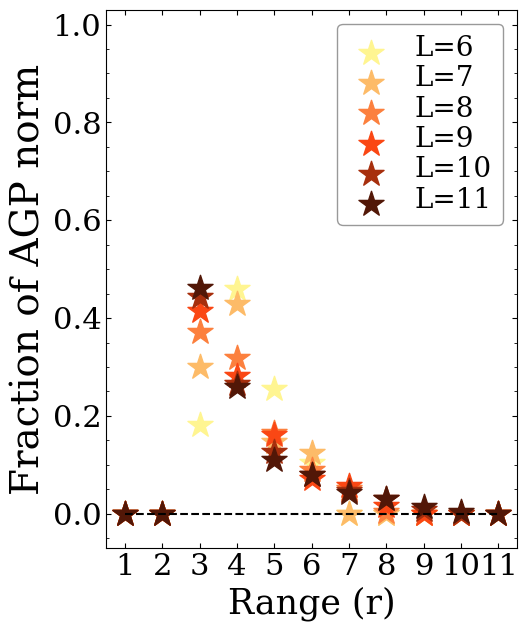}
\includegraphics[width=0.49\linewidth]{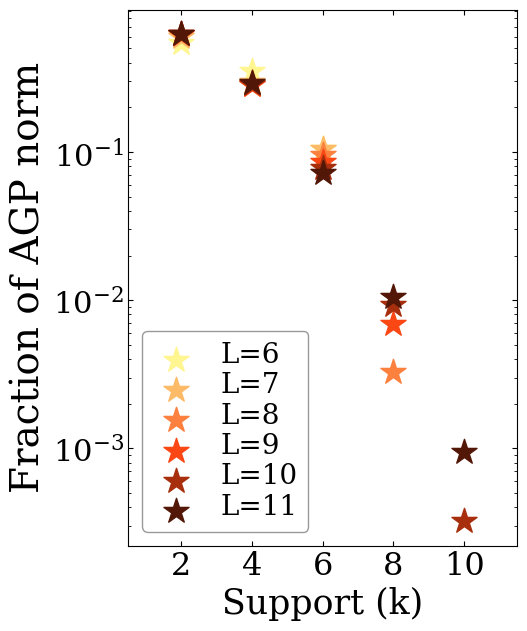}
\includegraphics[width=0.49\linewidth]{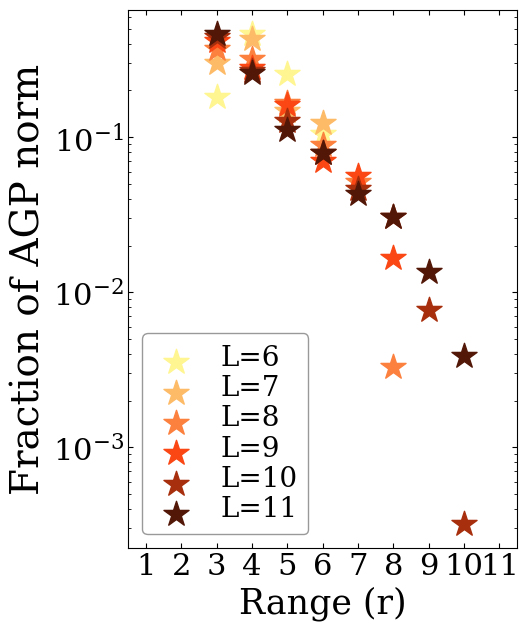}
\caption{Visualization of the structure of $\XAGP$ for $V_{\text{ex},1}$ in terms of the decomposition in the Pauli string basis.
Top panels: Left: Fraction of the total AGP norm in strings with support $k$ as defined in Eq.~(\ref{eq:wsuppk}).
Right: Fraction of the total AGP norm in strings with range $r$ as defined in Eq.~(\ref{eq:wranger}).
Bottom panels: Same date as in the top panels but displayed with logarithmic vertical scale.
Note that the fraction of the AGP norm on the lowest $k$ strings is, unlike in Fig.~\ref{fig:agp_vs1}, not changing with system size, and is in fact growing on the strings we know to make the exact $X$.
Similar data for $V_{\text{ex},3}$ can be found in Fig.~\ref{fig:Vex3} in App.~\ref{app:add_data_for_agp_global}.
For $V_{\text{ex},2}$ we do not show such plots as we recover the exact $X$ of Eq.~(\ref{eq:Xex2}).
Note: All the four-panel AGP visualization plots that follow represent the same information but for different perturbations.
We will refer readers to this plot for the information that applies to all of them and focus only on perturbation-specific details instead.}
\label{fig:Vex1}
\end{figure}

Perturbations generated by $X_{\text{ex}}$ have a known, well-defined generator in finite systems.
We can then ask the question of how faithfully our numerical procedure recovers it.
In Fig.~\ref{fig:Vex1} we show analysis of the AGP content for the perturbation $V_{\text{ex},1}$, while in Fig.~\ref{fig:Vex3} in App.~\ref{app:add_data_for_agp_global} we show similar analysis for $V_{\text{ex},3}$.
We find that we do not numerically recover the exactly known $X$ for $V_{\text{ex},1}$ and $V_{\text{ex},3}$ [we have checked that this is not because of the regulator, since the difference remains essentially unchanged for a range of cutoffs $\mu$ and for both regulators Eq.~(\ref{eq:freg}) or Eq.~(\ref{eq:fregstep})].
To understand why, it is useful to consider the symmetries of the operators in question. 
The generator of $V_{\text{ex},1}$, namely $X_{\text{ex},1}$, is even under both $\Theta$ and $I$, which are the same symmetries that IoMs $Q_\alpha^{(0)}$ have when $\alpha$ is even.
$X_{\text{ex},1}$ can therefore have non-trivial diagonal part which cannot be recovered by the corresponding $\XAGP$ [see the discussion in Sec.~\ref{subsec:propsAGP} after Eq.~(\ref{eq:XAGPdiag})].

Similar argument holds for $X_{\text{ex},3}$, which is odd under both $\Theta$ and $I$, which are the same as symmetries of the IoMs $Q_\alpha^{(0)}$ with $\alpha$ odd, allowing non-trival diagonal part for $X_{\text{ex},3}$ that our procedure cannot recover.
We can, however, check that for both $V_{\text{ex},1}$ and  $V_{\text{ex},3}$ what is missing in the $\XAGP$ is precisely the $X_{\text{diag}} = \sum_\nu P_\nu X P_\nu$.
In Appendix~\ref{app:brute-force-Xdiag} we outline a procedure for an unbiased recovery of the missing diagonal part by a minimization of a cost function that penalizes strings with larger support or range, which we have tested for $V_{\text{ex},1}$ and which works partially but can be systematically improved.

Even without any recovery, from Figs.~\ref{fig:Vex1} and \ref{fig:Vex3} we see that only strings with small supports and with small spatial ranges contribute to the AGP norm and that half of the norm is in the strings that make up the exact generator, e.g., support $k=2$ and range $k=3$ for $V_{\text{ex},1}$.
The log-scale plots in the bottom panels of the same figures additionally suggest exponential decay in contributions from strings with support and range larger than the ones in the generator.
Furthermore, the fraction of the AGP norm on the strings with the smallest support and range is either increasing or converging to a fixed value with increasing system size $L$.

On the other hand, we find that for the sizes in the Pauli string studies the $\XAGP$ for $V_{\text{ex},2}$ completely recovers its exact generator $X_{\text{ex},2}$.
For the regularized $\XAGP$ defined with the regularization in Eq.~(\ref{eq:freg}) there is a small error introduced by the cutoff $\mu$, while the $\XAGP$ obtained with the step regulator Eq.~(\ref{eq:fregstep}) is exactly $X_{\text{ex},2}$.
The ability to completely recover the generator of $V_{\text{ex},2}$ is again due to the symmetry argument explained above:
Now $X_{\text{ex},2}$ is odd under $\Theta$ and even under $I$, which is different from the symmetries of any $Q_\alpha^{(0)}$.
In App.~\ref{app:sim_diag} we argue that any diagonal operator can be generated in the algebra sense by $\{ Q_\alpha^{(0)} \}$ (in fact, for most of the sizes in the Pauli string studies just $Q_2^{(0)}$ and $Q_3^{(0)}$ are sufficient), and hence cannot have the same symmetries as $X_{\text{ex},2}$.
The diagonal part of $X_{\text{ex},2}$ is then zero, and hence the $V_{\text{ex},2}$ case does not suffer from the problems encountered by $V_{\text{ex},1}$ and $V_{\text{ex},3}$ cases.
[Nevertheless, we suspect that the exact recovery of the $X_{\text{ex},2}$ generator by the AGP is special for small sizes, while going to larger sizes one would need to use the generalized AGP approach described in App.~\ref{app:sim_diag}.]

\subsubsection{AGP for weak perturbations generated by $\Xbo$}
\label{subsubsec:AGPVbo}
\begin{figure}[ht]
\centering
\includegraphics[width=0.49\linewidth]{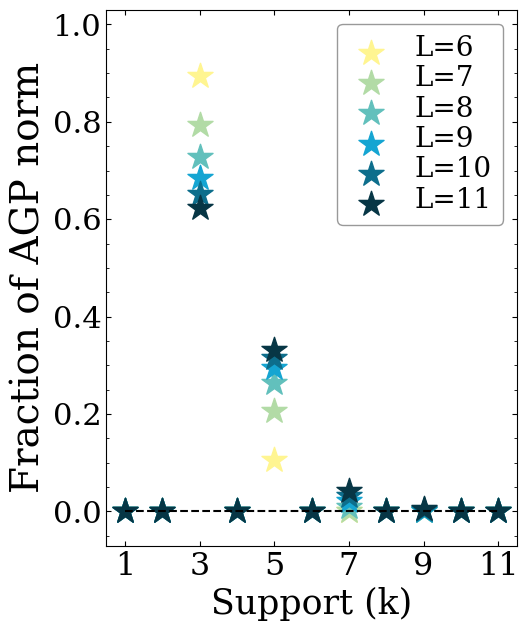}
\includegraphics[width=0.49\linewidth]{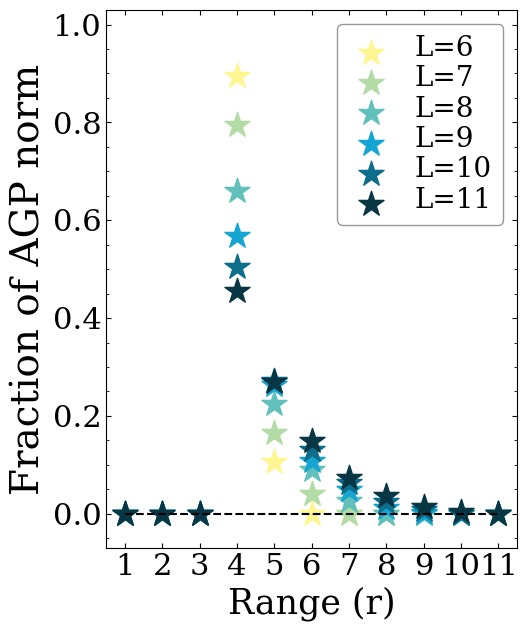}
\includegraphics[width=0.49\linewidth]{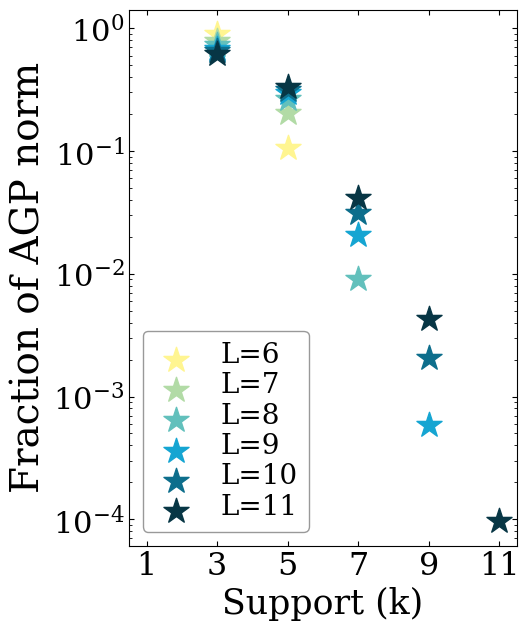}
\includegraphics[width=0.49\linewidth]{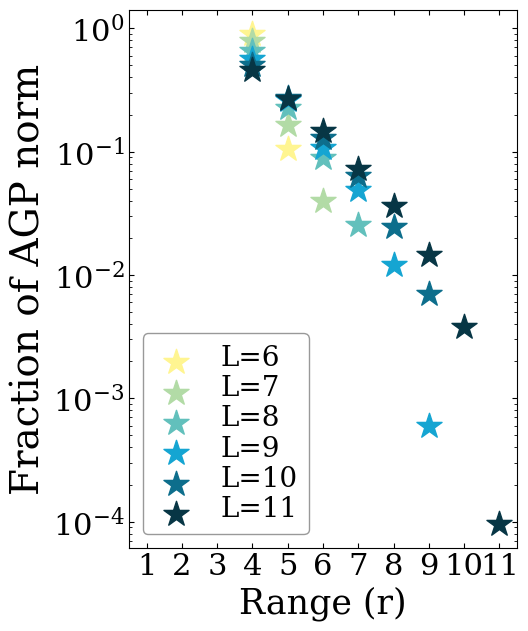}
\caption{Visualization of the structure of $X^{\text{AGP}}$ for $V_{\text{bo},1}$ in terms of the decomposition in the Pauli string basis origanized by support $k$ and range $r$ (see caption in Fig.~\ref{fig:Vex1} for details).
This object is our main interest.  
Similarly to the extensive local perturbations, the AGP here exhibits signs of quasi-locality, even though the AGP norm grows faster in this case.
The weight on the smallest $k$ and $r$ strings is decreasing, but this is balanced out by the next smallest $k$ and $r$, and it seems likely that the weights will converge rather than keep shifting towards the strings with larger $k$ and $r$ values (as was the case for the strong perturbations).}
\label{fig:Vbo1}
\end{figure}

Finite-size generators of $V_{\text{bo},1}$ and $V_{\text{bo},2}$ are our main objects of interest.
Here we present the Pauli string characterizations in the $V_{\text{bo},1}$ case shown in Fig.~\ref{fig:Vbo1}, while the $V_{\text{bo},2}$ case is shown in App.~\ref{app:add_data_for_agp_global} in Fig.~\ref{fig:Vbo2}.
We find that the distribution of the weight with support or range is very different from the strong perturbation cases like $V_{\text{s},1}$ and is closer to what we see for the cases with known extensive local generators like $V_{\text{ex},1}$. Nevertheless, there is a transfer of weight away from the smallest support onto strings with somewhat larger supports as we increase the system size $L$.
Both boost-generated perturbations show similar behavior with respect to the support and range of the Pauli strings that contribute to the norm squared, Figs.~\ref{fig:Vbo1} and \ref{fig:Vbo2}. 

Since $V_{\text{bo},1}$ is even under both $\Theta$ and $I$ symmetries, the corresponding $\XAGP$ is odd under $\Theta$ and even under $I$, while these symmetries are exactly opposite for $V_{\text{bo},2}$ and its AGP.
As argued in App.~\ref{app:sim_diag} and used in our discussion of the $V_{\text{ex}}$ cases in Sec.~\ref{subsubsec:AGPVex}, diagonal operators cannot have such $(\Theta, I) = (\text{odd}, \text{even})$ or $(\text{even}, \text{odd})$ symmetry numbers.
Hence, we consider the $V_{\text{bo},1}$ and $V_{\text{bo},2}$ cases as not suffering from the missing diagonal part in the recovery of $X$, in contrast to the $V_{\text{ex},1}$ and $V_{\text{ex},3}$ cases.
Furthermore, we note that $V_{\text{bo},1}$ has the same symmetries as $V_{\text{ex},2}$, where for our system sizes the AGP was able to recover the generator $X_{\text{ex},2}$ exactly.
Hence, for these sizes we can view $\XAGP[V_{\text{bo},1}]$ as the best possible recovery of the finite-size generator in this case as well.

As symmetry plays a big role in which strings will appear in the $\XAGP$, we note that the shortest range Pauli strings that are (odd, even) under $(\Theta, I)$ are proportional to $\sum_j (\vec{\sigma}_j \times \vec{\sigma}_{j+3}) \cdot (\vec{\sigma}_{j+1} - \vec{\sigma}_{j+2})$.
For $V_{\text{bo},1}$, these are the only strings of support $k=3$ and range $r=4$ that appear and most of the weight is on them.
This helps make sense of Fig.~\ref{fig:Vbo1} by noting that the data for the fraction of the norm from $r=4$ strings is really showing contributions from this particular operator.
The large contribution from this operator is not exclusive to $V_{\text{bo},1}$, but is shared by other weak perturbations having the same symmetries.
In fact, this operator is the one with the largest fraction of the AGP norm for $V_{\text{bi},1}$ and is actually the exact generator of $V_{\text{ex},2}$.

On sizes $L \leq 6$, there is only one more translationally invariant operator with these symmetries.
It is thus not surprising that the fraction of the AGP norm on this particular operator is largest for the smallest $L$ for both $V_{\text{bo},1}$ and $V_{\text{ex},2}$.
However, while for $V_{\text{ex},2}$ this operator is also the exact $X_{\text{ex},2}$, for $V_{\text{bo},1}$ the weight on this operator decreases with the system size.
The norm shifts to operators with $k=5$ and $7$ (some norm also stays on $k=3, r>4$), and at first they appear to grow significantly, but the weights seems to approach some limiting values as the system size increases.
As shown in the lower panels of Fig.~\ref{fig:Vbo1}, these limiting values appear to decrease exponentially with the support and range, suggesting some possible convergence in the large $L$ limit.
Nevertheless, we have not been able to capture the observed AGP with a closed-form analytic expression.

It should be noted that the exponential decrease of the weight for larger and larger supports and ranges is a very peculiar feature.
The number of Pauli strings of a given support $k$ or range $r$ grows exponentially with $k$ or $r$.
The ``quasilocality'' exhibited by the $\XAGP$ for the $V_{\text{bo}}$-type perturbations is therefore a very ``rare'' property in the space of operators.
This characterization is one of our main focuses and results in this work.

\subsubsection{AGP for weak perturbations generated by $X_{\text{bi}}$}
\label{subsubsec:AGPVbi}
\begin{figure}[ht]
\centering
\includegraphics[width=0.48\linewidth]{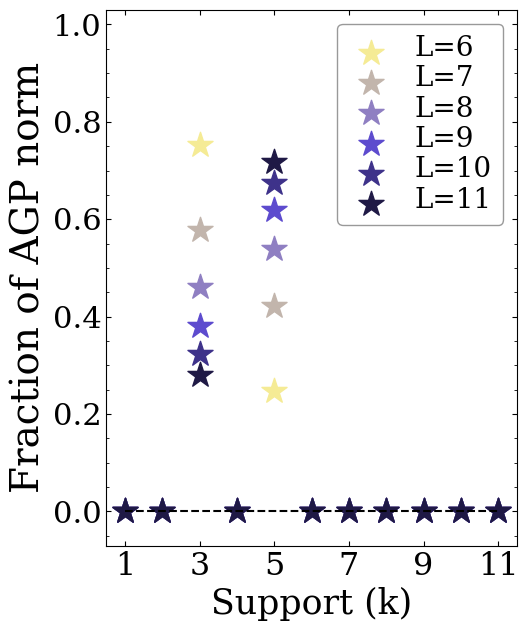}
\includegraphics[width=0.48\linewidth]{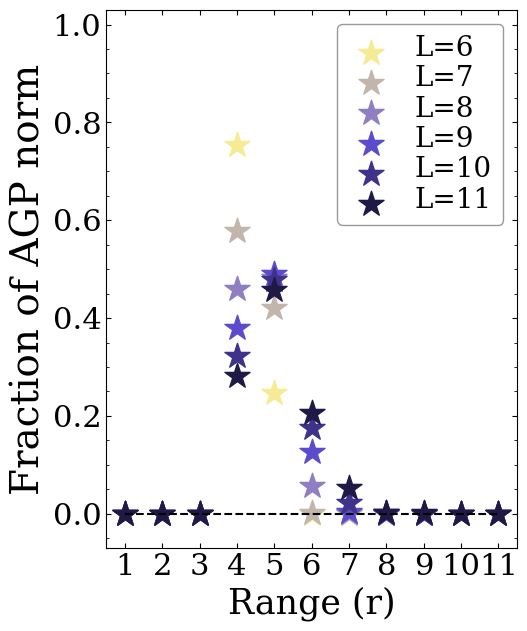}
\includegraphics[width=0.48\linewidth]{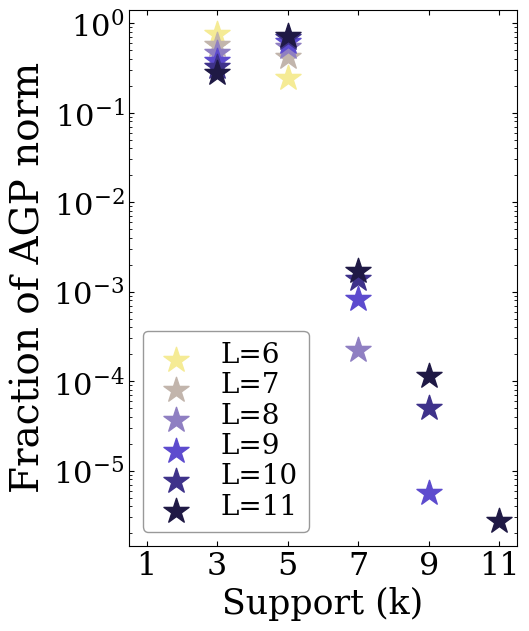}
\includegraphics[width=0.48\linewidth]{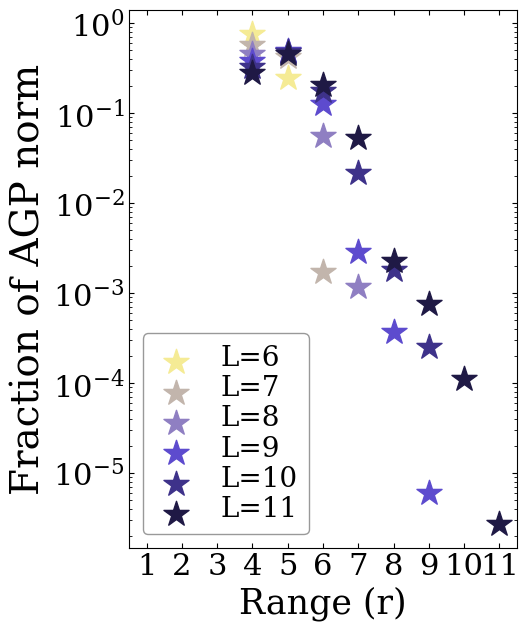}
\caption{Visualization of $X^{\text{AGP}}$ for $V_{\text{bi},1}$ in terms of the decomposition in the Pauli string basis organized by support $k$ and range $r$ (see caption in Fig.~\ref{fig:Vex1} for details). 
This figure, in particular the log-scale panels, was an early indicator that the AGP could be well approximated by an operator with a closed analytical form.
As the system size increases, the AGP norm, which in this case grows even faster than for the $V_{\text{bo}}$-type perturbations, does not keep extending over strings with larger support and range but is instead well contained on strings with support 3 and 5.}
\label{fig:Vbi1}
\end{figure}

We now consider AGPs for $V_{\text{bi},1}$ and $V_{\text{bi},2}$ that we view as finite-size PBC proxies of the corresponding infinite-system bilocal generators.
Examining the Pauli string visualization of the $\XAGP$ for $V_{\text{bi},1}$ in Fig.~\ref{fig:Vbi1}, it appears to converge to an operator with strings primarily of support $3$ and $5$; while the AGP for $V_{\text{bi},2}$ in Fig.~\ref{fig:Vbi2} in App.~\ref{app:add_data_for_agp_global} shows primarily support $4$ and $6$. 
Strings with larger supports are two orders of magnitude smaller, which seems significant.
This is highly suggestive of similarity to the infinite-chain expression Eq.~(\ref{eq:Xbi}), where we indeed have such exclusive Pauli string supports, and motivates us to look for a similar expression on finite chains.

Consider the following ansatz
\begin{equation}
\widetilde{X}^{\beta\gamma} = \sum_{j=1}^L \sum_{k=1}^L c_{j,k} \, \{ q_{\beta,j}^{(0)},\,  q_{\gamma,k}^{(0)} \}
\end{equation}
with
\begin{equation}
\label{eq:cjkPBC}
c_{j,k} = t(k-j) = t(k-j \pm L) ~,
\end{equation}
i.e., $L$-periodic function of the distance between the two points.
This has similar structure to the infinite-system generator in Eq.~(\ref{eq:Xbi}) and is well-defined on the finite PBC chain [the infinite-system case formally corresponds to a $\theta$-function $t_{L=\infty}(n) = 0$, $1/2$, and $1$ for $n<0$, $n=0$, and $n>0$].
Using the current-defining Eq.~(\ref{eq:current}), which is valid on a finite PBC chain, we can calculate
\begin{align}
& i[\widetilde{X}^{\beta\gamma}, Q_\alpha^{(0)}] = \sum_{j=1}^L \Big(\{ q_{\gamma,j}^{(0)}, \sum_{k=1}^L J_{\beta\alpha,k}^{(0)} (c_{k-1,j} - c_{k,j}) \} \notag\\
& \qquad\qquad\qquad~~ - \{ q_{\beta,j}^{(0)}, \sum_{k=1}^L J_{\gamma\alpha,k}^{(0)} (c_{j,k} - c_{j,k-1}) \} \Big),
\end{align}
where we have performed some resummations and reorganizations for the final form to resemble the structure in Eq.~(\ref{eq:bilocal}).
Since on the PBC chain we have, e.g., $\sum_k (c_{k-1,j} - c_{k,j}) = 0$, it is not possible to match the definition of $V_{\text{bi}}$ in Eq.~(\ref{eq:bilocal}).
The closest we can get is to require
\begin{align}
& c_{k-1,j} - c_{k,j} = \frac{1}{2} (\delta_{k,j} + \delta_{k,j+1}) - \frac{1}{L} ~, \\
& c_{j,k} - c_{j,k-1} = \frac{1}{2} (\delta_{k,j} + \delta_{k,j+1}) - \frac{1}{L} ~.
\end{align}
These two equations are compatible with the form in Eq.~(\ref{eq:cjkPBC}) and one condition
\begin{equation}
t(n) - t(n-1) = \frac{1}{2}(\delta_{n,0} + \delta_{n,1}) - \frac{1}{L} ~.  
\end{equation}
Its solution, up to an unimportant overall constant, reads
\begin{equation}
t(n) = 
\begin{cases}
0, & \text{if $n=0$} \\
\frac{1}{2} - \frac{n}{L}, & \text{if $n=1,2,\dots, L-1$} ~,
\end{cases}
\end{equation}
which is understood extended to other integers using PBC.
Thus, $t(n)$ shows step increases by $1/2 - 1/L$ between $n = -1 \equiv L-1$ and $n=0$, and between $n=0$ and $n=1$, while the values $t(1) = 1/2-1/L$ and $t(L-1) = t(-1) = -1/2 + 1/L$ are ``connected'' along the path $1, 2, \dots, L-1$ by a linear function with a small slope $(-1/L)$.
This is then the proxie form of the $\theta$ function $t_{L=\infty}(n)$ that can be realized in PBC.

From now on, we use $c_{j,k}$ as specified above.
We fix $\alpha = 2$ [since $H_0 = 2Q_2^{(0)}$] and write the perturbation in Eq.~(\ref{eq:Vbi_betagamma}) as $V_{\text{bi}}^{\beta\gamma}$.
With these conventions, we have
\begin{equation}
i[\widetilde{X}^{\beta\gamma}, H_0] = V_{\text{bi}}^{\beta\gamma} + \frac{2}{L} \left(\{ Q_\beta^{(0)}, J_{\gamma,2;\text{tot}}^{(0)} \} - \{ Q_\gamma^{(0)}, J_{\beta,2;\text{tot}}^{(0)} \} \right).
\end{equation}
Thus, $\widetilde{X}^{\beta\gamma}$ is not the exact AGP for $V_{\text{bi}}^{\beta\gamma}$ but captures it up to the exhibited non-local terms with $\sim 1/L$ amplitudes.
These non-local terms have conserved total charges as factors in them, which makes them rather special.
Furthermore, they contain total current operators that also define $V_{\text{bo}}$-type perturbations studied earlier, see Eq.~(\ref{eq:Vbo_beta}) which we write as $V_{\text{bo}}^\beta$.
Suppose we know the AGP for such a perturbation in our finite PBC system,
\begin{equation}
i\left[ \XAGP[V_{\text{bo}}^\beta], H_0 \right] = V_{\text{bo}}^\beta - (V_{\text{bo}}^\beta)_{\text{diag}} ~.
\end{equation}
Using commutation of $Q_\gamma^{(0)}$ and $H_0$, it is easy to check that
\begin{equation}
i[\{Q_\gamma^{(0)}, \XAGP[V_{\text{bo}}^\beta] \}, H_0] = \{Q_\gamma^{(0)}, V_{\text{bo}}^\beta - (V_{\text{bo}}^\beta)_{\text{diag}} \} ~.
\end{equation}
Similar equation holds with $\beta$ and $\gamma$ interchanged.
Hence we have
\begin{align}
& i\left[\widetilde{X}^{\beta\gamma} + \frac{1}{L}\{Q_\gamma^{(0)}, \XAGP[V_{\text{bo}}^\beta] \} - \frac{1}{L}\{Q_\beta^{(0)}, \XAGP[V_{\text{bo}}^\gamma] \}, H_0 \right] \nonumber \\
& = V_{\text{bi}}^{\beta\gamma} - \frac{1}{L} \{Q_\gamma^{(0)}, (V_{\text{bo}}^\beta)_{\text{diag}} \} + \frac{1}{L} \{Q_\beta^{(0)}, (V_{\text{bo}}^\gamma)_{\text{diag}} \}.
\end{align}
Since $Q_\beta^{(0)}$ and $Q_\gamma^{(0)}$ are (block)-diagonal with respect to the degenerate subspaces of $H_0$, this
equation implies
\begin{equation}
(V_{\text{bi}}^{\beta\gamma})_{\text{diag}} = \frac{1}{L} \{Q_\gamma^{(0)}, (V_{\text{bo}}^\beta)_{\text{diag}} \} - \frac{1}{L} \{Q_\beta^{(0)}, (V_{\text{bo}}^\gamma)_{\text{diag}} \}.
\end{equation}

We can obtain the AGP for the $V_{\text{bi}}$-type perturbation in terms of the $\tilde{X}$ and the AGP for the $V_{\text{bo}}$-type perturbations:
\begin{align}
\XAGP[V_{\text{bi}}^{\beta\gamma}] & = \widetilde{X}^{\beta\gamma} - (\widetilde{X}^{\beta\gamma})_{\text{diag}} + \frac{1}{L}\{Q_\gamma^{(0)}, \XAGP[V_{\text{bo}}^\beta] \} \notag\\
& -\frac{1}{L}\{Q_\beta^{(0)}, \XAGP[V_{\text{bo}}^\gamma]\} ~,
\end{align}
where the subtraction of the (block-)diagonal part of $\widetilde{X}^{\beta\gamma}$ appears since the $\XAGP$s are defined to have such parts equal to zero, and we have again used that $Q_\beta^{(0)}$ and $Q_\gamma^{(0)}$ are (block)-diagonal so that the block-diagonal parts of the $\sim 1/L$ terms are already zero.

Specializing to the cases we consider in this work, $V_{\text{bi},1} \equiv V_{\text{bi}}^{2,3}$ and $V_{\text{bi},2} \equiv V_{\text{bi}}^{2,4}$, we further note that $V_\text{bo}^{\beta=2} \sim J_{2,2;\text{tot}}^{(0)} \sim Q_3^{(0)}$ happens to be one of the IoMs in the Heisenberg chain, and hence $\XAGP[V_\text{bo}^{\beta=2}] = 0$.
In this case, the expressions simplify somewhat and involve the AGPs calculated for the already discussed $V_{\text{bo}}$-type perturbations, respectively $V_{\text{bo},1} \equiv V_{\text{bo}}^{\gamma=3}$ and $V_{\text{bo},2} \equiv V_{\text{bo}}^{\gamma=4}$:
\begin{align}
& \XAGP_{\text{bi},1} = \widetilde{X}^{2,3} - (\widetilde{X}^{2,3})_{\text{diag}} + \frac{1}{L} \{\XAGP_{\text{bo},1}, Q_2^{(0)} \} ~, \label{eq:Xbi1_Xbo1} \\
& \XAGP_{\text{bi},2} = \widetilde{X}^{2,4} - (\widetilde{X}^{2,4})_{\text{diag}} + \frac{1}{L} \{\XAGP_{\text{bo},2}, Q_2^{(0)} \} ~. \label{eq:Xbi2_Xbo2}
\end{align}
We can confirm that these relations are true for the $\XAGP$s we recover numerically for the $V_{\text{bi},1/2}$ perturbations and the corresponding $V_{\text{bo},1/2}$ perturbations, which we studied in Sec.~\ref{subsubsec:AGPVbo}.

Note that a direct verification requires calculating the diagonal part of $\widetilde{X}^{\beta\gamma}$, which we can do numerically.
On the other hand, if we formulate the task as checking the above relation effectively recovering $\widetilde{X}^{\beta\gamma}$ from just the $\XAGP[V_{\text{bi},1/2}]$ and $\XAGP[V_{\text{bo},1/2}]$ measurements, this is more non-trivial and in full completeness requires the generalized AGP formulation described in App.~\ref{app:sim_diag}.
For small sizes and with the right symmetries, the diagonal part of $\widetilde{X}^{\beta\gamma}$ may vanish already in the original AGP formulation and we can verify the corresponding simpler relations.
This is the case for $\widetilde{X}^{2,3}$ for the sizes in Fig.~\ref{fig:Vbi1} but not for $\widetilde{X}^{2,4}$ for Fig.~\ref{fig:Vbi2}, where we need to use the generalized AGP method.
In the former case, it is also important that we use a form of density of the $Q_3^{(0)}$ IoM that properly matches inversion properties of the $q_{2,j}^{(0)}$ density so that $\widetilde{X}^{2,3}$ has definite inversion symmetry; specifically, $\tilde{q}_{3,j}^{(0)} = (q_{3,j-1}^{(0)} + q_{3,j}^{(0)})/2$ satisfies this.
We refer readers to App.~\ref{app:sim_diag} for more details.

We can now discuss the relative importance of the different contributions to $\XAGP[V_{\text{bi}}]$ in the above equations.
Recall our finding in Sec.~\ref{sec:global-agp-scaling} that the AGP norm for a $V_{\text{bo}}$-type perturbation already has a suppressed norm compared to naive expectations, namely the numerical result $\| \XAGP_{\text{bo}} \|^{2} \sim L^{1+p}$ with $p < 1$ (in fact, $p$ is close to zero from the analysis in the table in Fig.~\ref{fig:agp-both-global}).
In the above expressions, the $\XAGP_{\text{bo}}$ is multiplied by an extensive operator $Q_2^{(0)}$ 
but is suppressed by the additional factor of $1/L$.
We can rigorously upper-bound the norm of this contribution by using the following bound on the Frobenius norm of a product of two operators $A$ and $B$:
\begin{equation}
\| A B \|_{\text{Frob.}} \leq \| A \|_{\text{op.}} \| B \|_{\text{Frob.}} ~,
\end{equation}
where the subscript ``Frob.''\ refers to the Frobenius norm (which is the default norm in this paper when no subscript is used), while ``op.'' refers to the usual operator norm.
Since $\| Q_2^{(0)} \|_{\text{op.}}/L = \text{const}$, we obtain
\begin{equation}
\left\| \frac{1}{L} \{\XAGP_{\text{bo}}, Q_2^{(0)} \} \right\|^{2} \leq a L^{1+p}
\label{eq:bound_norm_Xbi_correction}
\end{equation}
with some fixed number $a$ and $p<1$.

On the other hand, it is easy to see that
\begin{equation}
\| \widetilde{X}^{\beta\gamma} \|^2 = b L^2 + O(L)
\end{equation}
with some fixed number $b$.
Since we are able to construct $\widetilde{X}_{\text{bi}}$ we numerically confirm this scaling.
We note that if one only has access to smaller system sizes such as $L<20$, the power law fits to $\| \widetilde{X}^{\beta\gamma} \|^2 = c + d L^s$ give somewhat
exaggerated exponent closer to $s \approx 2.5$, and we suspect the scalings we report in the table in Fig.~\ref{fig:agp-both-global} have larger $p$ values than they would if we could access much larger sizes.
We additionally look at the scaling of $\| \widetilde{X}^{\beta\gamma} \|^2$ compared to the scaling of $\| \XAGP[V_{\text{bi}}^{\beta\gamma}] \|^2$ and find that our upper bounds in Eq.~(\ref{eq:bound_norm_Xbi_correction}) capture the correct power law.
Based on their scaling, contributions from $\widetilde{X}^{\beta\gamma}$ dominate, which is then a good approximation for $\XAGP[V_{\text{bi}}]$.
We can now say that we understand the origin of the behavior we saw in Fig.~\ref{fig:Vbi1} that initiated these considerations.

Finally, we note how the above formalism applies to $V_{\text{bi}}$-type weak integrability-breaking perturbations of free-fermion models with number conservation, such as those studied in Sec.~V~A in Ref.~\cite{Surace2023}.
In this case, each IoM $Q_\beta^{(0)}$ is a translationally invariant hopping ``Hamiltonian,'' with either real or imaginary hopping of some range, and the corresponding total current $J_{\beta,2;\text{tot}}$ is in fact another IoM.
Hence, the exact AGP in finite-size PBC systems is given by $\XAGP[V_{\text{bi}}^{\beta\gamma}] = \widetilde{X}^{\beta\gamma} - (\widetilde{X}^{\beta\gamma})_{\text{diag}}$.
Of course, if one just wants to satisfy $i[X, H_0] = V_{\text{bi}}^{\beta\gamma} - (V_{\text{bi}}^{\beta\gamma})_{\text{diag}}$, the full $\widetilde{X}^{\beta\gamma}$ will do.
This result was obtained independently by Bal\'azs Pozsgay~\cite{Pozsgay_notes2023}, and we are grateful to him for discussions and communications.

\section{AGP studies of strictly local perturbations}
\label{sec:local}
We now present AGP studies of strictly local perturbations of the Heisenberg chain.
We start by considering local terms such as the Hamiltonian density at a site $j_0$.
It is not true in general that taking a single local term from an extensive weak perturbation (such as the single term $\sv{0}\cdot \sv{2}$ from $V_{\text{bo},1}$) will be a weak integrability-breaking perturbation.
However, we can construct a strictly local weak perturbation using generators $X_{\text{loc}}$ and $X_{\text{step}}$ defined in Sec.~\ref{sec:generators}.
These two classes of generators can be thought of as local versions of the extensive and boosted operators respectively.

We refer to these perturbations as ``local'' to emphasize that they have finite support localized around some site $j_0$ and thus break the translational symmetry of the Hamiltonian.
Note that we do not know a procedure that can be applied to construct strictly local weak-integrability breaking perturbations that would be local versions of the bilocal generators.
On the other hand, it is really easy to pick a strong integrability-breaking perturbation, since generic strictly local perturbations, just like translationally invariant ones, are expected to break integrability strongly.

\subsubsection{Strong perturbations}
\label{sec:strong-local}
The choices of local perturbations in this section are meant to illustrate the nuance in defining strictly local weak integrability-breaking perturbations.
Taking the Hamiltonian density at a single site (more generally, a local part of an IoM) or the already mentioned second-neighbor interaction (i.e., a local part of the particular writing of the weak integrability-breaking $V_{\text{bo},1}$) might seem as a possible choice for a weak local perturbation, but they both break the integrability strongly.
Specifically, we consider the following three perturbations:
\begin{align}
& V_{\text{s},3} = \vec{\sigma}_0 \cdot \vec{\sigma}_1 ~, \label{eq:Vs3} \\
& V_{\text{s},4} = \vec{\sigma}_0 \cdot \vec{\sigma}_2 ~, \label{eq:Vs4} \\
& V_{\text{s},5} = \vec{\sigma}_0 \cdot \vec{\sigma}_3 ~. \label{eq:Vs5}
\end{align}
We kept the naming convention for strong perturbations and continue to use ``$V_{\text{s}}$'' since all strong perturbations break integrability in the same way.
This is unlike the weak perturbations, where each weak perturbation breaks integrability in its own way.

\subsubsection{$X_{\text{loc}}$-generated perturbations}
\label{sec:V-local}
The choices we made for the generators of extensive local perturbations in Sec.~\ref{sec:weak-Xex-global} naturally lend themselves to the construction of strictly local weak perturbations that we discuss in this section.
We construct weak perturbations in this section by taking the local terms of the previously defined generators of $V_{\text{ex}}$s at a single site around $j_0$ and commute them with the Hamiltonian.
This procedure generates strictly local weak integrability-breaking perturbations that we study next.
Note that one could have picked any other group of extensive local operators for $X$.
Our choice is convenient for two reasons:
First, we include different symmetry scenarios while matching the ones we already discussed at the same time; second, this choice is equivalent to saying that we use the smallest operators that satisfy given symmetry properties.
An additional benefit of this choice will become clear when we discuss the extracted $\XAGP_{\text{loc}}$ in the next section.
We now define the three strictly local perturbations we study in this section:
\begin{align}
&V_{\text{loc},1} = 2
(\vec{\sigma}_0 \times \vec{\sigma}_2) \cdot (\vec{\sigma}_3 - \vec{\sigma}_{-1}) ~,
\label{eq:Vloc1} \\
&V_{\text{loc},2} = 8 \vec{\sigma}_0 \cdot \vec{\sigma}_3 - 4 \vec{\sigma}_0 \cdot \vec{\sigma}_2 - 4 \vec{\sigma}_1 \cdot \vec{\sigma}_3 \label{eq:Vloc2} \\
&\quad + 2 ((\vec{\sigma}_{-1} \times \vec{\sigma}_0) \times \vec{\sigma}_1) \cdot \vec{\sigma}_3 + 2 ((\vec{\sigma}_0 \times \vec{\sigma}_2) \times \vec{\sigma}_3) \cdot \vec{\sigma}_4 \nonumber \\ 
&\quad - 2 ((\vec{\sigma}_{-1} \times \vec{\sigma}_0) \times \vec{\sigma}_2) \cdot \vec{\sigma}_3 - 2 ((\vec{\sigma}_0 \times \vec{\sigma}_1) \times \vec{\sigma}_3) \cdot \vec{\sigma}_4 \nonumber \\
&\quad + 4 ((\vec{\sigma}_0 \times \vec{\sigma}_2) \times \vec{\sigma}_1) \cdot \vec{\sigma}_3 ~, \nonumber \\
& V_{\text{loc},3} = 4 \vec{\sigma}_0 \cdot \vec{\sigma}_2 - 4 \vec{\sigma}_1 \cdot \vec{\sigma}_3 \label{eq:Vloc3} \\
&\quad + 2 ((\vec{\sigma}_{-1} \times \vec{\sigma}_0) \times \vec{\sigma}_1) \cdot \vec{\sigma}_3 - 2 ((\vec{\sigma}_0 \times \vec{\sigma}_2) \times \vec{\sigma}_3) \cdot \vec{\sigma}_4 \nonumber \\
&\quad + 2 ((\vec{\sigma}_{-1} \times \vec{\sigma}_0) \times \vec{\sigma}_2) \cdot \vec{\sigma}_3 - 2 ((\vec{\sigma}_0 \times \vec{\sigma}_1) \times \vec{\sigma}_3) \cdot \vec{\sigma}_4 ~, \nonumber
\end{align}
which are obtained using generators 
\begin{align}
X_{\text{loc},1} &=  \vec{\sigma}_0 \cdot \vec{\sigma}_2 ~, \\
X_{\text{loc},2} &= (\vec{\sigma}_0 \times \vec{\sigma}_3) \cdot (\vec{\sigma}_1 - \vec{\sigma}_2)  ~, \\
X_{\text{loc},3} &= (\vec{\sigma}_0 \times \vec{\sigma}_3) \cdot (\vec{\sigma}_1 + \vec{\sigma}_2) ~.
\end{align}
Note that each $V_{\text{loc}}$ has the same $\Theta$ and $I$ symmetry properties as the respective extensive perturbations from the previous section, with the exception of the inversion $I$ being defined with respect to a single point (site $1$ for $V_{\text{loc},1}$ and bond center between sites $1$ and $2$ for $V_{\text{loc},2}$ and $V_{\text{loc},3}$). 
One difference from $V_{\text{ex}}$s defined earlier is that expressions for $V_{\text{loc}}$s seem more complicated.
This is because if, e.g., we naively tried a strictly local perturbation $V = (\vec{\sigma}_0 \times \vec{\sigma}_3) \cdot (\vec{\sigma}_1 - \vec{\sigma}_2)$ motivated from the writing of $V_{\text{ex},1}$ in Eq.~(\ref{eq:Vex1}) instead of the above defined $V_{\text{loc},1}$, we would find it to be strong integrability-breaking.
While $\sum_j V_{\text{loc},1}(j) = V_{\text{ex},1}$, it is not true that simply taking a term in any writing of $V_{\text{ex},1}$ at a single site $j_0$ would give a weak perturbation as well. 
Note that we set $j_0 = 0$ in the cases defined above, but this is true no matter the choice of $j_0$.
As long as the perturbation is obtained by evaluating the commutator with an $X_{\text{loc}}$, it will be a weak perturbation.

\subsubsection{$X_{\text{step}}$-generated perturbations}
\label{sec:Xstep}
A more interesting class of strictly local perturbations are those generated by $X_{\text{step}}$, which is a local counterpart of the boost.
Given our definition of $X_{\text{step}}$ in Eq.~(\ref{eq:Xstep}) and $V_{\text{step}}$ in Eq.~(\ref{eq:Vstepbeta}), we construct the following three weak perturbations from the local currents associated with the first three IOMs defined in Sec.~\ref{subsec:Heischain}:
\begin{align}
& V_{\text{step},0} = 2J_{2,2;j_0} = (\vec{\sigma}_0 \times \vec{\sigma}_1) \cdot \vec{\sigma}_2 , \label{eq:Vstep0} \\
& V_{\text{step},1} = 2J_{3,2;j_0} = -((\vec{\sigma}_0 \times \vec{\sigma}_1) \times \vec{\sigma}_2) \cdot \vec{\sigma}_3 - 2\vec{\sigma}_1  \cdot \vec{\sigma}_2 , \label{eq:Vstep1} \\
& V_{\text{step},2} = 2J_{4,2;j_0} =  2(((\vec{\sigma}_0 \times \vec{\sigma}_1) \times \vec{\sigma}_2) \times \vec{\sigma}_3) \cdot \vec{\sigma}_4 \notag\\
& \quad + (\vec{\sigma}_0 \times \vec{\sigma}_1) \cdot \vec{\sigma}_3 + (\vec{\sigma}_1 \times \vec{\sigma}_3) \cdot \vec{\sigma}_4.
\label{eq:Vstep2}
\end{align}
where we specialized to $j_0 = 1$ in each case.

For the Heisenberg model, the $J_{2,2;j_0}$ is a local energy current that also happens to correspond to the local density of the IoM $Q_3^{(0)}$.
We already saw the importance of this fact when we discussed the relations for the bilocal generators, but note that this is not true for other IoMs.
$J_{3,2;j_0}$ and $J_{4,2;j_0}$ are local currents of $Q_3^{(0)}$ and $Q_4^{(0)}$ respectively, and while parts of them resemble local parts of the IoMs $Q_4^{(0)}$ and $Q_5^{(0)}$ respectively, they are not the same.
Furthermore, while $2J_{3,2;\text{tot}}$ is equal to $V_{\text{bo},1}$ up to true IoMs, cf.\ Eq.~(\ref{eq:Vbo1}), the latter rewriting as a sum of local terms reorganizes contributions from $2J_{3,2;j}$s.
As a result, the ``natural'' density of $V_{\text{bo},1}$ turns out to be the strong perturbation $V_{\text{s},4}$, while it is the actual local current $J_{3,2;j_0}$ that is the weak perturbation $V_{\text{step},2}$.

\begin{figure*}[ht]
    \centering
    \begin{minipage}[c]{0.5\linewidth}
        \includegraphics[width=\linewidth]{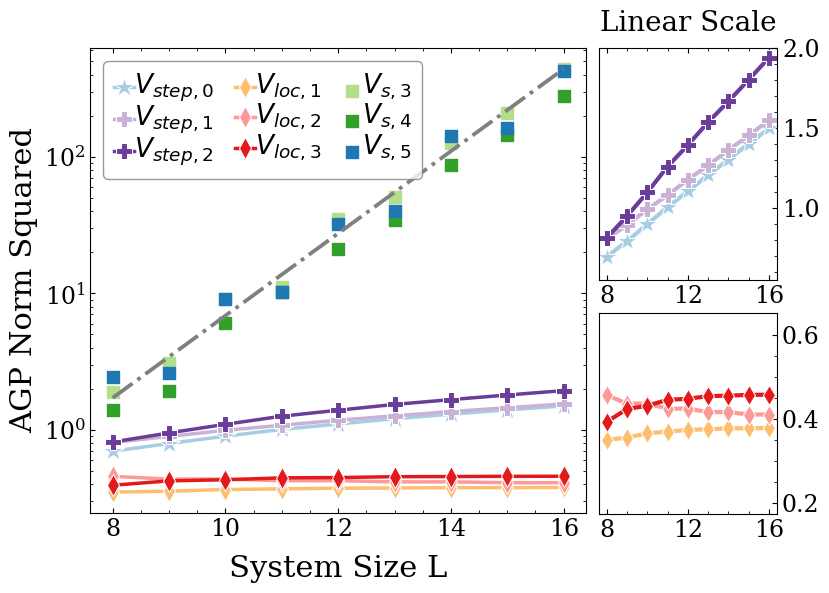}
    \end{minipage}
    \hfill
    \begin{minipage}[c]{0.49\linewidth}
        {\renewcommand{\arraystretch}{1.05}
        \begin{tabular}{c c c c c c}
        \centering
            $V$ & Eq.~reference & $\Theta$ & $I$ & ~$\|\XAGP\|^2$ & fit param. \\
           \midrule
            \midrule
            $V_{\text{step,0}} $  & Eq.~(\ref{eq:Vstep0}) & odd & odd & $a+bL^p $ &  $p=0.865$  \\
            $V_{\text{step},1}$ & Eq.~(\ref{eq:Vstep1}) & even &even & $a+bL^p$  & $p=1.032$ \\
            $V_{\text{step},2} $ &Eq.~(\ref{eq:Vstep2}) & odd & odd & $a+bL^p$ & $p=0.796$ \\
            \midrule
            \midrule
            $V_{\text{loc},1}$  & Eq.~(\ref{eq:Vloc1}) & odd & even & const. &  \\
            $V_{\text{loc},2}$ &Eq.~(\ref{eq:Vloc2})   & even & even & const. &  \\
            $V_{\text{loc},3}$ &Eq.~(\ref{eq:Vloc3})  & even& odd  & const. & \\
            \midrule
            \midrule
            $V_{\text{s},3}$ & Eq.~(\ref{eq:Vs3})& even & even& $a+be^{\kappa L}$ & $\kappa=0.68$  \\
            $V_{\text{s},4} $ & Eq.~(\ref{eq:Vs4})&even & even &  $a+be^{\kappa L}$ & $\kappa=0.72$\\
            $V_{\text{s},5} $ & Eq.~(\ref{eq:Vs5}) & even & even &   $a+be^{\kappa L}$ & $\kappa=0.64$ \\
           \midrule
        \end{tabular}}
    \end{minipage}
   \caption{Left, main panel: The AGP norm squared, $\| X^{\text{AGP}} \|^2$, as a function of the system size $L$ for the Heisenberg Hamiltonian perturbed by the local perturbations given in the table on the right.
   Dashed line shows the exponential function $\varpropto e^{\kappa L}$ with $\kappa = \ln{2}$ expected by ETH for strong perturbations, while detailed scaling forms and fits in all cases are listed in the table.
   Left, small panels: The weak integrability breaking perturbations only and on a linear scale; top: perturbations with known step generators in the infinite system; bottom: perturbations with known local generators on any finite PBC chain. As in Fig.~\ref{fig:agp-both-global}, some of the perturbations have been rescaled by a constant ($L$-independent) prefactor.
   Right, table: Listing of the perturbations studied with reference to equations; quantum numbers under the time reversal $\Theta$ and inversion $I$ symmetries; and their observed AGP norm scaling in the $S^z_{\text{tot}} = 0$ sector including key fit parameters.
   }
    \label{fig:agp-both-local}
\end{figure*}

\subsection{Scaling of the AGP norms}
Figure~\ref{fig:agp-both-local} shows the scaling of the AGP norms for the perturbations defined above.
In the previous section, when discussing extensive perturbations, we were interested in the scaling of $\| \XAGP \|^2/L$.
However, since the perturbations in this section are strictly local, we will not be rescaling the norm by $L$.
This is an important difference between Fig.~\ref{fig:agp-both-global} and Fig.~\ref{fig:agp-both-local}.
Note that similar arguments as the ones presented in Sec.~\ref{sec:global-agp-scaling} apply here as well. 
The AGP norm for the strong perturbations is expected to grow exponentially with the system size, while the AGP norm for $V_{\text{loc}}$-type perturbations is not expected to scale with $L$ at all and is expected to converge to a fixed value with the system size. 
We see in Fig.~\ref{fig:agp-both-local} that both of these predictions are confirmed by the numerics.

Naively placing an $X_{\text{step}}$-type generator onto a finite chain with $j_0$, e.g., somewhere in the middle of the chain, would give an operator whose squared norm would scale proportional to $L$.
We indeed observe that for each of the $V_{\text{step}}$-type perturbations their AGP norm squared scales with $L$.
What is more interesting is that the Pauli string characterization in the next section helps us to understand this result systematically.
Specifically, we are able to reduce the $\XAGP$ calculation problem to finding an $\widetilde{X}$ which we obtain analytically and which is a well-defined PBC proxy of the step function, and a part related to the $\XAGP$ of one of the $V_{\text{bo}}$-type perturbations from the previous section.

\begin{figure}[ht]
\centering
\includegraphics[width=0.49\linewidth]{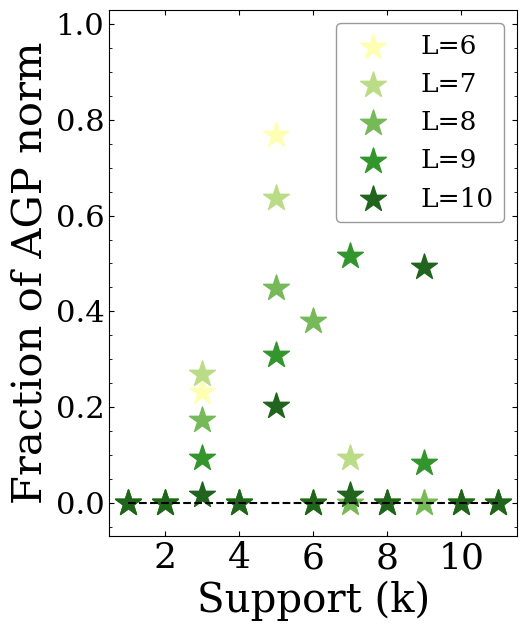}
\includegraphics[width=0.49\linewidth]{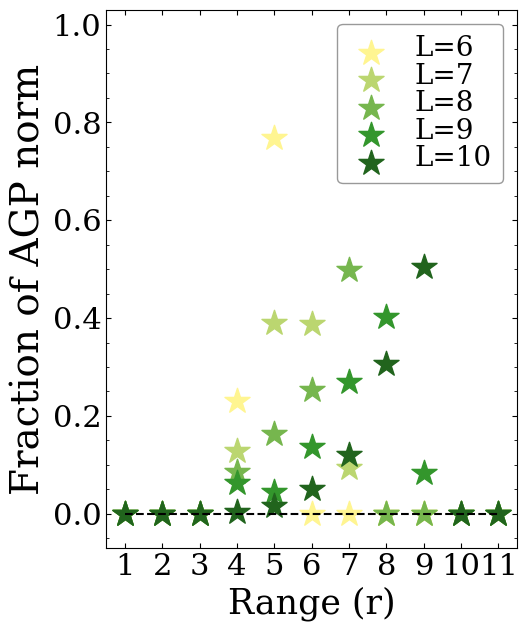}
\caption{Visualization of the structure of the AGP in terms of the decomposition in the Pauli string basis for $\XAGP$ when $H_0$ is perturbed by $V_{\text{s},4}$.
Similarly to the case of $V_{\text{s},1}$ in Fig.~\ref{fig:agp_vs1}, we observe that the AGP of a chaotic perturbation appears to be a highly non-local operator with an exponentially growing norm.}
\label{fig:agp_vs4}
\end{figure}

\subsection{Characterization of the AGPs in the Pauli string basis}
\label{subsec:agpPauliS_local}
\subsubsection{AGPs for strong perturbations}
Despite being strictly local perturbations, $V_{\text{s},3}$, $V_{\text{s},4}$, and $V_{\text{s},5}$ have the same dominant effect as extensive strong perturbations that we discussed in Sec.~\ref{subsec:agp_PauliS2}.
Breaking integrability ``strongly'' looks essentially the same regardless of the specific perturbation.
We illustrate this with the AGP for $V_{\text{s},4}$ in Fig.~\ref{fig:agp_vs4}, where we again see the weight of the AGP moving towards larger supports and longer ranges as we increase the system size. 
Thus we see the same non-locality of the AGP operator as for the extensive strong perturbations $V_{\text{s},1}$ and $V_{\text{s},2}$.

\begin{figure}[ht]
\centering
\includegraphics[width=0.49\linewidth]{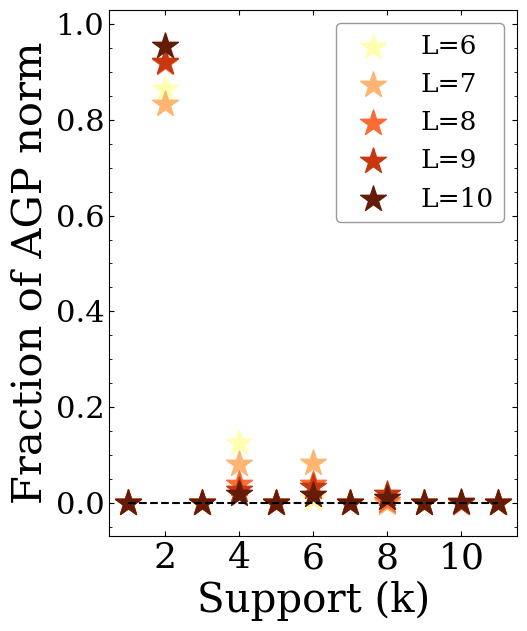}
\includegraphics[width=0.49\linewidth]{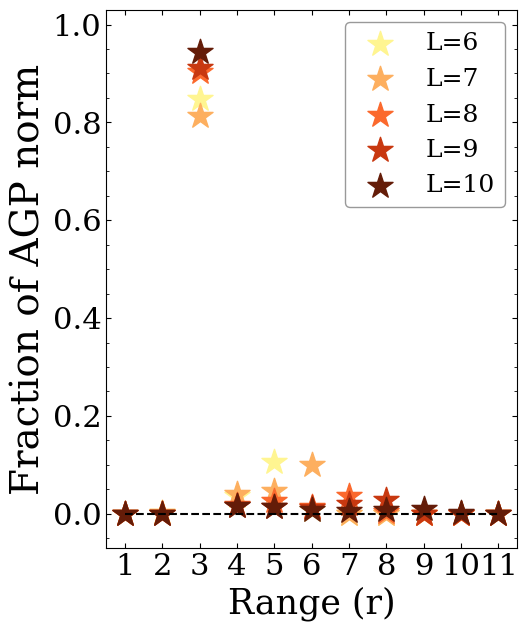}
\includegraphics[width=0.49\linewidth]{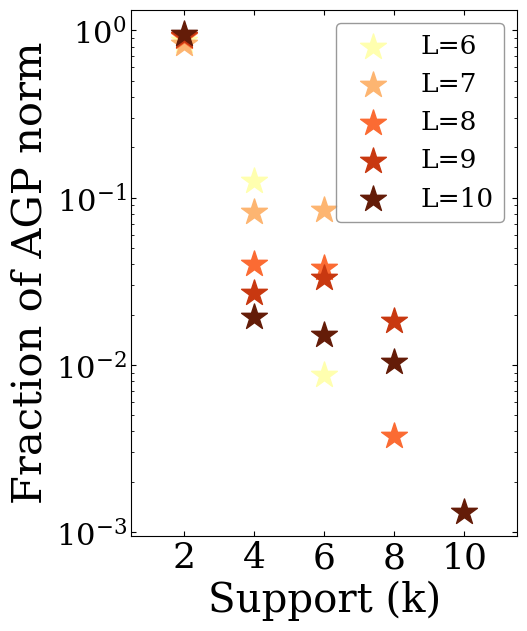}
\includegraphics[width=0.49\linewidth]{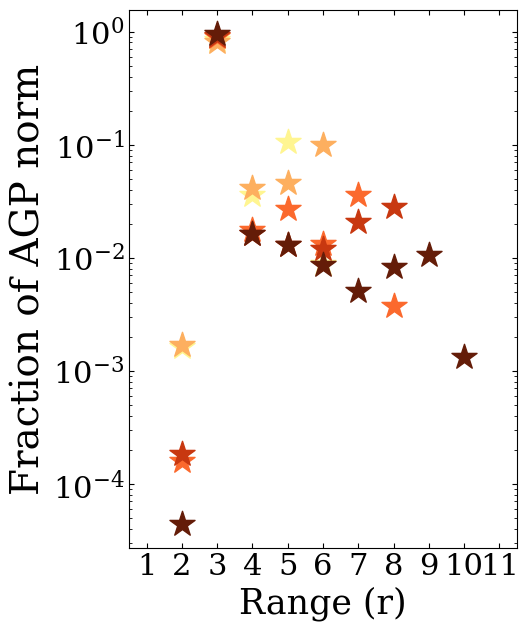}
\caption{Visualization of $\XAGP$ for $V_{loc,1}$ in terms of the decomposition in the Pauli string basis by support $k$ and range $r$ (see caption in Fig.~\ref{fig:Vex1} for details.) We note again that there seems to be a pattern here: an operator that is well described by support $k=2$ strings and a much smaller part that resembles the patterns observed for the extensive perturbations from the previous section. Again, we do not show data for $V_{loc,2}$ as it is exactly recovered and we leave $V_{loc,3}$ for the appendix (\ref{app:add_data_for_agp_global}).}
\label{fig:Vloc1}
\end{figure}

\subsubsection{AGPs for $X_{\text{loc}}$-generated perturbations}
For $V_{\text{loc},2}$, which is an appropriate local version of $V_{\text{ex},2}$ with the same symmetries, the computed $\XAGP$ recovers the exact $\Xloc$, just as one would expect.
As was the case with $V_{\text{ex},2}$, we do not show data for this complete recovery case.
For $V_{\text{loc},1}$ and $V_{\text{loc},3}$, on the other hand, we do not recover the exact $\Xloc$.
These cases are analogous to those of $V_{\text{ex},1}$ and $V_{\text{ex},3}$, where the AGPs could not be recovered exactly because of the missing diagonal part.
Figures~\ref{fig:Vloc1} and \ref{fig:Vloc3} illustrate the AGP decomposition in Pauli strings for $V_{\text{loc},1}$ and $V_{\text{loc},3}$ respectively.

The presence of Pauli strings with different supports and ranges (even supports in the AGP for $V_{\text{loc,1}}$ and odd supports in the AGP for $V_{\text{loc,3}}$ as required by their symmetries under $\Theta$) indicates that both $\XAGP_{\text{loc},1}$ and $\XAGP_{\text{loc},3}$ differ from $X_{\text{loc},1}$ and $X_{\text{loc},3}$ because of the missing diagonal part.
However, the dominant strings seen in the AGPs constitute precisely the support-2 range-3 operator $X_{\text{loc},1}$ and support-3 range-4 operator $X_{\text{loc},3}$ respectively, and their weight appears to grow towards 1 as the system size increases.

We note that, in any eigenstate $\ket{\Psi}$ of the translation operator $T$, $T \ket{\Psi} = e^{ik} \ket{\Psi}$, the expectation value $\bra{\Psi} X_{\text{loc},1} \ket{\Psi}$ is independent of the location $j_0$ and hence equals $(1/L) \bra{\Psi} X_{\text{ex},1} \ket{\Psi}$ (similar relation holds between $X_{\text{loc},3}$ and $X_{\text{ex},3}$).
Since we have noted the missing diagonal parts in $\XAGP_{\text{loc},1/3}$ and $\XAGP_{\text{ex},1/3}$, this suggests that we may be able to relate these AGPs even without knowing the diagonal parts.
And indeed, we numerically find that the following relations hold:
\begin{align}
& X_{\text{loc},1} - X_{\text{loc},1}^{\text{AGPgen}} = \frac{1}{L}( X_{\text{ex},1} - X_{\text{ex},1}^{\text{AGPgen}}) ~, \label{eq:diagrelXloc1Xex1} \\
& X_{\text{loc},3} - X_{\text{loc},3}^{\text{AGPgen}} = \frac{1}{L} (X_{\text{ex},3} - X_{\text{ex},3}^{\text{AGPgen}}) ~. \label{eq:diagrelXloc3Xex3}
\end{align}
However, to come up with these relations, we needed to use a more refined definition of AGP denoted here as $X^{\text{AGPgen}}$, which is the generalized AGP formulation described in App.~\ref{app:sim_diag}.
With this definition, the missing block-diagonal part of the exact $X$ that we are trying to recover from its $\XAGP$ proxy is partially fixed by using the information about the known corrections $Q_\alpha^{(1)}$ of the IoMs.
As explained in App.~\ref{app:sim_diag}, this assumes complete resolution of degeneracies with the help of additional IoMs and reduces the ambiguity in the recovery of $X$ to the strictly diagonal part in this basis, while at the same time the basis vectors are guaranteed to be momentum eigenstates ensuring applicability of $\bra{\Psi} X_{\text{loc}} \ket{\Psi} = (1/L) \bra{\Psi} X_{\text{ex}} \ket{\Psi}$ discussed earlier.

We illustrate the above relation in Appendix~\ref{app:sim_diag} in 
Figs.~\ref{fig:XAGPgen_Vex1} and \ref{fig:XAGPgen_Vloc1} where we compare $\XAGPgen_{\text{ex},1}$ and $\XAGPgen_{\text{loc},1}$
and see visual agreement in the patterns of the distribution of strings with $k$ and $r$ excluding the $k=2, r=3$ strings that define $X_{\text{loc},1}$ and $X_{\text{ex},1}$.
If one compares Figs.~\ref{fig:Vex1} and \ref{fig:Vloc1} instead, such relations are not immediately obvious.

\subsubsection{AGPs for $X_{\text{step}}$-generated perturbations}
\label{subsubsec:step}
\begin{figure}[ht]
\centering
\includegraphics[width=0.49\linewidth]{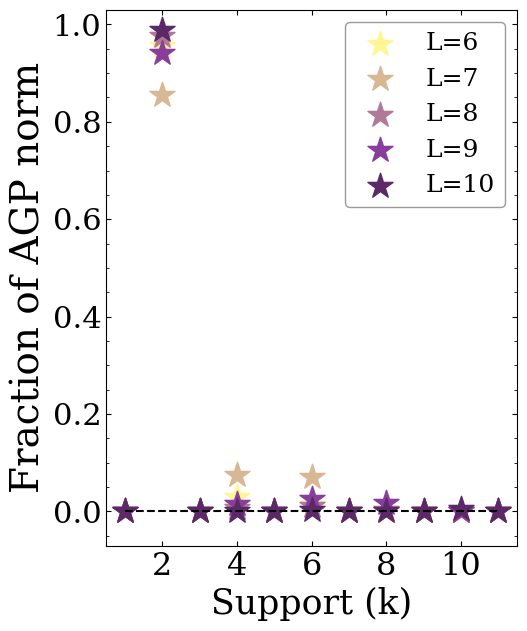}
\includegraphics[width=0.49\linewidth]{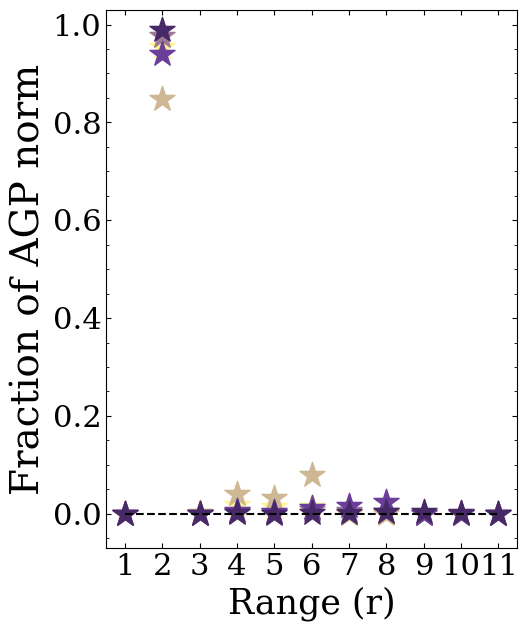}
\includegraphics[width=0.49\linewidth]{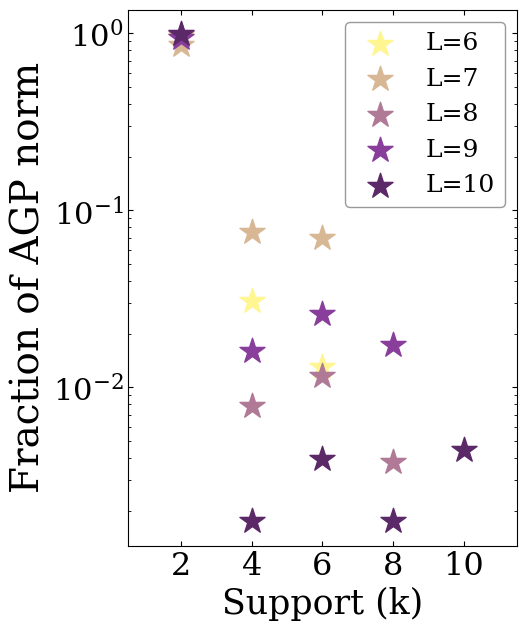}
\includegraphics[width=0.49\linewidth]{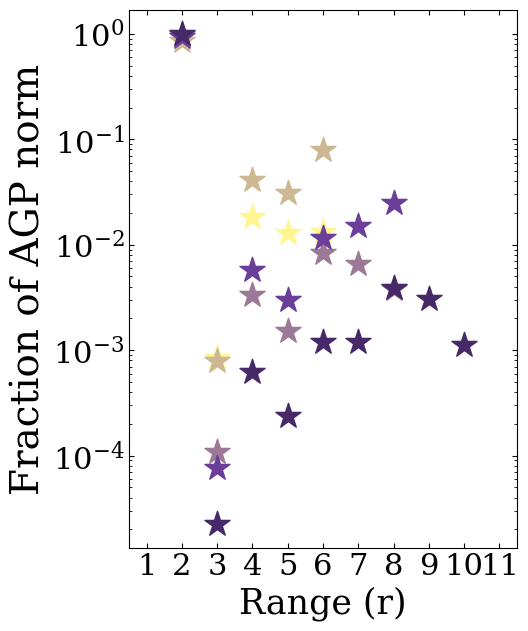}
\caption{Visualization of $X^{\text{AGP}}$ for $V_{\text{step},0}$ in terms of the decomposition in the Pauli string basis organized by support $k$ and range $r$.
Similarly to the case of $V_{\text{loc}}$s, we are encouraged by the plots to look for an approximate closed-form $X$, since unlike for the boost, weight on the higher support strings is quickly dropping with system size.}
\label{fig:vstep0}
\end{figure}

\begin{figure}[ht]
\centering
\includegraphics[width=0.49\linewidth]{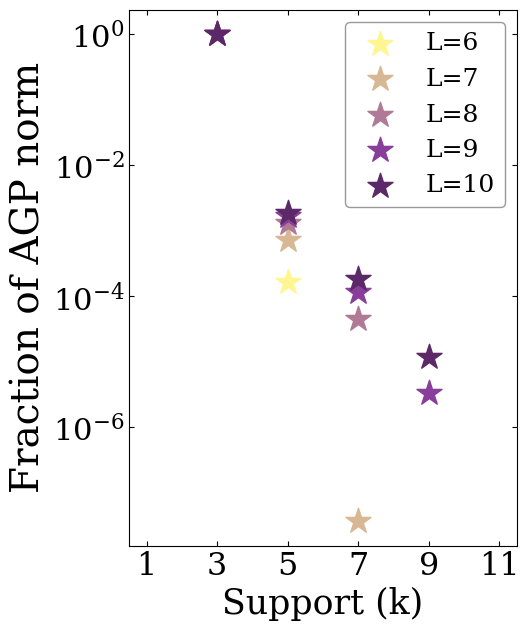}
\includegraphics[width=0.49\linewidth]{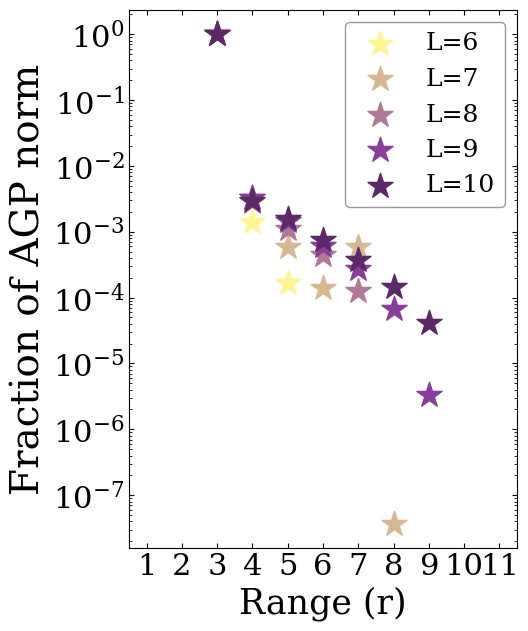}
\caption{Visualization of $X^{\text{AGP}}$ for $V_{\text{step},1}$ in terms of the decomposition in the Pauli string basis organized by support $k$ and range $r$.
We report only the log-scale plots here since the difference between the weight on the support $k=2$ and $k=4$ strings is too large and one does not get to observe the pattern seen above which Eq.~(\ref{eq:XAGP4Vstep}) fully captures.}
\label{fig:vstep1}
\end{figure}

Figures~\ref{fig:vstep0} and \ref{fig:vstep1} show Pauli string characterization of the $\XAGP$ for the $V_{\text{step},0}$ and $V_{\text{step},1}$ perturbations respectively.
In both cases, we see that the dominant weight is on the short Pauli strings of the type present in the corresponding infinite-system $\Xstep$ generators [Eq.~(\ref{eq:Xstep})]: strings $\vec{\sigma}_j \cdot \vec{\sigma}_{j+1}$ with $k=2, r=2$ in the $V_{\text{step},0}$ case and strings $(\vec{\sigma}_j \times \vec{\sigma}_{j+1}) \cdot \vec{\sigma}_{j+2}$ with $k=3, r=3$ in the $V_{\text{step},1}$ case.

The weight on all other types of strings is much smaller and decreases quickly with the system size (modulo even/odd effect) in the $V_{\text{step},0}$ case, in contrast to the pattern observed in the $V_{\text{bo}}$-type cases earlier.
In the $V_{\text{step},1}$ and $V_{\text{step},2}$ cases, the weight on the sub-dominant strings appears to saturate with the system size to values that decrease exponentially with $k$ and $r$, similar to the pattern in the boosted cases (the observed saturation may be due to small system sizes used in the Pauli string studies).

A more detailed examination of the spatial distribution of the amplitudes on the dominant strings (which cannot be seen from the aggregate information in Figs.~\ref{fig:vstep0} and \ref{fig:vstep1}) indicates an interesting saw-tooth pattern with location $j$, showing a jump near $j_0$ and a gradual nearly linear variation across the system that connects continuously across the PBC to the values on the two sides of the jump.

Motivated by these observations, we then attempt to capture the dominant part of the AGP by the following ansatz:
\begin{equation}
\widetilde{X}^\beta = \sum_{j=1}^L c_j q_{\beta,j}^{(0)} ~,
\label{eq:tildeXbeta}
\end{equation}
which resembles the infinite-system generator in Eq.~(\ref{eq:Xstep}) but is defined on the finite PBC chain.
Using Eq.~(\ref{eq:current}) we can calculate
\begin{align}
i[\widetilde{X}^\beta, Q_\alpha^{(0)}] = \sum_{j=1}^L J_{\beta\alpha,j} (c_{j-1} - c_j) ~,
\end{align}
where we have reorganized the sum to resemble the structure of the perturbation $V_{\text{step}}$ in Eq.~(\ref{eq:Vstepbeta}).
Since on the PBC chain $\sum_j (c_{j-1} - c_j) = 0$, we cannot match such $V_{\text{step}}$, but we can get close to it by requiring
\begin{equation}
c_{j-1} - c_j = \delta_{j,j_0} - \frac{1}{L} ~.
\end{equation}

We can solve for the amplitudes, e.g., as
\begin{equation}
c_j = \begin{cases}
-\frac{1}{2} + \frac{1}{2L} + \frac{1}{L}(j-j_0), & \text{if $j = j_0, j_0+1, \dots L$,} \\
\frac{1}{2} + \frac{1}{2L} + \frac{1}{L}(j - j_0), & \text{if $j = 1, 2, \dots, j_0 - 1$,}
\end{cases}
\label{eq:cj_tildeXbeta}
\end{equation}
where for a more clear illustration of the step between $j_0 - 1$ and $j_0$ we have assumed $j_0 > 1$ so that $1 \leq j_0 - 1 < L$.
This is then extended to other $j$ using PBC; e.g., $c_0 \equiv c_L = 1/2 + 1/(2L) - j_0/L$ and satisfies $c_0 - c_1 = -1/L$.
Note that an overall shift of $c_j$s does not change $[\widetilde{X}^\beta, Q_\alpha^{(0)}]$, but the above choice gives $\widetilde{X}^\beta$ a definite inversion quantum number with respect to one location, which will be helpful later.
This is the saw-tooth pattern of the amplitudes that we first noticed numerically in the AGP studies for $V_{\text{step},0}$ and $V_{\text{step},1}$ described earlier.
We can view it as the PBC proxy for the step function appearing in the infinite-system generator in Eq.~(\ref{eq:Xstep}).

From now on we use the above $c_j$ and obtain
\begin{equation}
i[\widetilde{X}^\beta, Q_\alpha^{(0)}] = J_{\beta\alpha,j_0} - \frac{1}{L} J_{\beta\alpha;\text{tot}} ~.
\label{eq:comm_tXbeta_Qalpha}
\end{equation}
Taking $\alpha = 2$ and recalling $H_0 = 2Q_2^{(0)}$, we have
\begin{equation}
i[\widetilde{X}^\beta, H_0] = 2J_{\beta,2;j_0} - \frac{1}{L} 2J_{\beta,2;\text{tot}} = V_{\text{step}}^\beta - \frac{1}{L} V_{\text{bo}}^\beta ~.
\label{eq:tildeXbeta_comm_H0}
\end{equation}
Thus, with the help of $\widetilde{X}^\beta$, we have effectively reduced the problem of calculating the AGP for $V_{\text{step}}^\beta$ to that of calculating the AGP for the translationally invariant $V_{\text{bo}}^\beta$ studied numerically in Sec.~\ref{sec:transl}.
More precisely, the above equation implies that $V_{\text{step}}^\beta$ and $V_{\text{bo}}^\beta/L$ have the same (block-)diagonal part with respect to the degenerate subspaces of $H_0$, and we obtain
\begin{equation}
i\left[\widetilde{X}^\beta + \frac{1}{L} \XAGP[V_{\text{bo}}^\beta], H_0 \right] = V_{\text{step}}^\beta - (V_{\text{step}}^\beta)_{\text{diag}} ~.
\end{equation}
We then conclude that the following relation holds exactly:
\begin{equation}
\XAGP[V_{\text{step}}^\beta] =  \widetilde{X}^\beta - (\widetilde{X}^\beta)_{\text{diag}} + \frac{1}{L} \XAGP[V_{\text{bo}}^\beta] ~,
\label{eq:XAGP4Vstep}
\end{equation}
where the subtraction of the (block-)diagonal part of $\widetilde{X}^\beta$ appears since the $\XAGP$s are defined to have such parts equal to zero.

Let us now consider our specific $V_{\text{step}}$-type perturbations.
In the case of $V_{\text{step},0} \equiv V_{\text{step}}^{\beta=2}$, the related $V_{\text{bo}}$-type perturbation, $V_{\text{bo}}^{\beta=2} \sim J_{2,2;\text{tot}} \sim Q_3^{(0)}$, happens to be an IoM.
Hence, $\XAGP[V_{\text{bo}}^{\beta=2}] = 0$, leading to
\begin{equation}
\XAGP[V_{\text{step},0}] = \widetilde{X}^{\beta=2} - (\widetilde{X}^{\beta=2})_{\text{diag}} ~.
\label{eq:XAGP4Vstep0}
\end{equation}
We can use these result as follows.
From Eq.~(\ref{eq:tildeXbeta_comm_H0}) we see that the simple generator $\widetilde{X}^{\beta=2}$ reproduces the $V_{\text{step},0}$ perturbation exactly on the PBC chain up to an IoM contribution $\sim J_{2,2;\text{tot}}/L$.
We can view $\XAGP[V_{\text{step},0}]$ as trying to capture this $\widetilde{X}^{\beta=2}$, which it does up to the (block-)diagonal part.
This is what we see in the Pauli string characterization of $\XAGP[V_{\text{step},0}]$ in Fig.~\ref{fig:vstep0}.
The non-zero weight on the sub-dominant Pauli strings in the figure comes from the $(\widetilde{X}^{\beta=2})_{\text{diag}}$ part in Eq.~(\ref{eq:XAGP4Vstep0}), and the numerical data suggests that the relative weight of this part goes to zero as we increase $L$.

For the following methodological discussion, let us pretend that we do not know the exact nice PBC generator $\widetilde{X}^{\beta=2}$, but our goal is to recover it from the measured $\XAGP[V_{\text{step},0}]$.
One way to proceed could be to consider $X' = \XAGP[V_{\text{step},0}] + Y_{\text{diag}}$ with arbitrary (block)-diagonal $Y_{\text{diag}}$ and to try to find the nicest (in some Pauli string locality measure) $X'$ by using this freedom, in the spirit described in App.~\ref{app:brute-force-Xdiag}.
Alternatively and more controllably, we can use  additional IoMs and apply the generalized AGP approach described in App.~\ref{app:sim_diag} to resolve some of the $H_0$ degeneracies and fill in some of the missing matrix elements of the sought-for generator that were set to zero in the original $\XAGP$ construction.
Specifically, here we have additional information about the weak integrability breaking perturbation that, besides $V_{\text{step},0} = 2Q_2^{(1)} = 2J_{2,2;j_0}$, the same infinite-system generator also gives us quasi-IoM corrections, $Q_\alpha^{(1)} = J_{2,\alpha;j_0}$, and we can feed this information to the generalized AGP approach.
As described in App.~\ref{app:sim_diag}, for essentially all system sizes used in Fig.~\ref{fig:vstep0}, $Q_3^{(0)}$ already resolves all degeneracies of $H_0$. 
When we calculate the corresponding $X^{\text{AGPgen}}$, Eq.~(\ref{eq:AGPgen}), using the quasi-IoM correction $Q_3^{(1)} = J_{2,3;j_0}$, we find that the $X^{\text{AGPgen}}$ matches exactly the analytical $\widetilde{X}^{\beta=2}$!
That is, an analog of Fig.~\ref{fig:vstep0} for the $X^{\text{AGPgen}}$ has all Pauli string amplitudes equal to zero except for those in $\widetilde{X}^{\beta=2}$.
We provide some additional details of understanding this case in App.~\ref{subapp:applications_XAGPgen}.

Turning to the other $V_{\text{step}}$-type perturbations, we no longer know $\XAGP[V_{\text{bo}}^\beta]$ analytically and need to use the numerically calculated AGP properties for the relevant $V_{\text{bo}}$-type perturbations from Sec.~\ref{sec:transl}.
In the case of $V_{\text{step},1} \equiv V_{\text{step}}^{\beta=3}$ from Fig.~\ref{fig:vstep1}, we find that the following relation holds to our numerical accuracy for these sizes:
\begin{equation}
\XAGP[V_{\text{step},1}] = \widetilde{X}^{\beta=3} + \frac{1}{L} \XAGP[V_{\text{bo},1}] ~.
\label{eq:XAGPVstep1}
\end{equation}
Comparing with the exact Eq.~(\ref{eq:XAGP4Vstep}), this finding means that for these sizes $(\widetilde{X}^{\beta=3})_{\text{diag}} = 0$, which again we verify as a check.
We suspect that this will eventually fail for larger $L$, requiring generalized AGP instead to have relation like the above, while in App.~\ref{subapp:applications_XAGPgen} we provide some thoughts why the original AGP may be sufficient in this case for the small sizes.

On the other hand, for the perturbation $V_{\text{step},2} \equiv V_{\text{step}}^{\beta=4}$ presented in Fig.~\ref{fig:Vstep2} in App.~\ref{app:add_data_for_agp_global},
such relation does not hold exactly with the original AGP, but it holds when we apply the generalized AGP.
We refer to App.~\ref{subapp:applications_XAGPgen} for more details.

Finally, we can now analytically understand the scaling of the AGP norm for the $V_{\text{step}}$-type perturbations.
Clearly, since $c_j \sim O(1)$ for all $j$ along the chain, we have
\begin{equation}
\|\widetilde{X}^\beta \|^2 \sim L ~.
\end{equation}
On the other hand, the numerical measurements (see table in Fig.~\ref{fig:agp-both-global}) give us
\begin{equation}
\left\| \frac{1}{L} \XAGP[V_{\text{bo}}^\beta] \right\|^2 \sim L^{-1+p} ~,
\label{eq:norm_XAGPbo_over_L}
\end{equation}
with $p$ close to zero. 
We conclude that the AGP norm of $V_{\text{step}}$-type perturbations is in fact dominated by the $\widetilde{X}$ contributions.
We note that the asymptotic scaling in Eq.~(\ref{eq:norm_XAGPbo_over_L}) is deduced from studies of the AGP norm up to sizes $L = 18$.
On the other hand, the Pauli string characterizations are performed for much smaller sizes up to $L = 11$, for which the l.h.s.\ in the above equation in fact still grows with $L$, and this may be the reason why we do not see decrease in the relative weight on the subdominant Pauli strings in Fig.~\ref{fig:vstep1} yet.

\section{Conclusions}
\label{sec:conclusions}
In this work, we have studied proxies for weak integrability-breaking perturbations on finite chains with periodic boundary conditions.
Inspired by previous results for infinite systems, where weak perturbations were constructed from various classes of generators, we have studied the properties of the AGP for these different classes.
The motivation comes from the intuition that the AGP plays a similar role as the infinite-size generator.
We first analyse the scaling of the norm of the AGP for various types of perturbations, finding an exponential scaling with the system size for strong perturbations, and a polynomial (or logarithmic) scaling for weak perturbations.
These scalings are obtained both for translationally-invariant (extensive) perturbations and for strictly local ones,
and different classes are characterized by distinct power laws: $\| \XAGP \|^2 \sim L$, $L^{1+p}$, and $L^2$ for $V_{\text{ex}}$, $V_{\text{bo}}$ (with $p$ close to zero), and $V_{\text{bi}}$ type translationally-invariant perturbations; $\| \XAGP \|^2 \sim \text{const}$ and $L$ for $V_{\text{loc}}$ and $V_{\text{step}}$ type local perturbations.

We then characterise the AGP by examining its decomposition in Pauli string operators.
This allows us to understand its locality properties. 
While for strong perturbations the AGP looks completely delocalized, for all the weak perturbations we consider we find that the weight on strings with large supports or ranges is strongly suppressed.

The observation of this quasilocality is particularly significant for weak perturbations that are generated by boosted operators since their generators are not well defined in such finite systems.
It would be interesting to understand whether there is an underlying structure that explains this quasilocality for finite chains.
Recent works have shown progress in this direction: Refs.~\cite{Gombor2022, Leeuw2023}, for example, showed how to generalize the definition of transfer matrices for integrable long-range spin chains, dealing with the ``wrapping corrections'' that appear for finite chains.

For other types of weak perturbations whose generators are similarly problematic in finite chains (namely, bilocal- and step-generated perturbations), we are able to find exact relations between their AGPs and the AGPs of boost-generated perturbations. With these relations, we can understand both the scaling of the AGP norm and the locality properties of such perturbations from the results found for boost-generated ones.

Our study shows how the scaling of the AGP norm is a useful proxy for weak perturbations.
While this result is in line with Ref.~\cite{Orlov_2023}, which also focused on perturbations of the Heisenberg spin chain, it would be interesting to explore possible limitations of this approach.
A recent work on perturbed free fermion models showed that a polynomial scaling of the AGP is possible even for some strong perturbations (those that are local in the fermionic representation) \cite{Pozsgay2024}.
Moreover, one can contemplate cases where only a subset of the original IoMs are quasi-conserved:
For example, Ref.~\cite{Pozsgay2024} showed that for some perturbations of the XX model half of the initial IoMs are quasi-conserved. 
This suggests that other types of weak perturbations (beyond the ones that are known to be generated via long-range deformations) might exist. These possibilities deserve further investigation.

Finally, the properties of the AGP studied in this work might have consequences on the transport properties and the relaxation rates of integrable models perturbed by integrability-breaking perturbations~\cite{Surace2023}.
Elucidating these connections is an interesting question that we leave for future works.

\section{Acknowledgments}
We thank Benjamin Doyon, Fabian Essler, Sarang Gopalakrishnan, Aditi Mitra, Anatoli Polkovnikov, and Shinsei Ryu for discussions, and particularly Bal\'azs Pozsgay for sharing his insights and unpublished results about weak integrability-breaking.
SV and OIM acknowledge support by the National Science Foundation through grant DMR-2001186.
FMS acknowledges support provided by the U.S.\ Department of Energy Office of Science, Office of Advanced Scientific Computing Research, (DE-SC0020290); DOE National Quantum Information Science Research Centers, Quantum Systems Accelerator; and by Amazon Web Services, AWS Quantum Program.
A part of this work was done at the Aspen Center for Physics, which is supported by National Science Foundation grant PHY-2210452. 
This work was partially supported by a grant from the Simons Foundation.

\bibliography{bib}

\begin{thebibliography}{55}%
\makeatletter
\providecommand \@ifxundefined [1]{%
 \@ifx{#1\undefined}
}%
\providecommand \@ifnum [1]{%
 \ifnum #1\expandafter \@firstoftwo
 \else \expandafter \@secondoftwo
 \fi
}%
\providecommand \@ifx [1]{%
 \ifx #1\expandafter \@firstoftwo
 \else \expandafter \@secondoftwo
 \fi
}%
\providecommand \natexlab [1]{#1}%
\providecommand \enquote  [1]{``#1''}%
\providecommand \bibnamefont  [1]{#1}%
\providecommand \bibfnamefont [1]{#1}%
\providecommand \citenamefont [1]{#1}%
\providecommand \href@noop [0]{\@secondoftwo}%
\providecommand \href [0]{\begingroup \@sanitize@url \@href}%
\providecommand \@href[1]{\@@startlink{#1}\@@href}%
\providecommand \@@href[1]{\endgroup#1\@@endlink}%
\providecommand \@sanitize@url [0]{\catcode `\\12\catcode `\$12\catcode
  `\&12\catcode `\#12\catcode `\^12\catcode `\_12\catcode `\%12\relax}%
\providecommand \@@startlink[1]{}%
\providecommand \@@endlink[0]{}%
\providecommand \url  [0]{\begingroup\@sanitize@url \@url }%
\providecommand \@url [1]{\endgroup\@href {#1}{\urlprefix }}%
\providecommand \urlprefix  [0]{URL }%
\providecommand \Eprint [0]{\href }%
\providecommand \doibase [0]{https://doi.org/}%
\providecommand \selectlanguage [0]{\@gobble}%
\providecommand \bibinfo  [0]{\@secondoftwo}%
\providecommand \bibfield  [0]{\@secondoftwo}%
\providecommand \translation [1]{[#1]}%
\providecommand \BibitemOpen [0]{}%
\providecommand \bibitemStop [0]{}%
\providecommand \bibitemNoStop [0]{.\EOS\space}%
\providecommand \EOS [0]{\spacefactor3000\relax}%
\providecommand \BibitemShut  [1]{\csname bibitem#1\endcsname}%
\let\auto@bib@innerbib\@empty
\bibitem [{\citenamefont {Deutsch}(1991)}]{Deutsch1991ETH}%
  \BibitemOpen
  \bibfield  {author} {\bibinfo {author} {\bibfnamefont {J.~M.}\ \bibnamefont
  {Deutsch}},\ }\bibfield  {title} {\bibinfo {title} {Quantum statistical
  mechanics in a closed system},\ }\href
  {https://doi.org/10.1103/PhysRevA.43.2046} {\bibfield  {journal} {\bibinfo
  {journal} {Phys. Rev. A}\ }\textbf {\bibinfo {volume} {43}},\ \bibinfo
  {pages} {2046} (\bibinfo {year} {1991})}\BibitemShut {NoStop}%
\bibitem [{\citenamefont {Srednicki}(1994)}]{Srednicki1994ETH}%
  \BibitemOpen
  \bibfield  {author} {\bibinfo {author} {\bibfnamefont {M.}~\bibnamefont
  {Srednicki}},\ }\bibfield  {title} {\bibinfo {title} {Chaos and quantum
  thermalization},\ }\href {https://doi.org/10.1103/PhysRevE.50.888} {\bibfield
   {journal} {\bibinfo  {journal} {Phys. Rev. E}\ }\textbf {\bibinfo {volume}
  {50}},\ \bibinfo {pages} {888} (\bibinfo {year} {1994})}\BibitemShut
  {NoStop}%
\bibitem [{\citenamefont {Papi{\'{c}}}(2022)}]{Papic2021review}%
  \BibitemOpen
  \bibfield  {author} {\bibinfo {author} {\bibfnamefont {Z.}~\bibnamefont
  {Papi{\'{c}}}},\ }\bibinfo {title} {{Weak Ergodicity Breaking Through the
  Lens of Quantum Entanglement}},\ in\ \href
  {https://doi.org/10.1007/978-3-031-03998-0_13} {\emph {\bibinfo {booktitle}
  {{Entanglement in Spin Chains: From Theory to Quantum Technology
  Applications}}}},\ \bibinfo {editor} {edited by\ \bibinfo {editor}
  {\bibfnamefont {A.}~\bibnamefont {Bayat}}, \bibinfo {editor} {\bibfnamefont
  {S.}~\bibnamefont {Bose}},\ and\ \bibinfo {editor} {\bibfnamefont
  {H.}~\bibnamefont {Johannesson}}}\ (\bibinfo  {publisher} {Springer
  International Publishing},\ \bibinfo {address} {Cham},\ \bibinfo {year}
  {2022})\ pp.\ \bibinfo {pages} {341--395}\BibitemShut {NoStop}%
\bibitem [{\citenamefont {Moudgalya}\ \emph {et~al.}(2022)\citenamefont
  {Moudgalya}, \citenamefont {Bernevig},\ and\ \citenamefont
  {Regnault}}]{Moudgalya2021review}%
  \BibitemOpen
  \bibfield  {author} {\bibinfo {author} {\bibfnamefont {S.}~\bibnamefont
  {Moudgalya}}, \bibinfo {author} {\bibfnamefont {B.~A.}\ \bibnamefont
  {Bernevig}},\ and\ \bibinfo {author} {\bibfnamefont {N.}~\bibnamefont
  {Regnault}},\ }\bibfield  {title} {\bibinfo {title} {Quantum many-body scars
  and {Hilbert} space fragmentation: a review of exact results},\ }\href
  {https://doi.org/10.1088/1361-6633/ac73a0} {\bibfield  {journal} {\bibinfo
  {journal} {Reports on Progress in Physics}\ }\textbf {\bibinfo {volume}
  {85}},\ \bibinfo {pages} {086501} (\bibinfo {year} {2022})}\BibitemShut
  {NoStop}%
\bibitem [{\citenamefont {Chandran}\ \emph {et~al.}(2023)\citenamefont
  {Chandran}, \citenamefont {Iadecola}, \citenamefont {Khemani},\ and\
  \citenamefont {Moessner}}]{Chandran2022review}%
  \BibitemOpen
  \bibfield  {author} {\bibinfo {author} {\bibfnamefont {A.}~\bibnamefont
  {Chandran}}, \bibinfo {author} {\bibfnamefont {T.}~\bibnamefont {Iadecola}},
  \bibinfo {author} {\bibfnamefont {V.}~\bibnamefont {Khemani}},\ and\ \bibinfo
  {author} {\bibfnamefont {R.}~\bibnamefont {Moessner}},\ }\bibfield  {title}
  {\bibinfo {title} {Quantum many-body scars: A quasiparticle perspective},\
  }\href {https://doi.org/10.1146/annurev-conmatphys-031620-101617} {\bibfield
  {journal} {\bibinfo  {journal} {Annual Review of Condensed Matter Physics}\
  }\textbf {\bibinfo {volume} {14}},\ \bibinfo {pages} {443} (\bibinfo {year}
  {2023})}\BibitemShut {NoStop}%
\bibitem [{\citenamefont {Polkovnikov}\ \emph {et~al.}(2011)\citenamefont
  {Polkovnikov}, \citenamefont {Sengupta}, \citenamefont {Silva},\ and\
  \citenamefont {Vengalattore}}]{Polkovnikov2011}%
  \BibitemOpen
  \bibfield  {author} {\bibinfo {author} {\bibfnamefont {A.}~\bibnamefont
  {Polkovnikov}}, \bibinfo {author} {\bibfnamefont {K.}~\bibnamefont
  {Sengupta}}, \bibinfo {author} {\bibfnamefont {A.}~\bibnamefont {Silva}},\
  and\ \bibinfo {author} {\bibfnamefont {M.}~\bibnamefont {Vengalattore}},\
  }\bibfield  {title} {\bibinfo {title} {Colloquium: Nonequilibrium dynamics of
  closed interacting quantum systems},\ }\href
  {https://doi.org/10.1103/RevModPhys.83.863} {\bibfield  {journal} {\bibinfo
  {journal} {Rev. Mod. Phys.}\ }\textbf {\bibinfo {volume} {83}},\ \bibinfo
  {pages} {863} (\bibinfo {year} {2011})}\BibitemShut {NoStop}%
\bibitem [{\citenamefont {Eisert}\ \emph {et~al.}(2015)\citenamefont {Eisert},
  \citenamefont {Friesdorf},\ and\ \citenamefont
  {Gogolin}}]{eisert2015quantum}%
  \BibitemOpen
  \bibfield  {author} {\bibinfo {author} {\bibfnamefont {J.}~\bibnamefont
  {Eisert}}, \bibinfo {author} {\bibfnamefont {M.}~\bibnamefont {Friesdorf}},\
  and\ \bibinfo {author} {\bibfnamefont {C.}~\bibnamefont {Gogolin}},\
  }\bibfield  {title} {\bibinfo {title} {Quantum many-body systems out of
  equilibrium},\ }\href {https://doi.org/10.1038/nphys3215} {\bibfield
  {journal} {\bibinfo  {journal} {Nature Physics}\ }\textbf {\bibinfo {volume}
  {11}},\ \bibinfo {pages} {124} (\bibinfo {year} {2015})}\BibitemShut
  {NoStop}%
\bibitem [{\citenamefont {D'Alessio}\ \emph {et~al.}(2016)\citenamefont
  {D'Alessio}, \citenamefont {Kafri}, \citenamefont {Polkovnikov},\ and\
  \citenamefont {Rigol}}]{DAlessio2016}%
  \BibitemOpen
  \bibfield  {author} {\bibinfo {author} {\bibfnamefont {L.}~\bibnamefont
  {D'Alessio}}, \bibinfo {author} {\bibfnamefont {Y.}~\bibnamefont {Kafri}},
  \bibinfo {author} {\bibfnamefont {A.}~\bibnamefont {Polkovnikov}},\ and\
  \bibinfo {author} {\bibfnamefont {M.}~\bibnamefont {Rigol}},\ }\bibfield
  {title} {\bibinfo {title} {From quantum chaos and eigenstate thermalization
  to statistical mechanics and thermodynamics},\ }\href
  {https://doi.org/10.1080/00018732.2016.1198134} {\bibfield  {journal}
  {\bibinfo  {journal} {Advances in Physics}\ }\textbf {\bibinfo {volume}
  {65}},\ \bibinfo {pages} {239} (\bibinfo {year} {2016})}\BibitemShut
  {NoStop}%
\bibitem [{\citenamefont {Deutsch}(2018)}]{Deutsch_2018}%
  \BibitemOpen
  \bibfield  {author} {\bibinfo {author} {\bibfnamefont {J.~M.}\ \bibnamefont
  {Deutsch}},\ }\bibfield  {title} {\bibinfo {title} {Eigenstate thermalization
  hypothesis},\ }\href {https://doi.org/10.1088/1361-6633/aac9f1} {\bibfield
  {journal} {\bibinfo  {journal} {Reports on Progress in Physics}\ }\textbf
  {\bibinfo {volume} {81}},\ \bibinfo {pages} {082001} (\bibinfo {year}
  {2018})}\BibitemShut {NoStop}%
\bibitem [{\citenamefont {Santos}\ and\ \citenamefont
  {Torres-Herrera}(2018)}]{santos2018nonequilibrium}%
  \BibitemOpen
  \bibfield  {author} {\bibinfo {author} {\bibfnamefont {L.~F.}\ \bibnamefont
  {Santos}}\ and\ \bibinfo {author} {\bibfnamefont {E.~J.}\ \bibnamefont
  {Torres-Herrera}},\ }\bibinfo {title} {Nonequilibrium quantum dynamics of
  many-body systems},\ in\ \href
  {https://doi.org/https://doi.org/10.1007/978-3-319-68109-2} {\emph {\bibinfo
  {booktitle} {Chaotic, Fractional, and Complex Dynamics: New Insights and
  Perspectives}}}\ (\bibinfo  {publisher} {Springer},\ \bibinfo {year} {2018})\
  pp.\ \bibinfo {pages} {231--260}\BibitemShut {NoStop}%
\bibitem [{\citenamefont {Rigol}\ \emph {et~al.}(2007)\citenamefont {Rigol},
  \citenamefont {Dunjko}, \citenamefont {Yurovsky},\ and\ \citenamefont
  {Olshanii}}]{Rigol2007}%
  \BibitemOpen
  \bibfield  {author} {\bibinfo {author} {\bibfnamefont {M.}~\bibnamefont
  {Rigol}}, \bibinfo {author} {\bibfnamefont {V.}~\bibnamefont {Dunjko}},
  \bibinfo {author} {\bibfnamefont {V.}~\bibnamefont {Yurovsky}},\ and\
  \bibinfo {author} {\bibfnamefont {M.}~\bibnamefont {Olshanii}},\ }\bibfield
  {title} {\bibinfo {title} {Relaxation in a completely integrable many-body
  quantum system: An ab initio study of the dynamics of the highly excited
  states of 1d lattice hard-core bosons},\ }\href
  {https://doi.org/10.1103/PhysRevLett.98.050405} {\bibfield  {journal}
  {\bibinfo  {journal} {Phys. Rev. Lett.}\ }\textbf {\bibinfo {volume} {98}},\
  \bibinfo {pages} {050405} (\bibinfo {year} {2007})}\BibitemShut {NoStop}%
\bibitem [{\citenamefont {Rigol}(2009)}]{Rigol2009}%
  \BibitemOpen
  \bibfield  {author} {\bibinfo {author} {\bibfnamefont {M.}~\bibnamefont
  {Rigol}},\ }\bibfield  {title} {\bibinfo {title} {Breakdown of thermalization
  in finite one-dimensional systems},\ }\href
  {https://doi.org/10.1103/PhysRevLett.103.100403} {\bibfield  {journal}
  {\bibinfo  {journal} {Phys. Rev. Lett.}\ }\textbf {\bibinfo {volume} {103}},\
  \bibinfo {pages} {100403} (\bibinfo {year} {2009})}\BibitemShut {NoStop}%
\bibitem [{\citenamefont {Cassidy}\ \emph {et~al.}(2011)\citenamefont
  {Cassidy}, \citenamefont {Clark},\ and\ \citenamefont {Rigol}}]{Cassidy2011}%
  \BibitemOpen
  \bibfield  {author} {\bibinfo {author} {\bibfnamefont {A.~C.}\ \bibnamefont
  {Cassidy}}, \bibinfo {author} {\bibfnamefont {C.~W.}\ \bibnamefont {Clark}},\
  and\ \bibinfo {author} {\bibfnamefont {M.}~\bibnamefont {Rigol}},\ }\bibfield
   {title} {\bibinfo {title} {Generalized thermalization in an integrable
  lattice system},\ }\href {https://doi.org/10.1103/PhysRevLett.106.140405}
  {\bibfield  {journal} {\bibinfo  {journal} {Phys. Rev. Lett.}\ }\textbf
  {\bibinfo {volume} {106}},\ \bibinfo {pages} {140405} (\bibinfo {year}
  {2011})}\BibitemShut {NoStop}%
\bibitem [{\citenamefont {Kollar}\ \emph {et~al.}(2011)\citenamefont {Kollar},
  \citenamefont {Wolf},\ and\ \citenamefont {Eckstein}}]{Kollar2011}%
  \BibitemOpen
  \bibfield  {author} {\bibinfo {author} {\bibfnamefont {M.}~\bibnamefont
  {Kollar}}, \bibinfo {author} {\bibfnamefont {F.~A.}\ \bibnamefont {Wolf}},\
  and\ \bibinfo {author} {\bibfnamefont {M.}~\bibnamefont {Eckstein}},\
  }\bibfield  {title} {\bibinfo {title} {Generalized gibbs ensemble prediction
  of prethermalization plateaus and their relation to nonthermal steady states
  in integrable systems},\ }\href {https://doi.org/10.1103/PhysRevB.84.054304}
  {\bibfield  {journal} {\bibinfo  {journal} {Phys. Rev. B}\ }\textbf {\bibinfo
  {volume} {84}},\ \bibinfo {pages} {054304} (\bibinfo {year}
  {2011})}\BibitemShut {NoStop}%
\bibitem [{\citenamefont {Gring}\ \emph {et~al.}(2012)\citenamefont {Gring},
  \citenamefont {Kuhnert}, \citenamefont {Langen}, \citenamefont {Kitagawa},
  \citenamefont {Rauer}, \citenamefont {Schreitl}, \citenamefont {Mazets},
  \citenamefont {Smith}, \citenamefont {Demler},\ and\ \citenamefont
  {Schmiedmayer}}]{Gring2012}%
  \BibitemOpen
  \bibfield  {author} {\bibinfo {author} {\bibfnamefont {M.}~\bibnamefont
  {Gring}}, \bibinfo {author} {\bibfnamefont {M.}~\bibnamefont {Kuhnert}},
  \bibinfo {author} {\bibfnamefont {T.}~\bibnamefont {Langen}}, \bibinfo
  {author} {\bibfnamefont {T.}~\bibnamefont {Kitagawa}}, \bibinfo {author}
  {\bibfnamefont {B.}~\bibnamefont {Rauer}}, \bibinfo {author} {\bibfnamefont
  {M.}~\bibnamefont {Schreitl}}, \bibinfo {author} {\bibfnamefont
  {I.}~\bibnamefont {Mazets}}, \bibinfo {author} {\bibfnamefont {D.~A.}\
  \bibnamefont {Smith}}, \bibinfo {author} {\bibfnamefont {E.}~\bibnamefont
  {Demler}},\ and\ \bibinfo {author} {\bibfnamefont {J.}~\bibnamefont
  {Schmiedmayer}},\ }\bibfield  {title} {\bibinfo {title} {Relaxation and
  prethermalization in an isolated quantum system},\ }\href
  {https://doi.org/10.1126/science.1224953} {\bibfield  {journal} {\bibinfo
  {journal} {Science}\ }\textbf {\bibinfo {volume} {337}},\ \bibinfo {pages}
  {1318} (\bibinfo {year} {2012})}\BibitemShut {NoStop}%
\bibitem [{\citenamefont {Ilievski}\ \emph {et~al.}(2015)\citenamefont
  {Ilievski}, \citenamefont {De~Nardis}, \citenamefont {Wouters}, \citenamefont
  {Caux}, \citenamefont {Essler},\ and\ \citenamefont {Prosen}}]{Ilievski2015}%
  \BibitemOpen
  \bibfield  {author} {\bibinfo {author} {\bibfnamefont {E.}~\bibnamefont
  {Ilievski}}, \bibinfo {author} {\bibfnamefont {J.}~\bibnamefont {De~Nardis}},
  \bibinfo {author} {\bibfnamefont {B.}~\bibnamefont {Wouters}}, \bibinfo
  {author} {\bibfnamefont {J.-S.}\ \bibnamefont {Caux}}, \bibinfo {author}
  {\bibfnamefont {F.~H.~L.}\ \bibnamefont {Essler}},\ and\ \bibinfo {author}
  {\bibfnamefont {T.}~\bibnamefont {Prosen}},\ }\bibfield  {title} {\bibinfo
  {title} {{Complete Generalized Gibbs Ensembles in an Interacting Theory}},\
  }\href {https://doi.org/10.1103/PhysRevLett.115.157201} {\bibfield  {journal}
  {\bibinfo  {journal} {Phys. Rev. Lett.}\ }\textbf {\bibinfo {volume} {115}},\
  \bibinfo {pages} {157201} (\bibinfo {year} {2015})}\BibitemShut {NoStop}%
\bibitem [{\citenamefont {Langen}\ \emph {et~al.}(2015)\citenamefont {Langen},
  \citenamefont {Erne}, \citenamefont {Geiger}, \citenamefont {Rauer},
  \citenamefont {Schweigler}, \citenamefont {Kuhnert}, \citenamefont
  {Rohringer}, \citenamefont {Mazets}, \citenamefont {Gasenzer},\ and\
  \citenamefont {Schmiedmayer}}]{Langen2015}%
  \BibitemOpen
  \bibfield  {author} {\bibinfo {author} {\bibfnamefont {T.}~\bibnamefont
  {Langen}}, \bibinfo {author} {\bibfnamefont {S.}~\bibnamefont {Erne}},
  \bibinfo {author} {\bibfnamefont {R.}~\bibnamefont {Geiger}}, \bibinfo
  {author} {\bibfnamefont {B.}~\bibnamefont {Rauer}}, \bibinfo {author}
  {\bibfnamefont {T.}~\bibnamefont {Schweigler}}, \bibinfo {author}
  {\bibfnamefont {M.}~\bibnamefont {Kuhnert}}, \bibinfo {author} {\bibfnamefont
  {W.}~\bibnamefont {Rohringer}}, \bibinfo {author} {\bibfnamefont {I.~E.}\
  \bibnamefont {Mazets}}, \bibinfo {author} {\bibfnamefont {T.}~\bibnamefont
  {Gasenzer}},\ and\ \bibinfo {author} {\bibfnamefont {J.}~\bibnamefont
  {Schmiedmayer}},\ }\bibfield  {title} {\bibinfo {title} {Experimental
  observation of a generalized gibbs ensemble},\ }\href
  {https://doi.org/10.1126/science.1257026} {\bibfield  {journal} {\bibinfo
  {journal} {Science}\ }\textbf {\bibinfo {volume} {348}},\ \bibinfo {pages}
  {207} (\bibinfo {year} {2015})}\BibitemShut {NoStop}%
\bibitem [{\citenamefont {Essler}\ and\ \citenamefont
  {Fagotti}(2016)}]{Essler_2016}%
  \BibitemOpen
  \bibfield  {author} {\bibinfo {author} {\bibfnamefont {F.~H.~L.}\
  \bibnamefont {Essler}}\ and\ \bibinfo {author} {\bibfnamefont
  {M.}~\bibnamefont {Fagotti}},\ }\bibfield  {title} {\bibinfo {title} {Quench
  dynamics and relaxation in isolated integrable quantum spin chains},\ }\href
  {https://doi.org/10.1088/1742-5468/2016/06/064002} {\bibfield  {journal}
  {\bibinfo  {journal} {Journal of Statistical Mechanics: Theory and
  Experiment}\ }\textbf {\bibinfo {volume} {2016}},\ \bibinfo {pages} {064002}
  (\bibinfo {year} {2016})}\BibitemShut {NoStop}%
\bibitem [{\citenamefont {Vidmar}\ and\ \citenamefont
  {Rigol}(2016)}]{Vidmar_2016}%
  \BibitemOpen
  \bibfield  {author} {\bibinfo {author} {\bibfnamefont {L.}~\bibnamefont
  {Vidmar}}\ and\ \bibinfo {author} {\bibfnamefont {M.}~\bibnamefont {Rigol}},\
  }\bibfield  {title} {\bibinfo {title} {{Generalized Gibbs ensemble in
  integrable lattice models}},\ }\href
  {https://doi.org/10.1088/1742-5468/2016/06/064007} {\bibfield  {journal}
  {\bibinfo  {journal} {Journal of Statistical Mechanics: Theory and
  Experiment}\ }\textbf {\bibinfo {volume} {2016}},\ \bibinfo {pages} {064007}
  (\bibinfo {year} {2016})}\BibitemShut {NoStop}%
\bibitem [{\citenamefont {Arnol'd}(2013)}]{arnol2013mathematical}%
  \BibitemOpen
  \bibfield  {author} {\bibinfo {author} {\bibfnamefont {V.~I.}\ \bibnamefont
  {Arnol'd}},\ }\href@noop {} {\emph {\bibinfo {title} {Mathematical methods of
  classical mechanics}}},\ Vol.~\bibinfo {volume} {60}\ (\bibinfo  {publisher}
  {Springer Science \& Business Media},\ \bibinfo {year} {2013})\BibitemShut
  {NoStop}%
\bibitem [{\citenamefont {Brandino}\ \emph {et~al.}(2015)\citenamefont
  {Brandino}, \citenamefont {Caux},\ and\ \citenamefont
  {Konik}}]{Brandino2015}%
  \BibitemOpen
  \bibfield  {author} {\bibinfo {author} {\bibfnamefont {G.~P.}\ \bibnamefont
  {Brandino}}, \bibinfo {author} {\bibfnamefont {J.-S.}\ \bibnamefont {Caux}},\
  and\ \bibinfo {author} {\bibfnamefont {R.~M.}\ \bibnamefont {Konik}},\
  }\bibfield  {title} {\bibinfo {title} {Glimmers of a quantum kam theorem:
  Insights from quantum quenches in one-dimensional bose gases},\ }\href
  {https://doi.org/10.1103/PhysRevX.5.041043} {\bibfield  {journal} {\bibinfo
  {journal} {Phys. Rev. X}\ }\textbf {\bibinfo {volume} {5}},\ \bibinfo {pages}
  {041043} (\bibinfo {year} {2015})}\BibitemShut {NoStop}%
\bibitem [{\citenamefont {Yu}\ \emph {et~al.}(2022)\citenamefont {Yu},
  \citenamefont {Ye}, \citenamefont {Liu},\ and\ \citenamefont
  {Zhang}}]{KAM2022}%
  \BibitemOpen
  \bibfield  {author} {\bibinfo {author} {\bibfnamefont {Y.-X.}\ \bibnamefont
  {Yu}}, \bibinfo {author} {\bibfnamefont {J.}~\bibnamefont {Ye}}, \bibinfo
  {author} {\bibfnamefont {W.-M.}\ \bibnamefont {Liu}},\ and\ \bibinfo {author}
  {\bibfnamefont {C.}~\bibnamefont {Zhang}},\ }\bibfield  {title} {\bibinfo
  {title} {{Quantum analog of the Kolmogorov-Arnold-Moser theorem in the
  anisotropic Dicke model and its possible implications in the hybrid
  Sachdev-Ye-Kitaev models}},\ }\href
  {https://doi.org/10.1103/PhysRevA.106.022213} {\bibfield  {journal} {\bibinfo
   {journal} {Phys. Rev. A}\ }\textbf {\bibinfo {volume} {106}},\ \bibinfo
  {pages} {022213} (\bibinfo {year} {2022})}\BibitemShut {NoStop}%
\bibitem [{\citenamefont {Rabson}\ \emph {et~al.}(2004)\citenamefont {Rabson},
  \citenamefont {Narozhny},\ and\ \citenamefont {Millis}}]{Rabson2004}%
  \BibitemOpen
  \bibfield  {author} {\bibinfo {author} {\bibfnamefont {D.~A.}\ \bibnamefont
  {Rabson}}, \bibinfo {author} {\bibfnamefont {B.~N.}\ \bibnamefont
  {Narozhny}},\ and\ \bibinfo {author} {\bibfnamefont {A.~J.}\ \bibnamefont
  {Millis}},\ }\bibfield  {title} {\bibinfo {title} {{Crossover from Poisson to
  Wigner-Dyson level statistics in spin chains with integrability breaking}},\
  }\href {https://doi.org/10.1103/PhysRevB.69.054403} {\bibfield  {journal}
  {\bibinfo  {journal} {Phys. Rev. B}\ }\textbf {\bibinfo {volume} {69}},\
  \bibinfo {pages} {054403} (\bibinfo {year} {2004})}\BibitemShut {NoStop}%
\bibitem [{\citenamefont {Jung}\ \emph
  {et~al.}(2006{\natexlab{a}})\citenamefont {Jung}, \citenamefont {Helmes},\
  and\ \citenamefont {Rosch}}]{Jung2006}%
  \BibitemOpen
  \bibfield  {author} {\bibinfo {author} {\bibfnamefont {P.}~\bibnamefont
  {Jung}}, \bibinfo {author} {\bibfnamefont {R.~W.}\ \bibnamefont {Helmes}},\
  and\ \bibinfo {author} {\bibfnamefont {A.}~\bibnamefont {Rosch}},\ }\bibfield
   {title} {\bibinfo {title} {Transport in almost integrable models: Perturbed
  {Heisenberg} chains},\ }\href {https://doi.org/10.1103/PhysRevLett.96.067202}
  {\bibfield  {journal} {\bibinfo  {journal} {Phys. Rev. Lett.}\ }\textbf
  {\bibinfo {volume} {96}},\ \bibinfo {pages} {067202} (\bibinfo {year}
  {2006}{\natexlab{a}})}\BibitemShut {NoStop}%
\bibitem [{\citenamefont {Mazets}\ \emph {et~al.}(2008)\citenamefont {Mazets},
  \citenamefont {Schumm},\ and\ \citenamefont {Schmiedmayer}}]{Mazets2008}%
  \BibitemOpen
  \bibfield  {author} {\bibinfo {author} {\bibfnamefont {I.~E.}\ \bibnamefont
  {Mazets}}, \bibinfo {author} {\bibfnamefont {T.}~\bibnamefont {Schumm}},\
  and\ \bibinfo {author} {\bibfnamefont {J.}~\bibnamefont {Schmiedmayer}},\
  }\bibfield  {title} {\bibinfo {title} {Breakdown of integrability in a
  quasi-1d ultracold bosonic gas},\ }\href
  {https://doi.org/10.1103/PhysRevLett.100.210403} {\bibfield  {journal}
  {\bibinfo  {journal} {Phys. Rev. Lett.}\ }\textbf {\bibinfo {volume} {100}},\
  \bibinfo {pages} {210403} (\bibinfo {year} {2008})}\BibitemShut {NoStop}%
\bibitem [{\citenamefont {Marcuzzi}\ \emph {et~al.}(2013)\citenamefont
  {Marcuzzi}, \citenamefont {Marino}, \citenamefont {Gambassi},\ and\
  \citenamefont {Silva}}]{Marcuzzi2013}%
  \BibitemOpen
  \bibfield  {author} {\bibinfo {author} {\bibfnamefont {M.}~\bibnamefont
  {Marcuzzi}}, \bibinfo {author} {\bibfnamefont {J.}~\bibnamefont {Marino}},
  \bibinfo {author} {\bibfnamefont {A.}~\bibnamefont {Gambassi}},\ and\
  \bibinfo {author} {\bibfnamefont {A.}~\bibnamefont {Silva}},\ }\bibfield
  {title} {\bibinfo {title} {Prethermalization in a nonintegrable quantum spin
  chain after a quench},\ }\href
  {https://doi.org/10.1103/PhysRevLett.111.197203} {\bibfield  {journal}
  {\bibinfo  {journal} {Phys. Rev. Lett.}\ }\textbf {\bibinfo {volume} {111}},\
  \bibinfo {pages} {197203} (\bibinfo {year} {2013})}\BibitemShut {NoStop}%
\bibitem [{\citenamefont {Essler}\ \emph {et~al.}(2014)\citenamefont {Essler},
  \citenamefont {Kehrein}, \citenamefont {Manmana},\ and\ \citenamefont
  {Robinson}}]{Essler2014}%
  \BibitemOpen
  \bibfield  {author} {\bibinfo {author} {\bibfnamefont {F.~H.~L.}\
  \bibnamefont {Essler}}, \bibinfo {author} {\bibfnamefont {S.}~\bibnamefont
  {Kehrein}}, \bibinfo {author} {\bibfnamefont {S.~R.}\ \bibnamefont
  {Manmana}},\ and\ \bibinfo {author} {\bibfnamefont {N.~J.}\ \bibnamefont
  {Robinson}},\ }\bibfield  {title} {\bibinfo {title} {Quench dynamics in a
  model with tuneable integrability breaking},\ }\href
  {https://doi.org/10.1103/PhysRevB.89.165104} {\bibfield  {journal} {\bibinfo
  {journal} {Phys. Rev. B}\ }\textbf {\bibinfo {volume} {89}},\ \bibinfo
  {pages} {165104} (\bibinfo {year} {2014})}\BibitemShut {NoStop}%
\bibitem [{\citenamefont {Bertini}\ \emph {et~al.}(2015)\citenamefont
  {Bertini}, \citenamefont {Essler}, \citenamefont {Groha},\ and\ \citenamefont
  {Robinson}}]{Bertini2015}%
  \BibitemOpen
  \bibfield  {author} {\bibinfo {author} {\bibfnamefont {B.}~\bibnamefont
  {Bertini}}, \bibinfo {author} {\bibfnamefont {F.~H.~L.}\ \bibnamefont
  {Essler}}, \bibinfo {author} {\bibfnamefont {S.}~\bibnamefont {Groha}},\ and\
  \bibinfo {author} {\bibfnamefont {N.~J.}\ \bibnamefont {Robinson}},\
  }\bibfield  {title} {\bibinfo {title} {Prethermalization and thermalization
  in models with weak integrability breaking},\ }\href
  {https://doi.org/10.1103/PhysRevLett.115.180601} {\bibfield  {journal}
  {\bibinfo  {journal} {Phys. Rev. Lett.}\ }\textbf {\bibinfo {volume} {115}},\
  \bibinfo {pages} {180601} (\bibinfo {year} {2015})}\BibitemShut {NoStop}%
\bibitem [{\citenamefont {Mierzejewski}\ \emph {et~al.}(2015)\citenamefont
  {Mierzejewski}, \citenamefont {Prosen},\ and\ \citenamefont
  {Prelovsek}}]{Mierzejewski2015}%
  \BibitemOpen
  \bibfield  {author} {\bibinfo {author} {\bibfnamefont {M.}~\bibnamefont
  {Mierzejewski}}, \bibinfo {author} {\bibfnamefont {T.}~\bibnamefont
  {Prosen}},\ and\ \bibinfo {author} {\bibfnamefont {P.}~\bibnamefont
  {Prelovsek}},\ }\bibfield  {title} {\bibinfo {title} {Approximate
  conservation laws in perturbed integrable lattice models},\ }\href
  {https://doi.org/10.1103/PhysRevB.92.195121} {\bibfield  {journal} {\bibinfo
  {journal} {Phys. Rev. B}\ }\textbf {\bibinfo {volume} {92}},\ \bibinfo
  {pages} {195121} (\bibinfo {year} {2015})}\BibitemShut {NoStop}%
\bibitem [{\citenamefont {Bertini}\ \emph {et~al.}(2016)\citenamefont
  {Bertini}, \citenamefont {Essler}, \citenamefont {Groha},\ and\ \citenamefont
  {Robinson}}]{Bertini2016}%
  \BibitemOpen
  \bibfield  {author} {\bibinfo {author} {\bibfnamefont {B.}~\bibnamefont
  {Bertini}}, \bibinfo {author} {\bibfnamefont {F.~H.~L.}\ \bibnamefont
  {Essler}}, \bibinfo {author} {\bibfnamefont {S.}~\bibnamefont {Groha}},\ and\
  \bibinfo {author} {\bibfnamefont {N.~J.}\ \bibnamefont {Robinson}},\
  }\bibfield  {title} {\bibinfo {title} {Thermalization and light cones in a
  model with weak integrability breaking},\ }\href
  {https://doi.org/10.1103/PhysRevB.94.245117} {\bibfield  {journal} {\bibinfo
  {journal} {Phys. Rev. B}\ }\textbf {\bibinfo {volume} {94}},\ \bibinfo
  {pages} {245117} (\bibinfo {year} {2016})}\BibitemShut {NoStop}%
\bibitem [{\citenamefont {Tang}\ \emph {et~al.}(2018)\citenamefont {Tang},
  \citenamefont {Kao}, \citenamefont {Li}, \citenamefont {Seo}, \citenamefont
  {Mallayya}, \citenamefont {Rigol}, \citenamefont {Gopalakrishnan},\ and\
  \citenamefont {Lev}}]{Tang2018}%
  \BibitemOpen
  \bibfield  {author} {\bibinfo {author} {\bibfnamefont {Y.}~\bibnamefont
  {Tang}}, \bibinfo {author} {\bibfnamefont {W.}~\bibnamefont {Kao}}, \bibinfo
  {author} {\bibfnamefont {K.-Y.}\ \bibnamefont {Li}}, \bibinfo {author}
  {\bibfnamefont {S.}~\bibnamefont {Seo}}, \bibinfo {author} {\bibfnamefont
  {K.}~\bibnamefont {Mallayya}}, \bibinfo {author} {\bibfnamefont
  {M.}~\bibnamefont {Rigol}}, \bibinfo {author} {\bibfnamefont
  {S.}~\bibnamefont {Gopalakrishnan}},\ and\ \bibinfo {author} {\bibfnamefont
  {B.~L.}\ \bibnamefont {Lev}},\ }\bibfield  {title} {\bibinfo {title}
  {Thermalization near integrability in a dipolar quantum {Newton}'s cradle},\
  }\href {https://doi.org/10.1103/PhysRevX.8.021030} {\bibfield  {journal}
  {\bibinfo  {journal} {Phys. Rev. X}\ }\textbf {\bibinfo {volume} {8}},\
  \bibinfo {pages} {021030} (\bibinfo {year} {2018})}\BibitemShut {NoStop}%
\bibitem [{\citenamefont {Friedman}\ \emph {et~al.}(2020)\citenamefont
  {Friedman}, \citenamefont {Gopalakrishnan},\ and\ \citenamefont
  {Vasseur}}]{Friedman2020}%
  \BibitemOpen
  \bibfield  {author} {\bibinfo {author} {\bibfnamefont {A.~J.}\ \bibnamefont
  {Friedman}}, \bibinfo {author} {\bibfnamefont {S.}~\bibnamefont
  {Gopalakrishnan}},\ and\ \bibinfo {author} {\bibfnamefont {R.}~\bibnamefont
  {Vasseur}},\ }\bibfield  {title} {\bibinfo {title} {Diffusive hydrodynamics
  from integrability breaking},\ }\href
  {https://doi.org/10.1103/PhysRevB.101.180302} {\bibfield  {journal} {\bibinfo
   {journal} {Phys. Rev. B}\ }\textbf {\bibinfo {volume} {101}},\ \bibinfo
  {pages} {180302} (\bibinfo {year} {2020})}\BibitemShut {NoStop}%
\bibitem [{\citenamefont {LeBlond}\ \emph {et~al.}(2021)\citenamefont
  {LeBlond}, \citenamefont {Sels}, \citenamefont {Polkovnikov},\ and\
  \citenamefont {Rigol}}]{LeBlond2021}%
  \BibitemOpen
  \bibfield  {author} {\bibinfo {author} {\bibfnamefont {T.}~\bibnamefont
  {LeBlond}}, \bibinfo {author} {\bibfnamefont {D.}~\bibnamefont {Sels}},
  \bibinfo {author} {\bibfnamefont {A.}~\bibnamefont {Polkovnikov}},\ and\
  \bibinfo {author} {\bibfnamefont {M.}~\bibnamefont {Rigol}},\ }\bibfield
  {title} {\bibinfo {title} {Universality in the onset of quantum chaos in
  many-body systems},\ }\href {https://doi.org/10.1103/PhysRevB.104.L201117}
  {\bibfield  {journal} {\bibinfo  {journal} {Phys. Rev. B}\ }\textbf {\bibinfo
  {volume} {104}},\ \bibinfo {pages} {L201117} (\bibinfo {year}
  {2021})}\BibitemShut {NoStop}%
\bibitem [{\citenamefont {Bastianello}\ \emph {et~al.}(2021)\citenamefont
  {Bastianello}, \citenamefont {Luca},\ and\ \citenamefont
  {Vasseur}}]{Bastianello2021}%
  \BibitemOpen
  \bibfield  {author} {\bibinfo {author} {\bibfnamefont {A.}~\bibnamefont
  {Bastianello}}, \bibinfo {author} {\bibfnamefont {A.~D.}\ \bibnamefont
  {Luca}},\ and\ \bibinfo {author} {\bibfnamefont {R.}~\bibnamefont
  {Vasseur}},\ }\bibfield  {title} {\bibinfo {title} {Hydrodynamics of weak
  integrability breaking},\ }\href {https://doi.org/10.1088/1742-5468/ac26b2}
  {\bibfield  {journal} {\bibinfo  {journal} {Journal of Statistical Mechanics:
  Theory and Experiment}\ }\textbf {\bibinfo {volume} {2021}},\ \bibinfo
  {pages} {114003} (\bibinfo {year} {2021})}\BibitemShut {NoStop}%
\bibitem [{\citenamefont {Bulchandani}\ \emph {et~al.}(2022)\citenamefont
  {Bulchandani}, \citenamefont {Huse},\ and\ \citenamefont
  {Gopalakrishnan}}]{Bulchandani}%
  \BibitemOpen
  \bibfield  {author} {\bibinfo {author} {\bibfnamefont {V.~B.}\ \bibnamefont
  {Bulchandani}}, \bibinfo {author} {\bibfnamefont {D.~A.}\ \bibnamefont
  {Huse}},\ and\ \bibinfo {author} {\bibfnamefont {S.}~\bibnamefont
  {Gopalakrishnan}},\ }\bibfield  {title} {\bibinfo {title} {Onset of many-body
  quantum chaos due to breaking integrability},\ }\href
  {https://doi.org/10.1103/PhysRevB.105.214308} {\bibfield  {journal} {\bibinfo
   {journal} {Phys. Rev. B}\ }\textbf {\bibinfo {volume} {105}},\ \bibinfo
  {pages} {214308} (\bibinfo {year} {2022})}\BibitemShut {NoStop}%
\bibitem [{\citenamefont {Durnin}\ \emph {et~al.}(2021)\citenamefont {Durnin},
  \citenamefont {Bhaseen},\ and\ \citenamefont {Doyon}}]{Durnin2021}%
  \BibitemOpen
  \bibfield  {author} {\bibinfo {author} {\bibfnamefont {J.}~\bibnamefont
  {Durnin}}, \bibinfo {author} {\bibfnamefont {M.~J.}\ \bibnamefont
  {Bhaseen}},\ and\ \bibinfo {author} {\bibfnamefont {B.}~\bibnamefont
  {Doyon}},\ }\bibfield  {title} {\bibinfo {title} {Nonequilibrium dynamics and
  weakly broken integrability},\ }\href
  {https://doi.org/10.1103/PhysRevLett.127.130601} {\bibfield  {journal}
  {\bibinfo  {journal} {Phys. Rev. Lett.}\ }\textbf {\bibinfo {volume} {127}},\
  \bibinfo {pages} {130601} (\bibinfo {year} {2021})}\BibitemShut {NoStop}%
\bibitem [{\citenamefont {Kuo}\ \emph {et~al.}(2023)\citenamefont {Kuo},
  \citenamefont {Ware}, \citenamefont {Lunts}, \citenamefont {Hafezi},\ and\
  \citenamefont {White}}]{kuo2023energy}%
  \BibitemOpen
  \bibfield  {author} {\bibinfo {author} {\bibfnamefont {E.-J.}\ \bibnamefont
  {Kuo}}, \bibinfo {author} {\bibfnamefont {B.}~\bibnamefont {Ware}}, \bibinfo
  {author} {\bibfnamefont {P.}~\bibnamefont {Lunts}}, \bibinfo {author}
  {\bibfnamefont {M.}~\bibnamefont {Hafezi}},\ and\ \bibinfo {author}
  {\bibfnamefont {C.~D.}\ \bibnamefont {White}},\ }\bibfield  {title} {\bibinfo
  {title} {Energy diffusion in weakly interacting chains with fermionic
  dissipation-assisted operator evolution},\ }\bibfield  {journal} {\bibinfo
  {journal} {arXiv preprint arXiv:2311.17148}\ }\href
  {https://doi.org/10.48550/arXiv.2311.17148} {10.48550/arXiv.2311.17148}
  (\bibinfo {year} {2023})\BibitemShut {NoStop}%
\bibitem [{\citenamefont {Szász-Schagrin}\ \emph {et~al.}(2021)\citenamefont
  {Szász-Schagrin}, \citenamefont {Pozsgay},\ and\ \citenamefont
  {Takács}}]{Szasz2021}%
  \BibitemOpen
  \bibfield  {author} {\bibinfo {author} {\bibfnamefont {D.}~\bibnamefont
  {Szász-Schagrin}}, \bibinfo {author} {\bibfnamefont {B.}~\bibnamefont
  {Pozsgay}},\ and\ \bibinfo {author} {\bibfnamefont {G.}~\bibnamefont
  {Takács}},\ }\bibfield  {title} {\bibinfo {title} {{Weak integrability
  breaking and level spacing distribution}},\ }\href
  {https://doi.org/10.21468/SciPostPhys.11.2.037} {\bibfield  {journal}
  {\bibinfo  {journal} {SciPost Phys.}\ }\textbf {\bibinfo {volume} {11}},\
  \bibinfo {pages} {037} (\bibinfo {year} {2021})}\BibitemShut {NoStop}%
\bibitem [{\citenamefont {Kurlov}\ \emph {et~al.}(2022)\citenamefont {Kurlov},
  \citenamefont {Malikis},\ and\ \citenamefont {Gritsev}}]{Kurlov2022}%
  \BibitemOpen
  \bibfield  {author} {\bibinfo {author} {\bibfnamefont {D.~V.}\ \bibnamefont
  {Kurlov}}, \bibinfo {author} {\bibfnamefont {S.}~\bibnamefont {Malikis}},\
  and\ \bibinfo {author} {\bibfnamefont {V.}~\bibnamefont {Gritsev}},\
  }\bibfield  {title} {\bibinfo {title} {Quasiconserved quantities in the
  perturbed spin-$\frac{1}{2}$ {XXX} model},\ }\href
  {https://doi.org/10.1103/PhysRevB.105.104302} {\bibfield  {journal} {\bibinfo
   {journal} {Phys. Rev. B}\ }\textbf {\bibinfo {volume} {105}},\ \bibinfo
  {pages} {104302} (\bibinfo {year} {2022})}\BibitemShut {NoStop}%
\bibitem [{\citenamefont {Surace}\ and\ \citenamefont
  {Motrunich}(2023)}]{Surace2023}%
  \BibitemOpen
  \bibfield  {author} {\bibinfo {author} {\bibfnamefont {F.~M.}\ \bibnamefont
  {Surace}}\ and\ \bibinfo {author} {\bibfnamefont {O.}~\bibnamefont
  {Motrunich}},\ }\bibfield  {title} {\bibinfo {title} {Weak integrability
  breaking perturbations of integrable models},\ }\href
  {https://doi.org/10.1103/PhysRevResearch.5.043019} {\bibfield  {journal}
  {\bibinfo  {journal} {Phys. Rev. Res.}\ }\textbf {\bibinfo {volume} {5}},\
  \bibinfo {pages} {043019} (\bibinfo {year} {2023})}\BibitemShut {NoStop}%
\bibitem [{\citenamefont {Jung}\ \emph
  {et~al.}(2006{\natexlab{b}})\citenamefont {Jung}, \citenamefont {Helmes},\
  and\ \citenamefont {Rosch}}]{Jung2006_2}%
  \BibitemOpen
  \bibfield  {author} {\bibinfo {author} {\bibfnamefont {P.}~\bibnamefont
  {Jung}}, \bibinfo {author} {\bibfnamefont {R.~W.}\ \bibnamefont {Helmes}},\
  and\ \bibinfo {author} {\bibfnamefont {A.}~\bibnamefont {Rosch}},\ }\bibfield
   {title} {\bibinfo {title} {Transport in almost integrable models: Perturbed
  heisenberg chains},\ }\href {https://doi.org/10.1103/PhysRevLett.96.067202}
  {\bibfield  {journal} {\bibinfo  {journal} {Phys. Rev. Lett.}\ }\textbf
  {\bibinfo {volume} {96}},\ \bibinfo {pages} {067202} (\bibinfo {year}
  {2006}{\natexlab{b}})}\BibitemShut {NoStop}%
\bibitem [{\citenamefont {Orlov}\ \emph {et~al.}(2023)\citenamefont {Orlov},
  \citenamefont {Tiutiakina}, \citenamefont {Sharipov}, \citenamefont
  {Petrova}, \citenamefont {Gritsev},\ and\ \citenamefont
  {Kurlov}}]{Orlov_2023}%
  \BibitemOpen
  \bibfield  {author} {\bibinfo {author} {\bibfnamefont {P.}~\bibnamefont
  {Orlov}}, \bibinfo {author} {\bibfnamefont {A.}~\bibnamefont {Tiutiakina}},
  \bibinfo {author} {\bibfnamefont {R.}~\bibnamefont {Sharipov}}, \bibinfo
  {author} {\bibfnamefont {E.}~\bibnamefont {Petrova}}, \bibinfo {author}
  {\bibfnamefont {V.}~\bibnamefont {Gritsev}},\ and\ \bibinfo {author}
  {\bibfnamefont {D.~V.}\ \bibnamefont {Kurlov}},\ }\bibfield  {title}
  {\bibinfo {title} {{Adiabatic eigenstate deformations and weak integrability
  breaking of Heisenberg chain}},\ }\href
  {https://doi.org/10.1103/PhysRevB.107.184312} {\bibfield  {journal} {\bibinfo
   {journal} {Phys. Rev. B}\ }\textbf {\bibinfo {volume} {107}},\ \bibinfo
  {pages} {184312} (\bibinfo {year} {2023})}\BibitemShut {NoStop}%
\bibitem [{\citenamefont {Pozsgay}(2020)}]{Pozsgay2020}%
  \BibitemOpen
  \bibfield  {author} {\bibinfo {author} {\bibfnamefont {B.}~\bibnamefont
  {Pozsgay}},\ }\bibfield  {title} {\bibinfo {title} {{Current operators in
  integrable spin chains: lessons from long range deformations}},\ }\href
  {https://doi.org/10.21468/SciPostPhys.8.2.016} {\bibfield  {journal}
  {\bibinfo  {journal} {SciPost Phys.}\ }\textbf {\bibinfo {volume} {8}},\
  \bibinfo {pages} {16} (\bibinfo {year} {2020})}\BibitemShut {NoStop}%
\bibitem [{\citenamefont {Bargheer}\ \emph {et~al.}(2008)\citenamefont
  {Bargheer}, \citenamefont {Beisert},\ and\ \citenamefont
  {Loebbert}}]{Bargheer_2008}%
  \BibitemOpen
  \bibfield  {author} {\bibinfo {author} {\bibfnamefont {T.}~\bibnamefont
  {Bargheer}}, \bibinfo {author} {\bibfnamefont {N.}~\bibnamefont {Beisert}},\
  and\ \bibinfo {author} {\bibfnamefont {F.}~\bibnamefont {Loebbert}},\
  }\bibfield  {title} {\bibinfo {title} {Boosting nearest-neighbour to
  long-range integrable spin chains},\ }\href
  {https://doi.org/10.1088/1742-5468/2008/11/L11001} {\bibfield  {journal}
  {\bibinfo  {journal} {Journal of Statistical Mechanics: Theory and
  Experiment}\ }\textbf {\bibinfo {volume} {2008}},\ \bibinfo {pages} {L11001}
  (\bibinfo {year} {2008})}\BibitemShut {NoStop}%
\bibitem [{\citenamefont {Bargheer}\ \emph {et~al.}(2009)\citenamefont
  {Bargheer}, \citenamefont {Beisert},\ and\ \citenamefont
  {Loebbert}}]{Bargheer_2009}%
  \BibitemOpen
  \bibfield  {author} {\bibinfo {author} {\bibfnamefont {T.}~\bibnamefont
  {Bargheer}}, \bibinfo {author} {\bibfnamefont {N.}~\bibnamefont {Beisert}},\
  and\ \bibinfo {author} {\bibfnamefont {F.}~\bibnamefont {Loebbert}},\
  }\bibfield  {title} {\bibinfo {title} {Long-range deformations for integrable
  spin chains},\ }\href {https://doi.org/10.1088/1751-8113/42/28/285205}
  {\bibfield  {journal} {\bibinfo  {journal} {Journal of Physics A:
  Mathematical and Theoretical}\ }\textbf {\bibinfo {volume} {42}},\ \bibinfo
  {pages} {285205} (\bibinfo {year} {2009})}\BibitemShut {NoStop}%
\bibitem [{\citenamefont {Kolodrubetz}\ \emph {et~al.}(2017)\citenamefont
  {Kolodrubetz}, \citenamefont {Sels}, \citenamefont {Mehta},\ and\
  \citenamefont {Polkovnikov}}]{KOLODRUBETZ20171}%
  \BibitemOpen
  \bibfield  {author} {\bibinfo {author} {\bibfnamefont {M.}~\bibnamefont
  {Kolodrubetz}}, \bibinfo {author} {\bibfnamefont {D.}~\bibnamefont {Sels}},
  \bibinfo {author} {\bibfnamefont {P.}~\bibnamefont {Mehta}},\ and\ \bibinfo
  {author} {\bibfnamefont {A.}~\bibnamefont {Polkovnikov}},\ }\bibfield
  {title} {\bibinfo {title} {Geometry and non-adiabatic response in quantum and
  classical systems},\ }\href
  {https://doi.org/https://doi.org/10.1016/j.physrep.2017.07.001} {\bibfield
  {journal} {\bibinfo  {journal} {Physics Reports}\ }\textbf {\bibinfo {volume}
  {697}},\ \bibinfo {pages} {1} (\bibinfo {year} {2017})}\BibitemShut {NoStop}%
\bibitem [{\citenamefont {Sierant}\ \emph {et~al.}(2019)\citenamefont
  {Sierant}, \citenamefont {Maksymov}, \citenamefont {Ku\ifmmode~\acute{s}\else
  \'{s}\fi{}},\ and\ \citenamefont {Zakrzewski}}]{Sierant2019}%
  \BibitemOpen
  \bibfield  {author} {\bibinfo {author} {\bibfnamefont {P.}~\bibnamefont
  {Sierant}}, \bibinfo {author} {\bibfnamefont {A.}~\bibnamefont {Maksymov}},
  \bibinfo {author} {\bibfnamefont {M.}~\bibnamefont {Ku\ifmmode~\acute{s}\else
  \'{s}\fi{}}},\ and\ \bibinfo {author} {\bibfnamefont {J.}~\bibnamefont
  {Zakrzewski}},\ }\bibfield  {title} {\bibinfo {title} {{Fidelity
  susceptibility in Gaussian random ensembles}},\ }\href
  {https://doi.org/10.1103/PhysRevE.99.050102} {\bibfield  {journal} {\bibinfo
  {journal} {Phys. Rev. E}\ }\textbf {\bibinfo {volume} {99}},\ \bibinfo
  {pages} {050102} (\bibinfo {year} {2019})}\BibitemShut {NoStop}%
\bibitem [{\citenamefont {Pandey}\ \emph {et~al.}(2020)\citenamefont {Pandey},
  \citenamefont {Claeys}, \citenamefont {Campbell}, \citenamefont
  {Polkovnikov},\ and\ \citenamefont {Sels}}]{Pandey2020}%
  \BibitemOpen
  \bibfield  {author} {\bibinfo {author} {\bibfnamefont {M.}~\bibnamefont
  {Pandey}}, \bibinfo {author} {\bibfnamefont {P.~W.}\ \bibnamefont {Claeys}},
  \bibinfo {author} {\bibfnamefont {D.~K.}\ \bibnamefont {Campbell}}, \bibinfo
  {author} {\bibfnamefont {A.}~\bibnamefont {Polkovnikov}},\ and\ \bibinfo
  {author} {\bibfnamefont {D.}~\bibnamefont {Sels}},\ }\bibfield  {title}
  {\bibinfo {title} {Adiabatic eigenstate deformations as a sensitive probe for
  quantum chaos},\ }\href {https://doi.org/10.1103/PhysRevX.10.041017}
  {\bibfield  {journal} {\bibinfo  {journal} {Phys. Rev. X}\ }\textbf {\bibinfo
  {volume} {10}},\ \bibinfo {pages} {041017} (\bibinfo {year}
  {2020})}\BibitemShut {NoStop}%
\bibitem [{\citenamefont {Gombor}(2022)}]{Gombor2022}%
  \BibitemOpen
  \bibfield  {author} {\bibinfo {author} {\bibfnamefont {T.}~\bibnamefont
  {Gombor}},\ }\bibfield  {title} {\bibinfo {title} {Wrapping corrections for
  long-range spin chains},\ }\href
  {https://doi.org/10.1103/PhysRevLett.129.270201} {\bibfield  {journal}
  {\bibinfo  {journal} {Phys. Rev. Lett.}\ }\textbf {\bibinfo {volume} {129}},\
  \bibinfo {pages} {270201} (\bibinfo {year} {2022})}\BibitemShut {NoStop}%
\bibitem [{\citenamefont {de~Leeuw}\ and\ \citenamefont
  {Retore}(2023)}]{Leeuw2023}%
  \BibitemOpen
  \bibfield  {author} {\bibinfo {author} {\bibfnamefont {M.}~\bibnamefont
  {de~Leeuw}}\ and\ \bibinfo {author} {\bibfnamefont {A.~L.}\ \bibnamefont
  {Retore}},\ }\bibfield  {title} {\bibinfo {title} {{Lifting integrable models
  and long-range interactions}},\ }\href
  {https://doi.org/10.21468/SciPostPhys.15.6.241} {\bibfield  {journal}
  {\bibinfo  {journal} {SciPost Phys.}\ }\textbf {\bibinfo {volume} {15}},\
  \bibinfo {pages} {241} (\bibinfo {year} {2023})}\BibitemShut {NoStop}%
\bibitem [{Note1()}]{Note1}%
  \BibitemOpen
  \bibinfo {note} {Compared to the standard AGP definition $\protect \mathcal
  {A}$ in the literature~\cite {Pandey2020}, our operator has the opposite
  sign, $X^{\protect \text {AGP}}= -\protect \mathcal {A}$.}\BibitemShut
  {Stop}%
\bibitem [{\citenamefont {De~Nardis}\ \emph {et~al.}(2021)\citenamefont
  {De~Nardis}, \citenamefont {Gopalakrishnan}, \citenamefont {Vasseur},\ and\
  \citenamefont {Ware}}]{DeNardis_2021}%
  \BibitemOpen
  \bibfield  {author} {\bibinfo {author} {\bibfnamefont {J.}~\bibnamefont
  {De~Nardis}}, \bibinfo {author} {\bibfnamefont {S.}~\bibnamefont
  {Gopalakrishnan}}, \bibinfo {author} {\bibfnamefont {R.}~\bibnamefont
  {Vasseur}},\ and\ \bibinfo {author} {\bibfnamefont {B.}~\bibnamefont
  {Ware}},\ }\bibfield  {title} {\bibinfo {title} {Stability of superdiffusion
  in nearly integrable spin chains},\ }\href
  {https://doi.org/10.1103/PhysRevLett.127.057201} {\bibfield  {journal}
  {\bibinfo  {journal} {Phys. Rev. Lett.}\ }\textbf {\bibinfo {volume} {127}},\
  \bibinfo {pages} {057201} (\bibinfo {year} {2021})}\BibitemShut {NoStop}%
\bibitem [{\citenamefont {Pozsgay}(2023)}]{Pozsgay_notes2023}%
  \BibitemOpen
  \bibfield  {author} {\bibinfo {author} {\bibfnamefont {B.}~\bibnamefont
  {Pozsgay}},\ }\href@noop {} {\bibinfo {title} {Unpublished notes}} (\bibinfo
  {year} {2023})\BibitemShut {NoStop}%
\bibitem [{\citenamefont {Pozsgay}\ \emph {et~al.}(2024)\citenamefont
  {Pozsgay}, \citenamefont {Sharipov}, \citenamefont {Tiutiakina},\ and\
  \citenamefont {Vona}}]{Pozsgay2024}%
  \BibitemOpen
  \bibfield  {author} {\bibinfo {author} {\bibfnamefont {B.}~\bibnamefont
  {Pozsgay}}, \bibinfo {author} {\bibfnamefont {R.}~\bibnamefont {Sharipov}},
  \bibinfo {author} {\bibfnamefont {A.}~\bibnamefont {Tiutiakina}},\ and\
  \bibinfo {author} {\bibfnamefont {I.}~\bibnamefont {Vona}},\ }\bibfield
  {title} {\bibinfo {title} {Adiabatic gauge potential and integrability
  breaking with free fermions},\ }\bibfield  {journal} {\bibinfo  {journal}
  {arXiv preprint arXiv:2402.12979}\ }\href
  {https://doi.org/10.48550/arXiv.2402.12979} {10.48550/arXiv.2402.12979}
  (\bibinfo {year} {2024})\BibitemShut {NoStop}%
\bibitem [{\citenamefont {Grabowski}\ and\ \citenamefont
  {Mathieu}(1994)}]{Grabowski1994}%
  \BibitemOpen
  \bibfield  {author} {\bibinfo {author} {\bibfnamefont {M.~P.}\ \bibnamefont
  {Grabowski}}\ and\ \bibinfo {author} {\bibfnamefont {P.}~\bibnamefont
  {Mathieu}},\ }\bibfield  {title} {\bibinfo {title} {Quantum integrals of
  motion for the {Heisenberg} spin chain},\ }\href
  {https://doi.org/10.1142/S0217732394002057} {\bibfield  {journal} {\bibinfo
  {journal} {Modern Physics Letters A}\ }\textbf {\bibinfo {volume} {09}},\
  \bibinfo {pages} {2197} (\bibinfo {year} {1994})}\BibitemShut {NoStop}%
\end{thebibliography}%

\appendix

\section{Resolving energy degeneracies and using known quasi-IoMs to improve $X$ recovery}
\label{app:sim_diag}
In Eq.~(\ref{eq:AGPod}) in Sec.~\ref{sec:agp}, the matrix elements of the AGP are defined to be $0$ within each degenerate subspace of $H_0$.
This choice is rather arbitrary, since any block-diagonal part (i.e., with the block structure defined by the degenerate subspaces of $H_0$) commutes with $H_0$ and can be added to $\XAGP$ without altering the desired property $i[X, H_0] = V - V_{\text{diag}}$.

For weak integrability-breaking perturbations, we would like to interpret $\XAGP$ as a generator, such that corrections to IoMs can be defined as in Eq.~(\ref{eq:Q1}).
In the case of perturbations generated from long-range integrable deformations including the ones using $\Xbi$ or $\Xbo$, we can calculate corrections to all IoMs in an infinite system by using $Q_\alpha^{(1)} = i[X, Q_\alpha^{(0)}]$, and these corrections are extensive local operators that are well-defined also on finite PBC chains and satisfy the quasi-IoM property Eq.~(\ref{eq:quasiIom}).
In this Appendix we show that we can use this information to obtain a better finite-size proxy for $X$.

The main idea is to impose additional requirements that $\XAGP$ satisfies
\begin{equation}
\label{eq:Xchcorr}
i[\XAGP, Q_\alpha^{(0)}] = Q_\alpha^{(1)} - Q_{\alpha,\text{diag}}^{(1)}
\end{equation}
for some additional $\alpha \geq 3$ besides $\alpha = 2$, where $Q_{\alpha,\text{diag}}^{(1)}$ is the (block)-diagonal part of $Q_\alpha^{(1)}$ with respect to the eigenspaces of $Q_\alpha^{(0)}$, cf.~Eq.~(\ref{eq:offdiag}).

We now show how to achieve this adding $\alpha = 3$.
To this end, it is useful to consider a basis $\{ \ket{n} \}$ that simultaneously diagonalizes $Q_2^{(0)}$ and $Q_3^{(0)}$:
\begin{equation}
Q^{(0)}_2 \ket{n} = E_n\ket{n} ~,
\qquad Q^{(0)}_3 \ket{n} = F_n\ket{n} ~.
\end{equation}
In the numerics, since $Q_2^{(0)}$ has degeneracies, diagonalizing $Q_2^{(0)}$ with a naive diagonalization routine does not generally give eigenstates of $Q_3^{(0)}$.
A common eigenbasis can be obtained, for example, by finding the eigenstates of $Q_2^{(0)} + x Q_3^{(0)}$, where $x$ is a real number [$x$ has to be sufficiently generic to prevent accidental degeneracies of the type $E_n + x F_n = E_m + x F_m$ when $(E_n, F_n) \neq (E_m, F_m)$].
Equation~(\ref{eq:Xchcorr}) implies that 
\begin{equation}
\label{eq:XnmQ3}
\XAGP_{nm}=\frac{(Q_3^{(1)})_{nm}}{i(F_m -F_n)} \quad \text{for } F_n \neq F_m ~.
\end{equation}
Let us compare these matrix elements with the ones in Eq.~(\ref{eq:AGPod}).
We first note that by using the property $[Q_2^{(0)}, Q_3^{(1)}]= [Q_3^{(0)}, Q_2^{(1)}]$ [cf.\ the quasi-IoM property Eq.~(\ref{eq:quasiIom})], we get
\begin{equation}
\frac{(Q_3^{(1)})_{nm}}{i(F_m - F_n)} = \frac{(Q_2^{(1)})_{nm}}{i(E_m - E_n)} \quad \text{for } F_n \neq F_m, E_n \neq E_m.
\end{equation}
This equality (replacing $Q_2^{(0)} \to H_0/2$, $Q_2^{(1)} \to V/2$, $E_n \to \epsilon_n/2$) guarantees that the $\XAGP$ defined in Eq.~(\ref{eq:AGPod}) already satisfies the condition Eq.~(\ref{eq:XnmQ3}) for $E_n \neq E_m$.
The interesting case is for $E_n = E_m$, $F_n \neq F_m$: for these matrix elements, Eq.~(\ref{eq:XnmQ3}) prescribes a non-trivial value, in contrast with the original definition in Eq.~(\ref{eq:AGPod}).
To summarize, this more refined definition of $\XAGP$ for the weak perturbation with given $Q_2^{(1)} = V/2$ and $Q_3^{(1)}$ has the form
\begin{equation}
\label{eq:AGPgen}
X^{\text{AGPgen}}_{nm} = 
\begin{cases}
\frac{(Q_2^{(1)})_{nm}}{i(E_m - E_n)}, & \text{if $E_n \neq E_m$}, \\
\frac{(Q_3^{(1)})_{nm}}{i(F_m - F_n)}, & \text{if $F_n \neq F_m$}, \\
0, & \text{if $E_n = E_m$ and $F_n =F_m$},
\end{cases}
\end{equation}
where the first two lines agree if both $E_n \neq E_m$ and $F_n \neq F_m$, and from now on we will use a new label to indicate this generalized AGP.
With this definition, $X^{\text{AGPgen}}$ satisfies Eq.~(\ref{eq:Xchcorr}) for $\alpha = 2$ and $3$.
We can proceed similarly including other IoMs $\alpha = 4, 5, \dots$ and information about the corresponding corrections.
This is useful whenever there are simultaneous degeneracies $E_n = E_m$, $F_n = F_m$ with $n \neq m$ that can be resolved by additional IoMs.

For the Heisenberg chain, which is the case of primary interest in this work, we find that  $Q_2^{(0)}$ and $Q_3^{(0)}$ are sufficient to resolve all the degeneracies [i.e., $n \neq m \implies (E_n, F_n) \neq (E_m, F_m)$ as 2D points] in the $S^z_{\text{tot}} = 0$ sector for even $L = 6, 8, 14$ and in the $S^z_{\text{tot}} = 1/2$ sector for all odd $L$ between and including $7$ and $21$.
This has some immediate consequences for these sizes.
Thus, all the other IoMs $\{ Q_\alpha^{(0)}, \alpha \geq 4 \}$ must be already diagonal in this basis. 
Furthermore, since $Q_2^{(0)}$ and $Q_3^{(0)}$ commute with the translation unitary and with $\vec{S}_{\text{tot}}^2$, these basis states must have definite momentum and total spin quantum numbers.
We can then argue that $Q_2^{(0)}$ and $Q_3^{(0)}$ are sufficient to resolve all the degeneracies in the entire Hilbert space that are not related to the $SU(2)$ spin symmetry, since all the other $S^z_{\text{tot}}$ sectors can be obtained from the present sector by repeatedly acting with $S^{\pm}_{\text{tot}}$. [As another check, we can also add $Q_1^{(0)} = \sum_j \sigma_j^z \sim S^z_{\text{tot}}$ to $Q_2^{(0)}$ and $Q_3^{(0)}$ and find numerically that this resolves all degeneracies in the entire Hilbert space for these sizes.] 
Hence, in the Heisenberg chains with these sizes we do not need to  add any further IoMs.

Let us continue discussing the system sizes where for the lowest $S^z_{\text{tot}}$ sector the $(Q_2^{(0)}, Q_3^{(0)})$ eigenvalue pairs $\{(E_n, F_n)\}$ form a non-degenerate set.
This implies that the corresponding two diagonal matrices can generate an arbitrary real-valued diagonal matrix in the algebra sense, where we can form arbitrary powers and products of the two matrices (implicitly allowing also $0$-th powers which then includes the identity matrix) and take linear combinations of such products with real-valued coefficients.
[Another way to see this is to first find an $x$ such that $Q_2^{(0)} + x Q_3^{(0)}$ is a non-degenerate diagonal matrix and note that such a matrix can generate the entire algebra of arbitrary diagonal matrices.]
In particular, this implies that any other $SU(2)$-symmetric IoM is diagonal in the same basis and can be algebraically generated from $Q_2^{(0)}$ and $Q_3^{(0)}$ as described above.
Since states in all the other $S^z_{\text{tot}}$ sectors can be obtained by acting with $S^{\pm}_{\text{tot}}$, the same algebraic expression then holds in the entire Hilbert space.
We conclude that for these system sizes, the rest of the IoMs, $\{ Q_\alpha^{(0)}, \alpha \geq 4 \}$, are functionally dependent on $Q_2^{(0)}$ and $Q_3^{(0)}$ (but the specific functional expressions depend on $L$).
It is interesting that for this large range of sizes (including all odd sizes that we could access) the $Q_2^{(0)}$ and $Q_3^{(0)}$ are sufficient to capture all the other IoMs; we are not aware of this observation in prior literature and are wondering if it can have some additional, perhaps quantitative, consequences.

Returning to general system sizes, among our accessed systems, for $L = 10, 12, 16, 20$ in the $S^z_{\text{tot}} = 0$ sector we find degeneracies that are not resolved by the pair $Q_2^{(0)}$ and $Q_3^{(0)}$.
Specifically, for $L = 10, 16$ we find two such degeneracies (one in the momentum $k=0$ sector and one in the $k=\pi$ sector) that get resolved when we add $Q_4^{(0)}$.
For $L = 12$ and  $L = 20$ we find one degeneracy in the $k=0$ sector that gets resolved once we add $Q_5^{(0)}$.

It appears that for any system size, we can resolve all the degeneracies if we add a sufficient number of the IoMs from $\{Q_\alpha^{(0) } \}$.
Once this is achieved, all discussions from the case when $Q_2^{(0)}$ and $Q_3^{(0)}$ were sufficient readily generalize, e.g., on the functional dependence of the other IoMs.
We can obtain one more interesting and useful corollary by first recalling that the IoMs with even $\alpha$ are even under both the time reversal $\Theta$ and inversion $I$ symmetries, while the IoMs with odd $\alpha$ are odd under both; hence all the IoMs are even under the combination $\Theta I$.
As a consequence of the preceding algebraic generation arguments, any real-valued diagonal matrix in the degeneracy-resolved basis must then be even under $\Theta I$.
Hence any Hermitian operator that is (even, odd) or (odd, even) under $(\Theta, I)$ must have zero diagonal in this basis, and we can use this for the AGPs whenever these happen to be their natural symmetries.

\subsection{Discussion of applications of the $X^{\text{AGPgen}}$}
\label{subapp:applications_XAGPgen}
We now discuss in more detail cases from the main text using the $X^{\text{AGPgen}}$ approach.
While it is easy to generalize Eq.~(\ref{eq:AGPgen}) to include information about more IoMs and their corrections, the formulation using only $\alpha = 2$ and $3$ is sufficient for most of the cases we consider.
For example, for sizes used in the AGP characterizations in terms of Pauli strings, it provides complete resolution for $L=6,7,8,9,11$, while it does not resolve only the two degeneracies for $L=10$ (and we checked that resolving this with $Q_4^{(0)}$ does not have any effect in our examples).
For a given weak integrability-breaking perturbation where we also know the correction $Q_3^{(1)}$ for the IoM $Q_3^{(0)}$ (e.g., from the infinite-system formulation, which is how we deduce it in the $V_{\text{bo}}$, $V_{\text{bi}}$, and $V_{\text{step}}$ cases and feed into the finite-size  $X^{\text{AGPgen}}$ calculations), we can thus reduce ambiguities in extracting the finite-size generator $X$ down to just the strictly diagonal part (this is a rigorous statement when we have complete resolution of degeneracies).

Furthermore, when the natural $(\Theta, I)$ symmetry quantum numbers for the generator are $(\text{odd}, \text{even})$ or $(\text{even}, \text{odd})$, according to the preceding algebraic arguments we can set the strictly diagonal matrix elements of the sought-for $X$ to zero, removing also this ambiguity.
This is the case for the perturbations $V_{\text{ex},2}$, $V_{\text{bo},1/2}$, $V_{\text{bi},1/2}$, $V_{\text{loc},2}$, and $V_{\text{step},0/1/2}$ (hence the generalized AGP allows complete recovery of the corresponding $X$), but not for the perturbations $V_{\text{ex},1/3}$ and $V_{\text{loc},1/3}$ that need separate treatment as in App.~\ref{app:brute-force-Xdiag}.

\subsubsection{Remarks on cases with $\XAGP = \XAGPgen$ for small sizes}
\label{subsubapp:AGPeqAGPgen}
Interestingly, in the cases of 
$V_{\text{ex},2}$, $V_{\text{bo},1}$, $V_{\text{bi},1}$, $V_{\text{loc},2}$, and $V_{\text{step},1}$, the $\XAGP$ extracted using the original method and the new $X^{\text{AGPgen}}$ agree to our numerical accuracy for the sizes $L=6$ to $11$ used in the Pauli string characterizations (this is the reason we did not need to present the generalized AGP formalism in the main text).
This means that the corresponding $X^{\text{AGPgen}}$s have zero block-diagonal parts with respect to the degenerate eigenspaces of $H_0$.
We are not able to argue this on general grounds and suspect that this is limited to such rather small system sizes.
Symmetries appear to play an important role, namely all of these cases have the AGP symmetries $(\Theta,I) = (\text{odd}, \text{even})$, which helps with the following partial understanding:
Most of the degenerate eigenspaces of $H_0$ are two-dimensional corresponding to degeneracy between $k$ and $-k \neq k$ (i.e., $k \neq 0, \pi$) momentum states -- these degeneracies are due to the non-commutation between the translation and inversion symmetries.
In such a subspace, we can find one basis vector with inversion eigenvalue $+1$ and the other with $-1$.
Working in this basis, an operator that is even under inversion cannot have off-diagonal matrix elements.
Furthermore, an operator that is odd under the time reversal has diagonal matrix elements equal to zero in this basis [we can see this, e.g., by using that $H_0$ is real-valued in the $\sigma^z$ basis and can be diagonalized as such, and that $\Theta$ in the presence of $SU(2)$ spin symmetry (assumed implicitly for all operators discussed) is related to $\Theta'$ which is simply complex conjugation in this basis, while odd-$\Theta'$ operators are pure imaginary in this basis].
Hence, Hermitian operators with $(\Theta,I) = (\text{odd}, \text{even})$ are identically zero within such two-dimensional subspaces.
Some special arguments like this may work also for other types of degenerate eigenspaces of $H_0$ for these small sizes, but we have not examined this in details.

On the other hand, for the cases of $V_{\text{bo},2}$, $V_{\text{bi},2}$, and $V_{\text{step},0/2}$ that all have $(\Theta,I) = (\text{even}, \text{odd})$, we find that $X^{\text{AGPgen}} \neq \XAGP$.
The difference is usually quite small numerically, as can be seen by comparing the second ($\XAGP$) and third ($X^{\text{AGPgen}}$) rows of panels in Fig.~\ref{fig:Vbo2} for $V_{\text{bo},2}$ and in Fig.~\ref{fig:Vbi2} for $V_{\text{bi},2}$.
Nevertheless, the generalized AGP provides important improvements which allow testing some analytical predictions.

Finally, we note that equality between $\XAGP$ and $X^{\text{AGPgen}}$ within two-dimensional degenerate eigenspaces of $H_0$ containing $k$ and $-k \neq k$ momentum states does not require definite $\text{even}$ or $\text{odd}$ inversion symmetry of $X$ if we are considering a translationally invariant system.
Indeed, the common basis of $Q_2^{(0)}$ and $Q_3^{(0)}$ assumed resolving all degeneracies must be contain precisely the $k$ and $-k$ momentum states (which must have $Q_3^{(0)}$ eigenvalues $\pm F \neq 0$).
Hence, any translationally invariant $Q_3^{(1)}$ will have zero matrix elements between these basis states, so the generalized AGP does not introduce anything new within this block compared to the original AGP.

We find numerically that $\XAGP = X^{\text{AGPgen}}$ for the perturbation $V_{\text{ex},3}$ for our small sizes, similarly to $V_{\text{bo/bi/step,1}}$ and unlike $V_{\text{bo/bi/step,2}}$; since $V_{\text{ex},3}$ has the same inversion symmetry but opposite time reversal compared to the $V_{\text{bi/bo/step},2}$ perturbations, the time reversal must be playing differentiating role here.
Thus we see that in the $V_{\text{ex},3}$ case, which has the opposite inversion to the $V_{\text{bo/bi/step,1}}$ cases, the translational invariance takes care of most of the instances of degenerate eigenspaces of $H_0$, while the time reversal and likely interplay with inversion somehow make other instances work out as well to achieve $\XAGP = X^{\text{AGPgen}}$.

\subsubsection{Exact generator for $V_{\text{step},0}$}
Let us now turn to applications of the generalized AGP approach.
One such application we mention in the main text is the recovery of the analytic finite-size generator for $V_{\text{step},0}$ (Sec.~\ref{subsubsec:step}) providing complete match (hence not needed to be shown in any figures), in contrast to the original AGP in Fig.~\ref{fig:vstep0}.
The rest of this paragraph provides some more technical details why this works for interested readers.
First, we note from Eq.~(\ref{eq:comm_tXbeta_Qalpha}) that for all $\alpha$ the same finite-size PBC generator $\widetilde{X}^{\beta=2}$ reproduces the desired $Q_\alpha^{(1)} = J_{2,\alpha;j_0}$ up to $J_{2,\alpha;\text{tot}}/L$.
The latter happens to be proportional to the IoM $Q_{\alpha+1}^{(0)}$, since in the Heisenberg chain we have recursive generation of the IoMs given in the infinite-system formalism by~\cite{Grabowski1994} $Q_{\alpha+1}^{(0)} \sim i[\mathcal{B}[Q_2^{(0)}], Q_\alpha^{(0)}] \sim J_{2,\alpha;\text{tot}}$, where the final expression is well-defined on finite PBC chains.
Thus, the term $J_{2,\alpha;\text{tot}}/L$ is already block-diagonal with respect to degenerate subspaces of any chosen set of IoMs in the generalized AGP setup.
We also note that when we consider strictly local perturbations or the corresponding generators, it is sufficient to have one inversion symmetry point for all symmetry arguments to work; this remark applies to all local perturbations and is implicit in all of the preceding discussions.
In the $V_{\text{step},0}$ case, both the perturbation Eq.~(\ref{eq:Vstep0}) and the corresponding exact generator $\widetilde{X}^{\beta=2}$ are odd under the inversion in site $j_0$ [the latter can be readily checked by examining Eq.~(\ref{eq:tildeXbeta}) with the amplitudes Eq.~(\ref{eq:cj_tildeXbeta}) and our $q_{2,j}^{(0)}$, and this is where the specific inversion-odd form of the amplitudes is important; such careful inversion symmetry considerations are often implicit in our discussions].
Thus, in this somewhat non-trivial example, we have a complete understanding of  ``missing parts'' when trying to find a ``simple'' generator using the AGP measurements and how to fix this using the generalized AGP.

\subsubsection{Exact relation between generalized AGPs for $V_{\text{bi}}$ and $V_{\text{bo}}$}
Another application of the $X^{\text{AGPgen}}$ approach in the main text is for checking analytic relations between generators for $V_{\text{bi},a}$ and $V_{\text{bo},a}$, $a=1,2$, in  Sec.~\ref{subsubsec:AGPVbi} [Eq.~(\ref{eq:Xbi1_Xbo1}) for $a=1$ and Eq.~(\ref{eq:Xbi2_Xbo2}) for $a=2$], where we would like to have simpler relations not involving knowledge of $(\widetilde{X}^{2,3})_{\text{diag}}$ and $(\widetilde{X}^{2,4})_{\text{diag}}$ respectively -- ideally where such terms would be absent, e.g., if our task were to recover the simple $\widetilde{X}^{\beta\gamma}$ from just the AGP-type measurements.
As mentioned in the main text, this indeed works for the original AGP in the $a=1$ case for our small sizes, where we now know that $\XAGP = X^{\text{AGPgen}}$.
On the other hand, we do need the generalized AGP in the $a=2$ case where $\XAGP \neq X^{\text{AGPgen}}$, and the precise relation confirmed numerically in this case reads
\begin{align}
X^{\text{AGPgen}}_{\text{bi},2} = \widetilde{X}^{2,4} + \frac{1}{L} \{X^{\text{AGPgen}}_{\text{bo},2}, Q_2^{(0)} \} ~.
\label{eq:XAGPgen_relation_Vbi2_Vbo2}
\end{align}

\subsubsection{Exact relation between generalized AGPs for $V_{\text{step}}$ and $V_{\text{bo}}$}
A similar application is for checking analytic relations between generators for $V_{\text{step},a}$ and $V_{\text{bo},a}$, $a=1,2$, in Sec.~\ref{subsubsec:step}: Eq.~(\ref{eq:XAGP4Vstep}) with $\beta=3$ for $a=1$ and with $\beta=4$ for $a=2$, where again we would like to have simpler relations not involving knowledge of $(\widetilde{X}^{\beta=3})_{\text{diag}}$ and $(\widetilde{X}^{\beta=4})_{\text{diag}}$, i.e., where the corresponding terms would be absent.
This indeed works for the original AGP in the $a=1$ case as exhibited in Eq.~(\ref{eq:XAGPVstep1}), where we now know $\XAGP = X^{\text{AGPgen}}$ for our small sizes.
On the other hand, the $a=2$ case does require the $X^{\text{AGPgen}}$, and the precise relation checked numerically reads
\begin{equation}
X^{\text{AGPgen}}[V_{\text{step},2}] = \widetilde{X}^{\beta=4} + \frac{1}{L} X^{\text{AGPgen}}[V_{\text{bo},2}] ~.
\label{eq:XAGPgen_relation_Vstep2_Vbo2}
\end{equation}

\subsubsection{Exact relation between generalized AGPs for $V_{\text{loc}}$ and $V_{\text{ex}}$}

\begin{figure}[ht]
\centering
\includegraphics[width=0.47\linewidth]{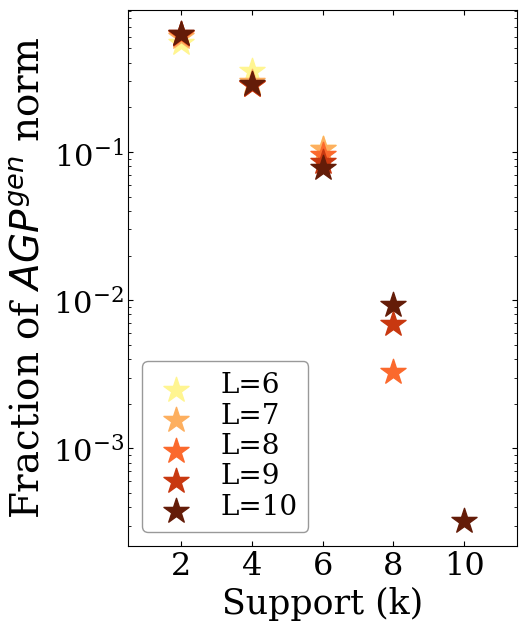}
\includegraphics[width=0.47\linewidth]{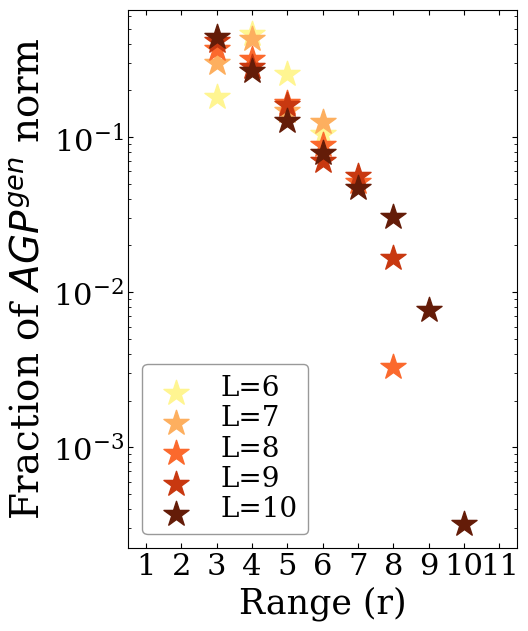}
\caption{Pauli string decomposition of the generalized AGP $X^{\text{AGPgen}}$ for $V_{\text{ex},1}$, showing only log-scale for the $y$-axis.
We see small numerical difference from the original AGP shown in Fig.~\ref{fig:Vex1}.
Note that here we use step regulator such that the $X^{\text{AGPgen}}$ is cutoff-independent for subsequent use to check exact relation Eq.~(\ref{eq:diagrelXloc1Xex1}), while in the main text we used the smooth regulator, but we have checked that the conclusion $\XAGP \neq X^{\text{AGPgen}}$ does not depend on this.
}
\label{fig:XAGPgen_Vex1}
\end{figure}

\begin{figure}[ht]
\centering
\includegraphics[width=0.47\linewidth]{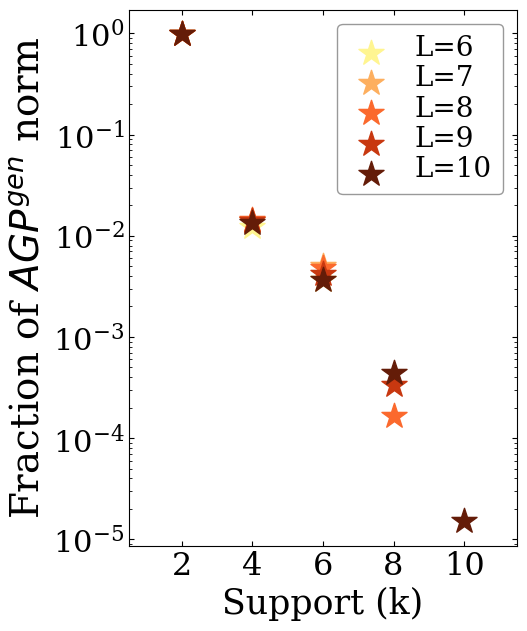}
\includegraphics[width=0.47\linewidth]{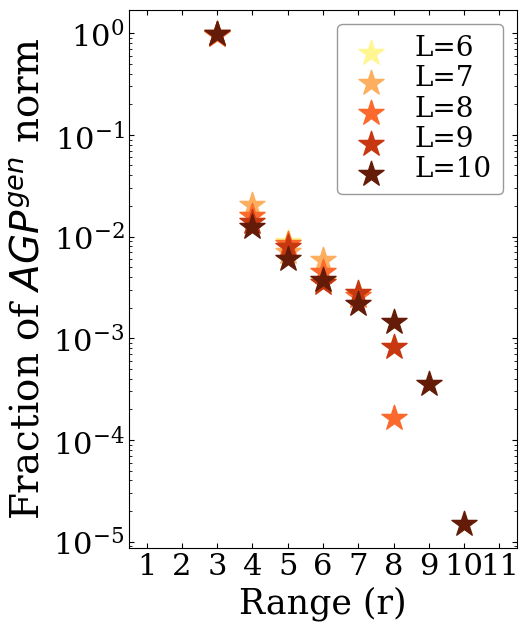}
\caption{Pauli string decomposition of the $X^{\text{AGPgen}}$ for $V_{\text{loc},1}$, showing only log-scale for the $y$-axis.
We see small numerical difference from the original AGP shown in Fig.~\ref{fig:Vloc1} (note that we use step regulator here vs smooth regulator in the main text, but this is not important).
Note close similarity of the patterns here and in Fig.~\ref{fig:XAGPgen_Vex1} for the $X^{\text{AGPgen}}[V_{\text{ex},1}]$; in fact, these satisfy exact relation Eq.~(\ref{eq:diagrelXloc1Xex1}), and the generalized AGP was crucial to achieve this.}
\label{fig:XAGPgen_Vloc1}
\end{figure}

One more application of the $X^{\text{AGPgen}}$ approach in the main text is to verify exact relations Eqs.~(\ref{eq:diagrelXloc1Xex1}) and (\ref{eq:diagrelXloc3Xex3}).
The explanations following these equations in the main text should be much more clear now following the developments in this Appendix.
Figure~\ref{fig:XAGPgen_Vex1} shows the $X^{\text{AGPgen}}[V_{\text{ex},1}]$, which is indeed different from $\XAGP[V_{\text{ex},1}]$ in Fig.~\ref{fig:Vex1}, though the difference is quantitatively fairly small.
We similarly see difference between $\XAGPgen[V_{\text{loc},1}]$ in Fig.~\ref{fig:XAGPgen_Vloc1} and $\XAGP[V_{\text{loc},1}]$ in Fig.~\ref{fig:Vloc1}.
Note that Figs.~\ref{fig:XAGPgen_Vex1} and \ref{fig:XAGPgen_Vloc1} are calculated using the step regulator to avoid any cutoff dependence and essentially work with the exact $X^{\text{AGPgen}}$ (up to roundoff errors).
These are then used to verify that Eq.~(\ref{eq:diagrelXloc1Xex1}) is satisfied within our numerical accuracy.
This identity can be glimpsed by comparing Figs.~\ref{fig:XAGPgen_Vex1} and \ref{fig:XAGPgen_Vloc1} where, we see the same patterns for weights on strings other than the dominant $k=2, r=3$ strings from the exact $X_{\text{ex},1}$ and $X_{\text{loc},1}$ as expected, while the exact relation also involves an $L$-dependent factor.

\section{Brute force recovery of  $X_{\text{diag}}$}
\label{app:brute-force-Xdiag}
For perturbations $V_{\text{ex},1}$ and $V_{\text{ex},3}$, the generator $X$ is known, but the $\XAGP$ we find is not exactly this $X$. 
This is because of the arbitrariness in the diagonal part in the definition of AGP: Since in defining $\XAGP$ in Eq.~(\ref{eq:AGPod}) we set the (block-)diagonal part to zero, we recover $X$ up to its diagonal part.
This is true even when using our method for improving the recovery of $X$ by partially fixing its block-diagonal parts arising from degeneracies in $H_0$, described in App.~\ref{app:sim_diag}:
While this procedure can fix all the elements $X^{\text{AGPgen}}_{nm}$ with $n \neq m$, the strictly diagonal ones $n = m$ are still set to $0$.

Here we explore if one can recover the diagonal from a simple minimization scheme.
The scheme aims at making the operator $X^{\text{AGPgen}}$ as local as possible by adding terms to the diagonal part, where locality is loosely defined as having the AGP contain Pauli strings with small support or range.
One can define 
\begin{equation}
X' = X^{\text{AGPgen}} + \sum_n c_n \ket{E_n}\bra{E_n} ~,  
\end{equation}
where the diagonal part is written in the common eigenbasis of the IoMs.
We can express the $X^{\text{AGPgen}}$ and the eigenstate projectors in the Pauli string basis as
\begin{align}
& X^{\text{AGPgen}} = \sum_S a_S O_S ~, \\
& \ket{E_n}\bra{E_n} = \sum_S b_{S,n} O_S ~,
\end{align}
where the sum goes over all Pauli strings labelled by $S$ as in Eq.~(\ref{eq:PauliSbasis}), with $O_S$ being shorthand for the corresponding operator.

We get 
\begin{equation}
X' = \sum_S \left(a_S + \sum_n b_{S,n} c_n \right) O_S ~,
\end{equation}
where $\{c_n \}$ is the vector of parameters to optimize.
Since we want Pauli strings with large support to be costly, we can choose the cost function $C$ to be, e.g.,
\begin{equation}
C = \sum_S \left(a_S + \sum_n b_{S,n} c_n \right)^2 k_S ~
\end{equation}
where $k_S = \mathrm{supp}(O_S)$.
Minimizing $C$ is equivalent to solving a system of linear equations since $C$ is quadratic in
$\{c_n \}$.

In our numerical studies, this procedure allows us to approximately recover $X_{\text{ex},1}$ from $\XAGP[V_{\text{ex},1}]$; the recovery is not complete and is a sizeable fraction off because of the soft character of the penalty on the larger strings. 
Better results can be obtained by defining the cost function with higher powers of the support.
For example, one can replace $k_S$ with the $m$-th power $(k_S)^m$ (and $m = 2, 3, \dots$) to further penalize the presence of strings with higher support.
With this modification and taking high powers $m$, we are able to recover almost the exact $X_{\text{ex},1}$.
Clearly, there is some arbitrariness in such cost function choices and uncertainty about the right one unless we know the character of the contributing Pauli strings, but with some experimentation this approach can indeed be used to address the missing diagonal problem in the AGPs.

While we have focused on the missing strictly diagonal matrix elements in the $X^{\text{AGPgen}}$, we can in principle apply a similar procedure to the original $\XAGP$ allowing addition of arbitrary block-diagonal matrices (with respect to eigenspaces of $H_0$) to construct $X'$ and trying to optimize over those to obtain as local $X'$ as possible.
This could be a black-box alternative to the approach in App.~\ref{app:sim_diag}, but it also increases the number of free parameters and again suffers from the above-mentioned uncertainties choosing a good cost function.
Hence, in most cases we use the generalized AGP approach when discussing recovering exact generators or exact relations among AGPs in the main text.

\section{Operator Participation Ratio (OPR) study of the AGPs}
\label{app:oipr}
\begin{figure}[ht]
\centering
\includegraphics[width=0.98\linewidth]{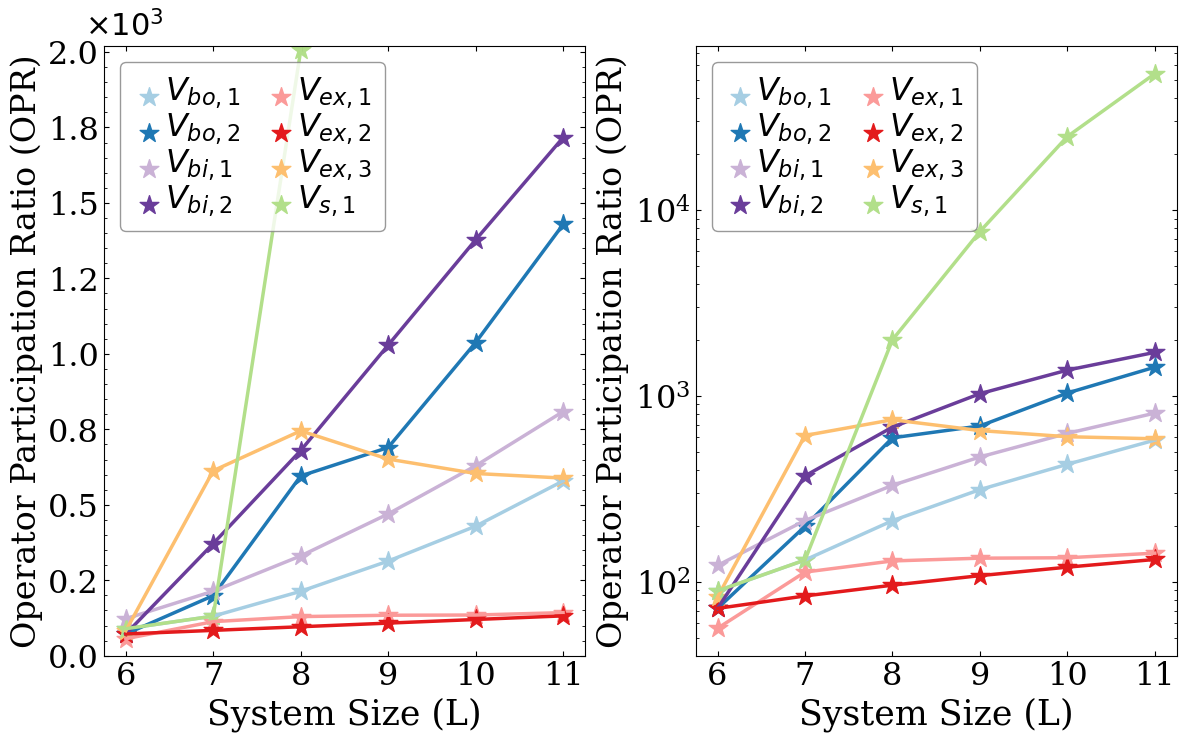}
\caption{Comparison of the operator participation ratio as defined in Eq.~(\ref{eq:oipr})  for the AGPs of all translationally invariant perturbations considered in Fig.~\ref{fig:agp-both-global}. 
These results additionally support our discussion and findings.}
\label{fig:OIPR}
\end{figure}

An additional metric we consider when looking at the $\XAGP$s in the Pauli string basis is the Operator Participation Ratio (OPR), which we define as 
\begin{equation}
\text{OPR}(\XAGP) = \left(\sum_S |w_\text{S}|^2 \right)^{-1}~,
\label{eq:oipr}
\end{equation}
where $w_{\text{S}}$ are the weights in the Pauli string basis, as defined in Eq.~(\ref{eq:wS}) with the total weight normalized to one. 
The OPR quantifies the spreading of the $\XAGP$ in the Pauli string basis.
It provides additional information compared to the AGP visualizations in the main text where different Pauli strings with the same size of the support or with the same length of the range are lumped together, while all different Pauli strings are distinguished in the OPR.
Figure~\ref{fig:OIPR} shows this metric of the AGPs for all of the translationally invariant perturbations, which allows us to compare and contrast them on one plot and provides a summative to our examinations of their structure in the Pauli string basis.

As we see in Fig.~\ref{fig:OIPR}, the strong integrability breaking perturbations are characterized by a fast exponential growth of the OPR with the system size, indicating that the $\XAGP$ is spread across many Pauli string operators. 
For weak integrability breaking perturbations, on the other hand, the OPRs grow as power laws, consistent with our results of Sec.~\ref{subsec:agp_PauliS2} and supporting a quasilocal character of the $\XAGP$ for such perturbations.
For the simplest such perturbations, namely $V_{\text{ex}}$-type, we expect the AGP OPR to grow as $L$, and in the $V_{\text{ex},2}$ case where the AGP recovers the exact $X_{\text{ex},2}$ the corresponding OPR in the figure is exactly $12 L$.
On the other hand, the instances $V_{\text{ex},1}$ and $V_{\text{ex},3}$ show strong and not fully systematic finite-size behaviors, presumably because of the missing diagonal problem in their $\XAGP$s which we kept as-is in the main text and here.

Finally, the $V_{\text{bo}}$- and $V_{\text{bi}}$-type perturbations show a faster power law increase of the corresponding AGP OPRs, but the available size range is clearly too small for accurate power law extraction.
To put this in perspective, the operators $\tilde{X}^{\beta\gamma}$ that are good approximations to the AGPs for $V_{\text{bi}}$-type perturbations have the OPR scaling as $\sim L^2$, and our numerical results are roughly consistent with such scaling (e.g., could be fit to $a + bL^2$ but also to more general $a + bL^p$).
We also see that the AGPs for the $V_{\text{bo}}$-type perturbations have the OPR that grows more weakly than for the $V_{\text{bi}}$-type perturbations, suggesting stronger quasi-localization in the former case compared to the latter case.

\section{Additional data for AGP studies of extensive and strictly local perturbations }
\label{app:add_data_for_agp_global}

\begin{figure}[ht]
\centering
\includegraphics[width=0.45\linewidth]{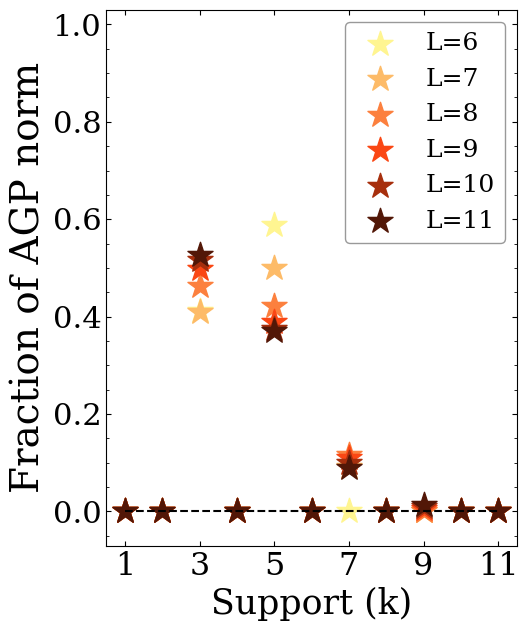}
\includegraphics[width=0.45\linewidth]{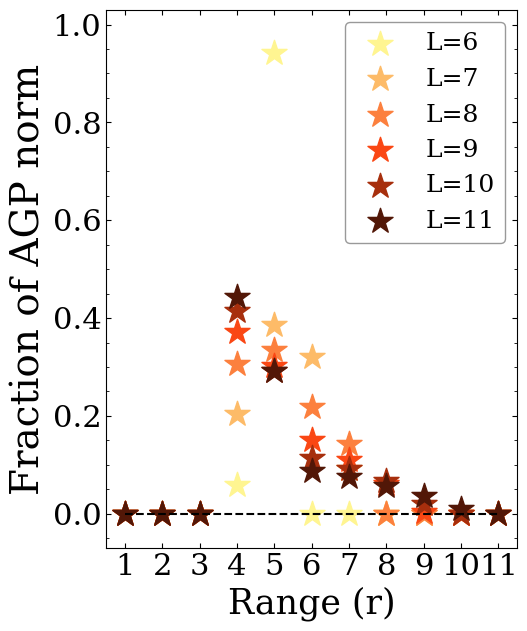}
\includegraphics[width=0.45\linewidth]{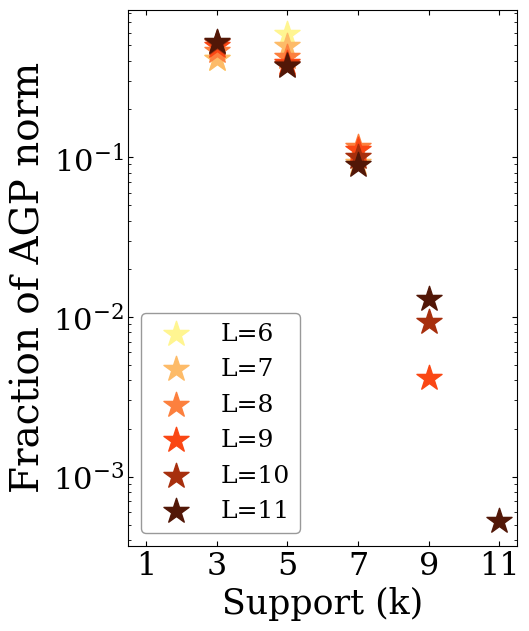}
\includegraphics[width=0.45\linewidth]{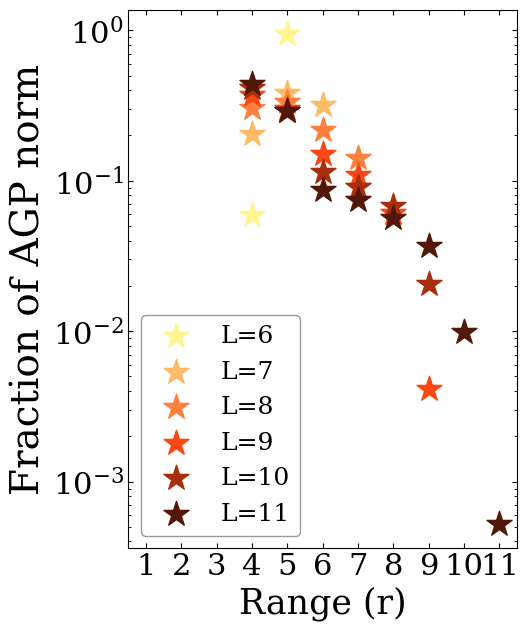}
\caption{Visualization of the structure of the $\XAGP$ in terms of the decomposition in the Pauli string basis for the perturbation $V_{\text{ex},3}$.}
\label{fig:Vex3}
\end{figure}

\begin{figure}[ht]
\centering
\includegraphics[width=0.45\linewidth]{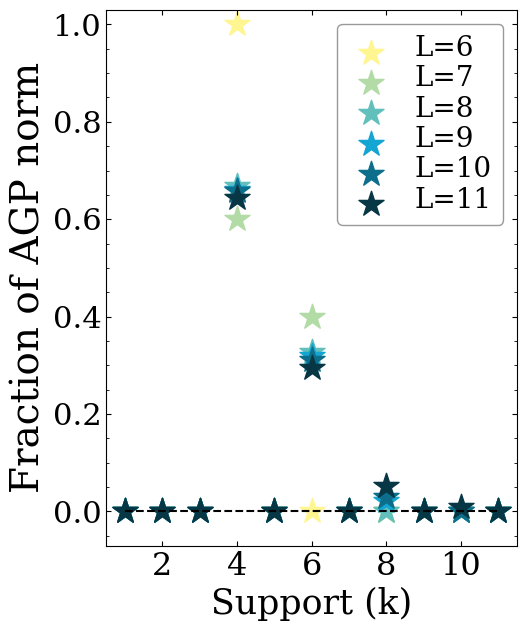}
\includegraphics[width=0.45\linewidth]{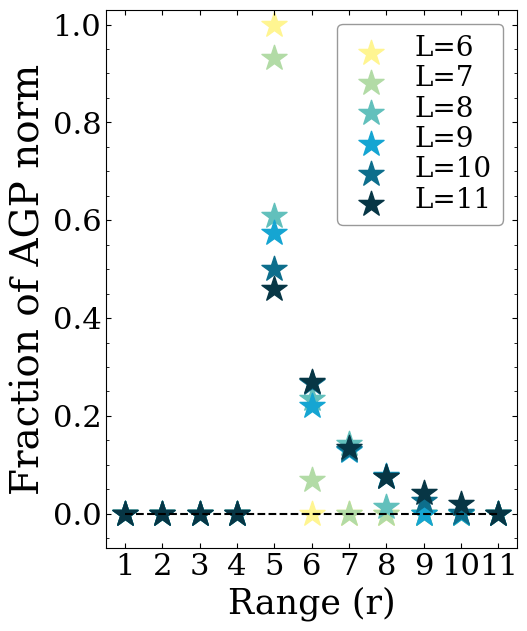}
\includegraphics[width=0.45\linewidth]{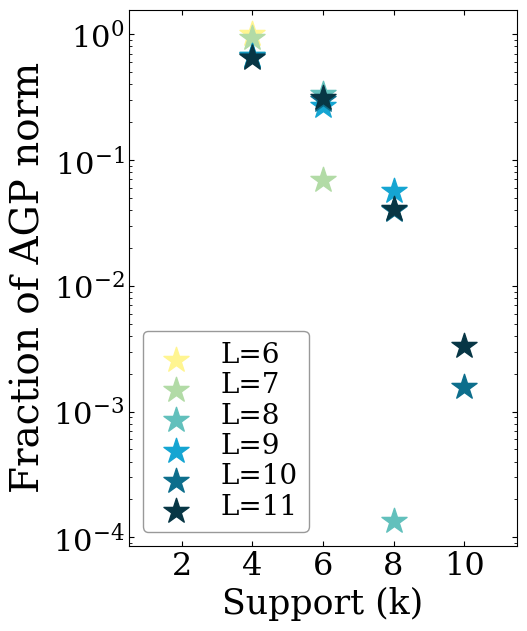}
\includegraphics[width=0.45\linewidth]{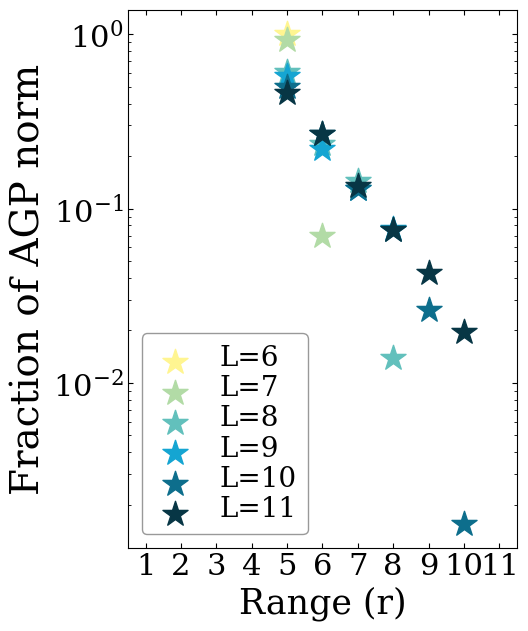}
\includegraphics[width=0.45\linewidth]{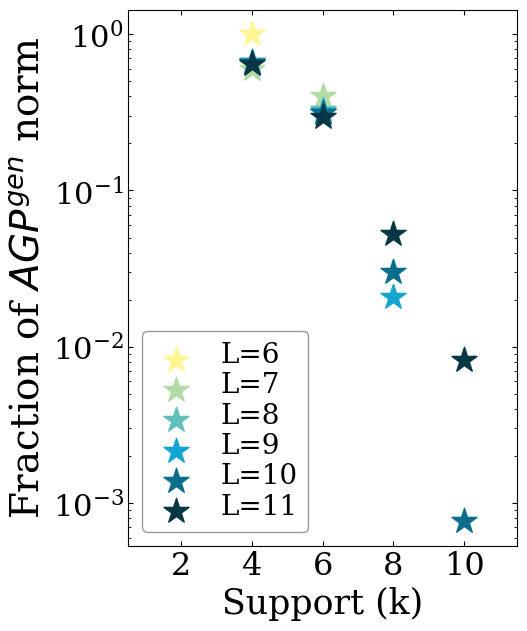}
\includegraphics[width=0.45\linewidth]{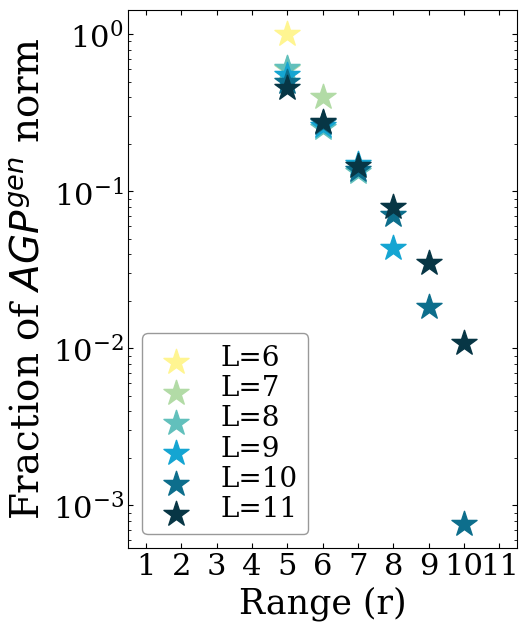}
\caption{Visualization of the structure of the $\XAGP$ (first and second rows, linear and log scales respectively) and the $X^\text{AGPgen}$ (third row, log scale) in terms of the decomposition in the Pauli string basis for the perturbation $V_{\text{bo},2}$.
The $X^\text{AGPgen}[V_{\text{bo},2}]$ from this figure together with $X^\text{AGPgen}[V_{\text{bi},2}]$ from Fig.~\ref{fig:Vbi2} satisfy relation Eq.~(\ref{eq:XAGPgen_relation_Vbi2_Vbo2}), while together with $X^\text{AGPgen}[V_{\text{step},2}]$ from Fig.~\ref{fig:Vstep2} satisfy Eq.~(\ref{eq:XAGPgen_relation_Vstep2_Vbo2}).
}
\label{fig:Vbo2}
\end{figure}

\begin{figure}[ht]
\centering
\includegraphics[width=0.45\linewidth]{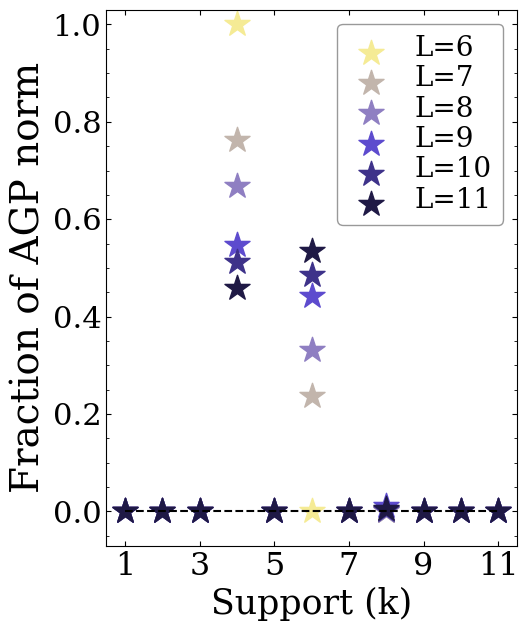}
\includegraphics[width=0.45\linewidth]{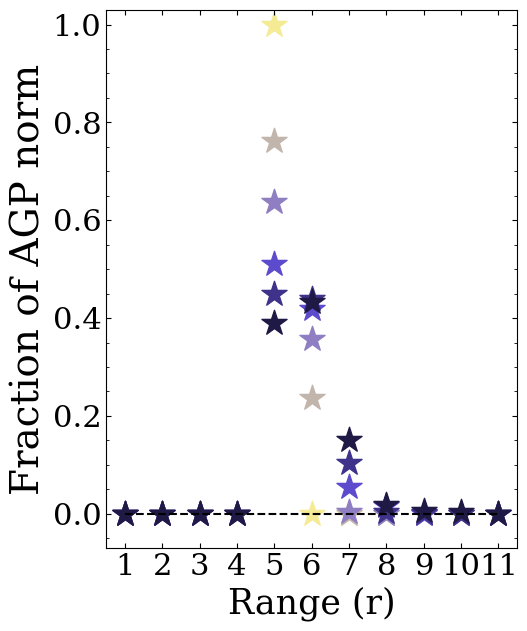}
\includegraphics[width=0.45\linewidth]{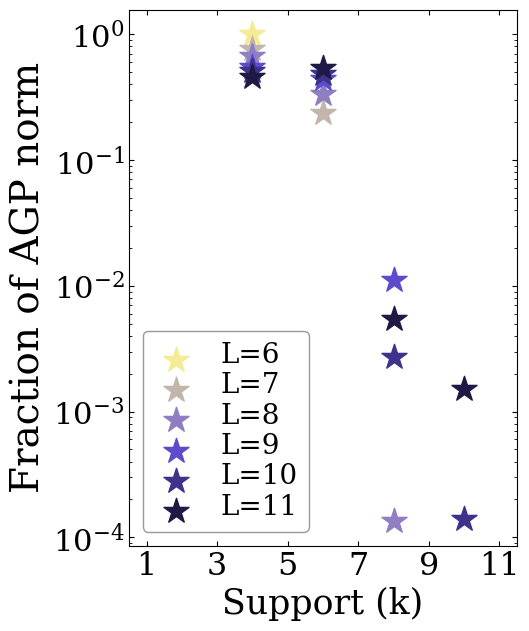}
\includegraphics[width=0.45\linewidth]{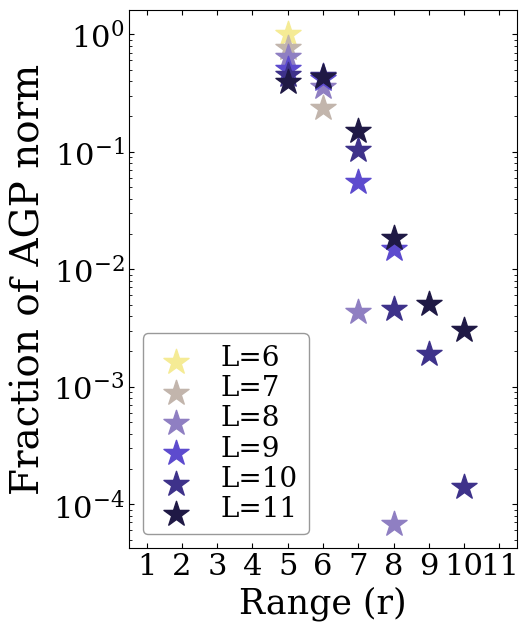}
\includegraphics[width=0.45\linewidth]{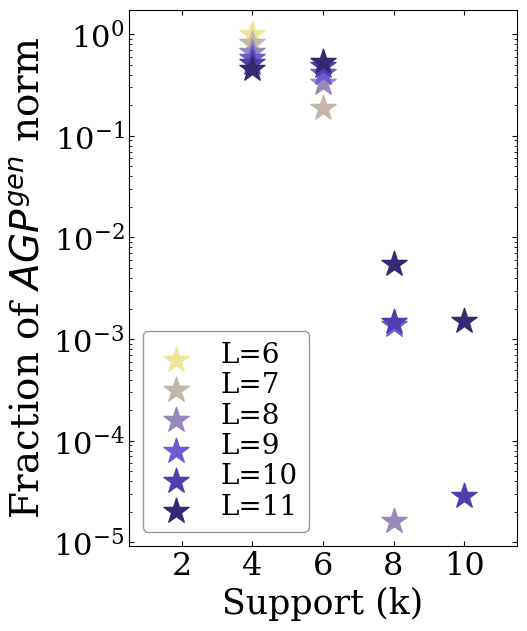}
\includegraphics[width=0.45\linewidth]{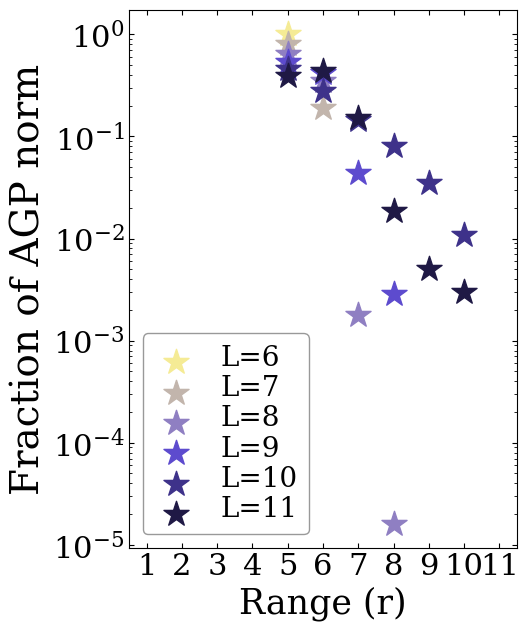}
\caption{Visualization of the structure of the $\XAGP$ (first and second rows) and the $X^\text{AGPgen}$ (third row) in terms of the decomposition in the Pauli string basis for the perturbation $V_{\text{bi},2}$.
The $X^\text{AGPgen}[V_{\text{bi},2}]$ from this figure together with $X^\text{AGPgen}[V_{\text{bo},2}]$ from Fig.~\ref{fig:Vbo2} satisfy relation Eq.~(\ref{eq:XAGPgen_relation_Vbi2_Vbo2}).
}
\label{fig:Vbi2}
\end{figure}

\begin{figure}[h]
\centering
\includegraphics[width=0.45\linewidth]{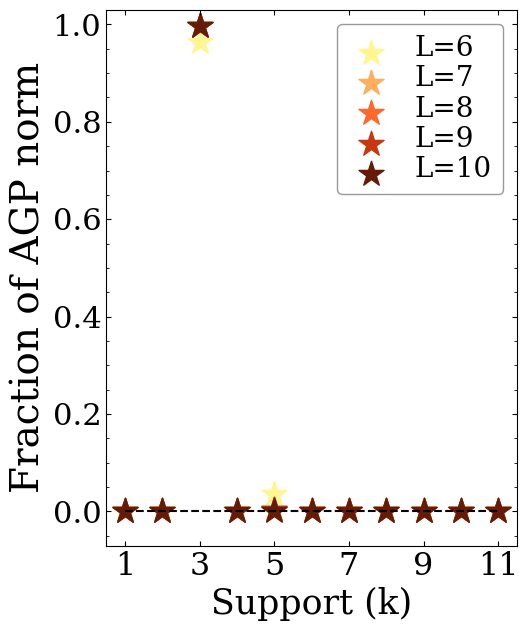}
\includegraphics[width=0.45\linewidth]{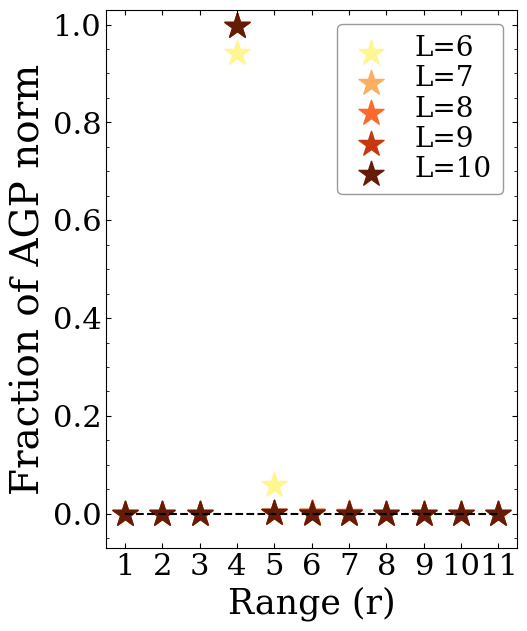}
\includegraphics[width=0.45\linewidth]{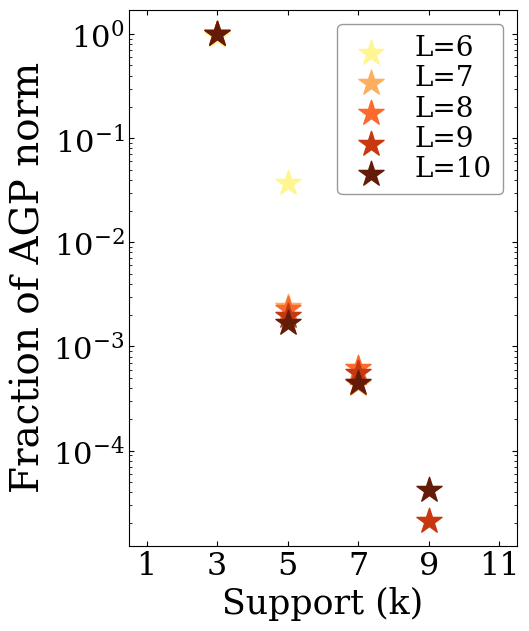}
\includegraphics[width=0.45\linewidth]{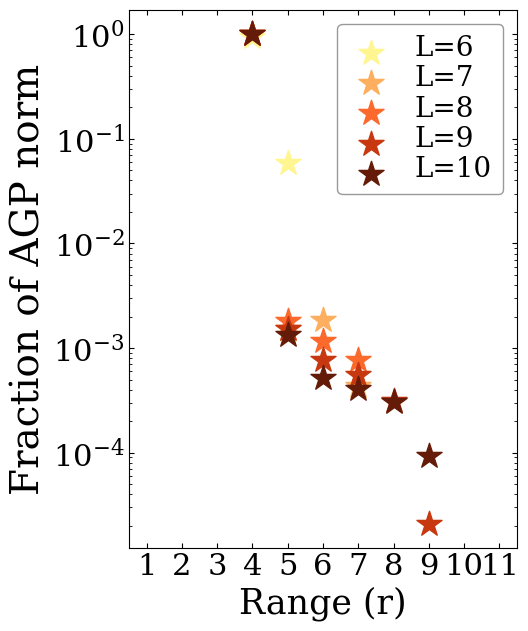}
\caption{Visualization of the structure of the $\XAGP$ in terms of the decomposition in the Pauli string basis for the perturbation $V_{\text{loc},3}$.
}
\label{fig:Vloc3}
\end{figure}

\begin{figure}[ht]
\centering
\includegraphics[width=0.45\linewidth]{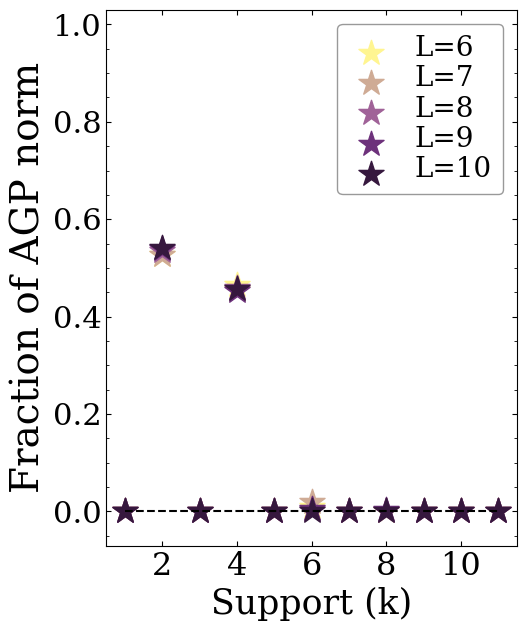}
\includegraphics[width=0.45\linewidth]{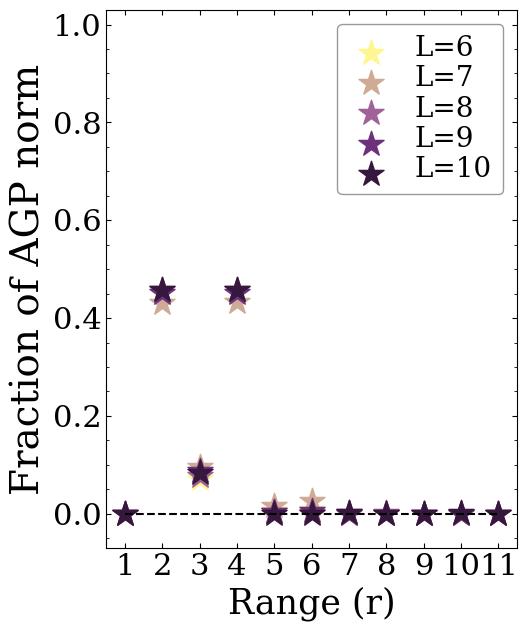}
\includegraphics[width=0.45\linewidth]{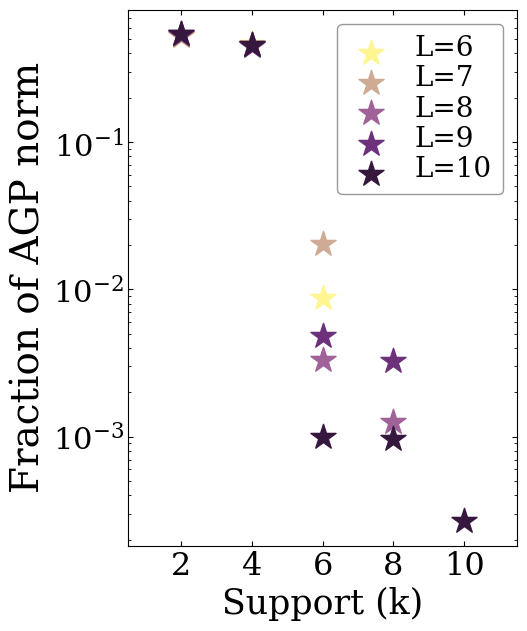}
\includegraphics[width=0.45\linewidth]{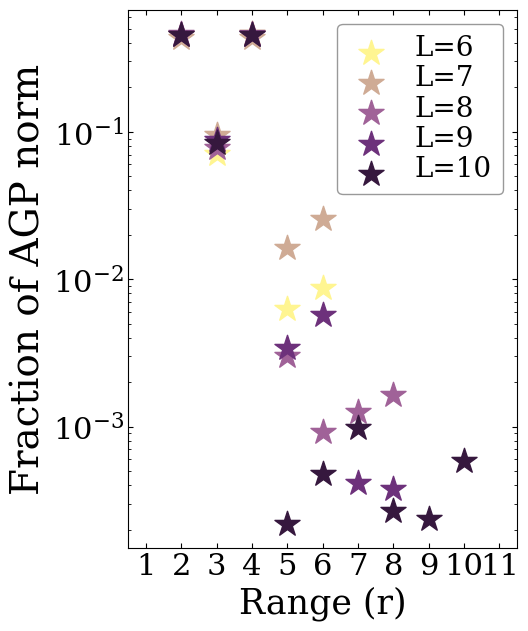}
\includegraphics[width=0.45\linewidth]{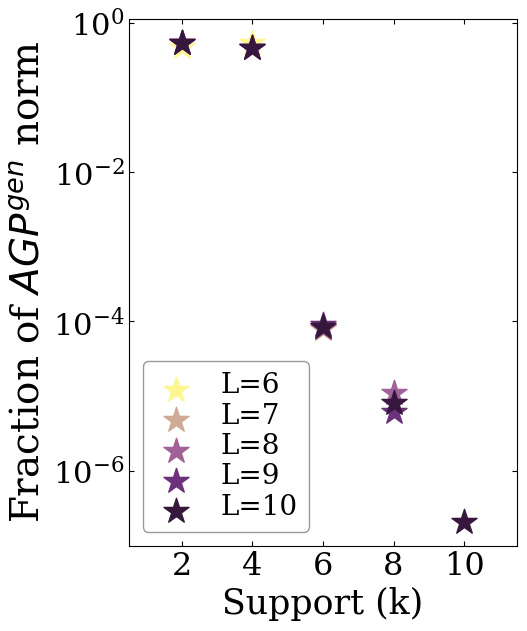}
\includegraphics[width=0.45\linewidth]{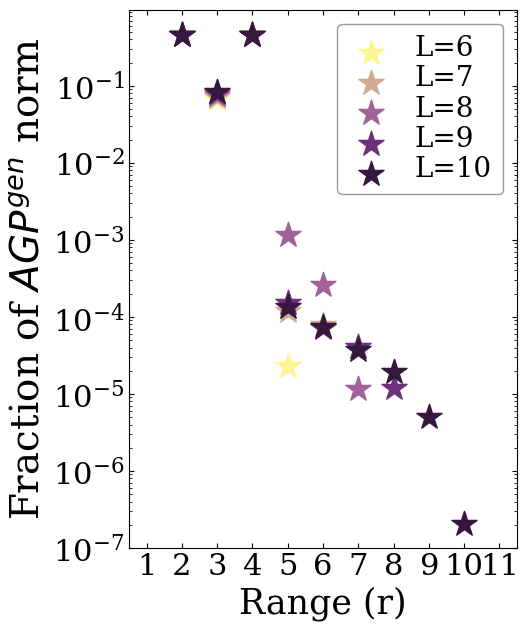}
\caption{Visualization of the structure of the $\XAGP$ (first and second rows) and the $X^\text{AGPgen}$ (third row) in terms of the decomposition in the Pauli string basis for the perturbation $V_{\text{step},2}$.
The $X^\text{AGPgen}[V_{\text{step},2}]$ from this figure together with $X^\text{AGPgen}[V_{\text{bo},2}]$ from Fig.~\ref{fig:Vbo2} satisfy relation Eq.~(\ref{eq:XAGPgen_relation_Vstep2_Vbo2}).
}
\label{fig:Vstep2}
\end{figure}

In Sec.~\ref{subsec:agp_PauliS2}, we presented the data of the characterization of $\XAGP$ in the Pauli string basis for some of the extensive perturbations we consider and in Sec.~\ref{subsec:agpPauliS_local} for some of the strictly local ones. 
In Figs.~\ref{fig:Vex3}, \ref{fig:Vbo2}, and \ref{fig:Vbi2},
we additionally present data for the perturbations $V_{\text{ex},3}$, $V_{\text{bo},2}$, and $V_{\text{bi},2}$ respectively. Coincidentally, the last two are exactly the ones that benefit from the $X^{\text{AGPgen}}$ approach, so in addition to the four panels we usually show in the figures ($\XAGP$ on linear and log scale, vs support $k$ and range $r$) we add the Pauli string characterization on the logarithmic scale of $X^{\text{AGPgen}}$. As discussed in App.~\ref{subsubapp:AGPeqAGPgen}, for $V_{\text{ex},3}$ the AGP and generalized AGP are the same for our sizes. We also present data for $V_{\text{loc,3}}$ and $V_{\text{step,2}}$ in Fig.~\ref{fig:Vloc3} and Fig.~\ref{fig:Vstep2}, which both include also the $\XAGPgen$ approach. In all cases where we show $\XAGPgen$, we can see that $X^{\text{AGPgen}} \neq \XAGP$, but the difference is rather small numerically.
Nevertheless, this improvement is needed to satisfy some analytical relations, listed individually in the figure captions where applicable.

\vfill\eject

\newpage

\vfill\eject

\newpage

\end{document}